\documentclass[aps,onecolumn,groupedaddress,showpacs,notitlepage,nofootinbib]{revtex4-1}

\usepackage{graphicx}
\usepackage{dcolumn}
\usepackage{bm}
\usepackage{amssymb}
\usepackage{mathrsfs}
\usepackage{amsmath}
\usepackage{epsfig}
\usepackage{color}
\usepackage{hhline}
\usepackage{hyperref}
\usepackage{comment} 
\usepackage{slashed} 

\allowdisplaybreaks[4]

\begin{document}

\title{Constraining the interaction strength between dark matter and
visible matter: II. scalar, vector and spin-3/2 dark matter}

\author{Zhao-Huan Yu$^{1,2}$}
\author{Jia-Ming Zheng$^{1,3}$}
\author{Xiao-Jun Bi$^{2}$}
\email[Email:~]{bixj@mail.ihep.ac.cn}
\author{Zhibing Li$^1$}
\author{Dao-Xin Yao$^1$}
\author{Hong-Hao Zhang$^1$}
\email[Email:~]{zhh98@mail.sysu.edu.cn}

\affiliation{$^1$School of Physics and Engineering, Sun Yat-Sen University,
Guangzhou 510275, China\\ $^2$Key Laboratory of Particle Astrophysics,
Institute of High Energy Physics, Chinese Academy of Sciences, Beijing
100049, China\\ $^3$School of Physics and Astronomy, University of Minnesota,
MN 55455, USA}

\begin{abstract}
We investigate the constraints on the scalar, vector and spin-3/2 dark matter
interaction with the standard model particles, from the observations
of dark matter relic density, the direct detection experiments of
CDMS and XENON, and the indirect detection of the $\bar p/p$ ratio by
PAMELA. A model independent way is adopted
by constructing general 4-particle operators up to dimension 6 for
the effective interaction between dark matter and standard model particles.
We find that the constraints from different experiments
are complementary with each other. Comparison among these constraints
may exclude some effective models of dark matter and
limit some parameters of others.
The spin-independent direct detection gives strong
constraints for some operators, while the indirect detection
of $\bar p/p$ data can be more sensitive than direct detection
or relic density for light dark matter (whose mass $\lesssim 70$~GeV)
in some cases. The constraints on some operators for spin-3/2 dark matter
are shown to be similar to those on their analogous operators for Dirac
fermionic dark matter. There are still some operators not sensitive to
the current dark matter direct and indirect search experiments.
\end{abstract}


\maketitle

\section{Introduction}

It is by now established~\cite{Tegmark:2006az,Komatsu:2008hk,Komatsu:2010fb}
that about 23\% of the constituents of the Universe are composed of
dark matter (DM)~\cite{kolb-turner,Jungman:1995df,Bertone:2004pz,
Murayama:2007ek,Feng:2010gw}.
However, its nature remains unclear. A well-motivated
candidate for DM is the weakly interacting massive particle
(WIMP), which must be stable, nonrelativistic, electrically
neutral and colorless. If the WIMP mass is from a few GeV to
TeV while their interaction strength is of the weak scale, they can
naturally yield the observed relic density of DM~\cite{Feng:2010gw}.
Although there is no stable WIMP in the Standard Model (SM),
WIMP candidates exist in various theoretical models trying to solve the
SM problems at the weak scale, such as supersymmetric models
\cite{Jungman:1995df,Goldberg:1983nd,Ellis:1983ew,Arnowitt:1992aq,Nath:1992ty,
Kane:1993td}, extra dimensional models~\cite{Kolb:1983fm,Cheng:2002ej,
Hooper:2007qk,Servant:2002aq,Servant:2002hb,Agashe:2004ci,
Agashe:2004bm,Agashe:2007jb}, little Higgs models
\cite{Cheng:2004yc,Low:2004xc,Birkedal:2006fz,Freitas:2009jq,Kim:2009dr},
left-right symmetric models~\cite{Dolle:2007ce,Guo:2008hy,Guo:2008si,
Guo:2010vy}, and some other models (e.g.~\cite{Khlopov:2008ty,Li:2010rz,
Cui:2011wk,Kanemura:2011mw}).

The specific models mentioned above are very attractive, but still lack
experimental support. The well-running LHC experiment may find some important
signals of these models in the near future. However, if other new particle
species are all so heavy that the DM particle is the only new particle
within the reach of LHC, it will be very difficult to know which model
the DM particle belongs to. In addition, it is
possible that the DM particle may be first observed by direct or indirect
detection experiments. These early observations may only provide
information about some general properties of the DM particle, and
may not be able to distinguish the underlying theories. Therefore,
the model-independent studies of the DM phenomenology can play an
important role as they may avoid theoretical bias
\cite{Birkedal:2004xn,Giuliani:2004uk,Kurylov:2003ra,Beltran:2008xg,
Cirelli:2008pk,Shepherd:2009sa}.
Recently there have been quite a few papers to study
various phenomenologies related with DM
in the model-independent way~\cite{Cao:2009uv,Cao:2009uw,
Beltran:2010ww,Fitzpatrick:2010em,Goodman:2010yf,Bai:2010hh,
Goodman:2010ku,Goodman:2010qn,Bell:2010ei,Zheng:2010js,Cheung:2010ua,
Cheung:2011nt,Mambrini:2011pw,Rajaraman:2011wf,Fox:2011pm,Goodman:2011jq,
Kamenik:2011vy,Shoemaker:2011vi}.
Especially the relic density measured by WMAP~\cite{Komatsu:2010fb},
direct detection from CDMS~\cite{Ahmed:2009zw,Akerib:2005za} and XENON
\cite{Aprile:2011hi,Angle:2008we}, and possible collider signals from
Tevatron (e.g.~\cite{Aaltonen:2008hh,CDF:monojet}) and from LHC (e.g.
\cite{Chatrchyan:2011nd,ATLAS:monojet}) are considered in these studies.

In our previous work~\cite{Zheng:2010js}, we investigated a general set
of 4-fermion operators for the effective interaction between
the spin-1/2 fermionic DM and the SM particles,
and gave the phenomenological constraints from
the observed DM relic density, the direct detection experiments by
CDMS and XENON, and the indirect detection of the $\bar{p}/p$
ratio by PAMELA~\cite{Adriani:2010rc}. It was found that the constraints
from different observations are quite complementary.
Besides the possibility of spin-1/2 WIMPs, it is also possible that DM is
composed of scalar, vector, or spin-3/2 WIMPs, which belong
to different representations of Lorentz group. In these cases,
the forms of possible effective operators are different.
These differences may lead to distinguishable phenomenological results.
In this work we
extend our previous analysis to the cases of scalar, vector and spin-3/2 DM.
We will consider a general set of 4-particle operators up to dimension 6
for the effective interaction between the WIMPs and the SM fermions and
compute their phenomenological constraints. We will use the updated limit
of XENON100 SI direct detection \cite{Aprile:2011hi}, which
is stronger than that adopted in our previous work~\cite{Zheng:2010js}.

This paper is organized as follows. In Sections \ref{sec-scal} and
\ref{sec-vect}, the effective models of scalar and vector DM are discussed,
respectively. In the subsections of these two sections, we explore
the constraints on these models from the DM relic density, direct and
indirect detection searches and the validity of effective
theory, and then present the combined constraints on the effective coupling
constants of these models.
In Section \ref{sec-3/2}, the study on the effective models of
spin-3/2 DM are carried out briefly.
The conclusions are given in Section \ref{sec-con}.

\section{Scalar dark matter\label{sec-scal}}

Let us begin with the case that DM consists of complex scalar
WIMPs ($\phi$ and its antiparticle $\phi^\dag$).
We will add remarks when there is a notable difference
between this case and the case of real scalar WIMPs.
In order to study phenomenologies in a model-independent way,
we construct effective interaction operators between the WIMPs and
the SM particles. These interaction operators are limited only by
the requirements of Hermiticity, Lorentz invariance and CPT invariance.

We make the following assumptions similar to those in
Ref.~\cite{Beltran:2008xg,Zheng:2010js}: (1) The WIMP is the only new
particle species at the electroweak scale, and any other new particle
species is much heavier than the WIMP. This implies
that the thermal relic density of the WIMP is not affected by resonances or
coannihilations. Thus it is possible to describe the interaction
between the WIMPs and the SM particles in terms of an effective
field theory. (2) The WIMP only interacts with the SM fermions
through a 4-particle effective interaction of
$\phi^\dag$-$\phi$-$\bar f$-$f$ type, but not with gauge or Higgs bosons.
For simplicity, this interaction is assumed to be dominated by only
one form in the set of 4-particle operators. (3) The WIMP-antiWIMP
annihilation channels to the SM fermion-antifermion pairs dominate
over other possible channels. In other words, the possible channels
to final states that include gauge or Higgs bosons are assumed
to be negligible.

The effective interaction term of Lagrangian between two complex scalar WIMPs
($\phi$ and $\phi^\dagger$) and two SM fermions ($f$ and $\bar f $)
is given by only one of the following expressions:
\begin{eqnarray}
\text{Scalar int. (S)}:&&\qquad \mathcal{L} _\mathrm{S} =
\sum_f\frac{F_{\mathrm{S},f}}{\sqrt 2} \phi ^\dag \phi \bar ff,
\label{lag:scalar:scal} \\
\text{Vector int. (V)}:&&\qquad \mathcal{L} _\mathrm{V} =
\sum_f\frac{F_{\mathrm{V},f}}{\sqrt 2} (\phi^\dag i\overleftrightarrow
{\partial_\mu}\phi) \bar f \gamma^\mu f,
\label{lag:scalar:vect} \\
\text{Scalar-pseudoscalar int. (SP)}:&&\qquad \mathcal{L}_\mathrm{SP} =
\sum_f\frac{F_{\mathrm{SP},f}}{\sqrt 2} \phi^\dag \phi \bar f i \gamma_5 f,
\label{lag:scalar:s_Ps} \\
\text{Vector-axialvector int. (VA)}:&&\qquad \mathcal{L}_\mathrm{VA}
= \sum_f\frac{F_{\mathrm{VA},f}}{\sqrt 2}(\phi^\dag i\overleftrightarrow
{\partial_\mu}\phi) \bar f \gamma^\mu \gamma_5 f,
\label{lag:scalar:v_Av}
\end{eqnarray}
where the sum of $f$ is over all the SM fermions, the symbol
$\overleftrightarrow{\partial_\mu}$ is short for $\overrightarrow
{\partial_\mu}-\overleftarrow{\partial_\mu}$, that is, $\phi^\dag
\overleftrightarrow{\partial_\mu}\phi\equiv \phi^\dag(\partial_\mu\phi)
-(\partial_\mu\phi^\dag)\phi$,
and the effective coupling constants $F_f$ are real-valued numbers.
$F_f$ have mass dimension of $-1$ for the dimension-5 operators (S and SP),
and that of $-2$ for the dimension-6 operators (V and VA).
We do not consider operators with more derivatives on
the fields since these operators have higher mass dimensions
such that they are more suppressed at low energy scale.
Note that for the case of a real scalar WIMP, the WIMP vector current
$\phi\,\overleftrightarrow{\partial_\mu}\phi$ vanishes so that the V and
VA effective operators vanish, leaving only the S and SP interactions.
Since $F_f$ are real-valued, the transformation
properties of the 4-particle operators, $\mathcal{L}_{\mathrm{S}}$,
$\mathcal{L}_{\mathrm{V}}$, $\mathcal{L}_{\mathrm{SP}}$ and $\mathcal{L}
_{\mathrm{VA}}$ under C, P and T are the same as the corresponding
4-fermion operators of Ref.~\cite{Zheng:2010js}.
All the operators are CPT invariant.

Each form of the interaction operators listed above describes an
effective model in which the WIMPs couple to the SM fermions.
For each case, we will calculate the corresponding
annihilation and scattering cross sections, which depend on the
WIMP mass $M_\phi$ and the coupling constants $F_f$. Associated
with the recent experimental results of the DM relic density, direct and
indirect detection experiments, the phenomenological constraints
on $F_f$ can be derived, respectively. Since the sensitive regions of
different kinds of experiments vary in different effective models,
it would be interesting and meaningful to compare these constraints.

\begin{figure}[!htbp]
\centering
\includegraphics[width=0.44\textwidth]{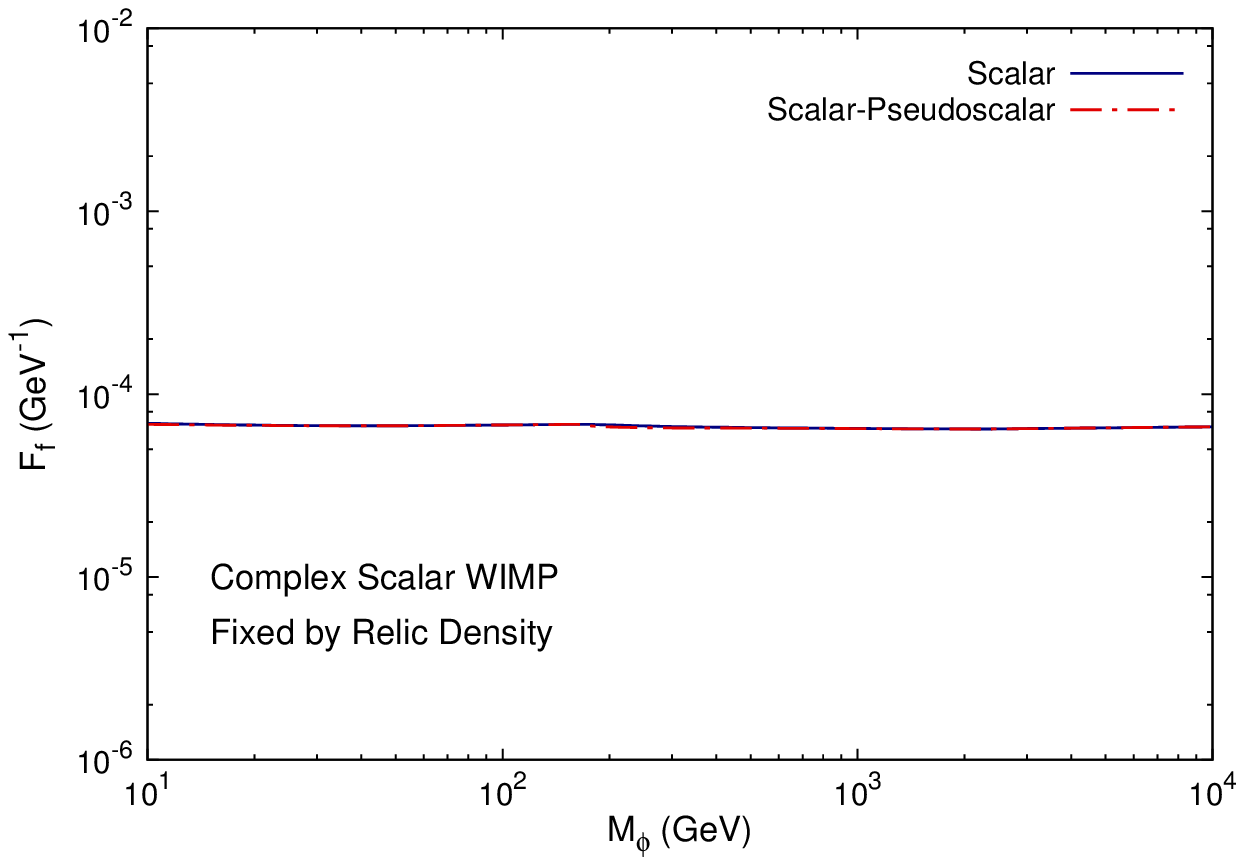}%
\hspace{0.01\textwidth}%
\includegraphics[width=0.44\textwidth]{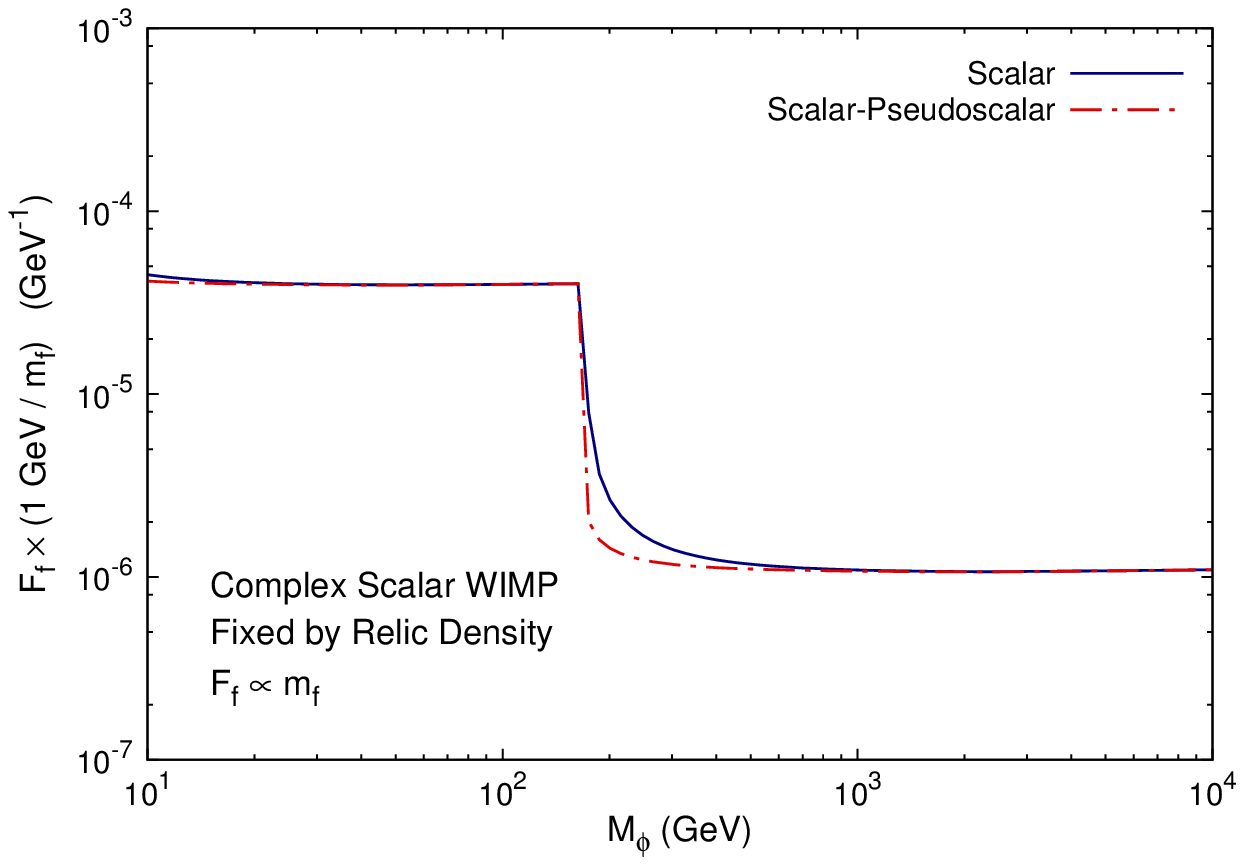}
\\
\includegraphics[width=0.44\textwidth]{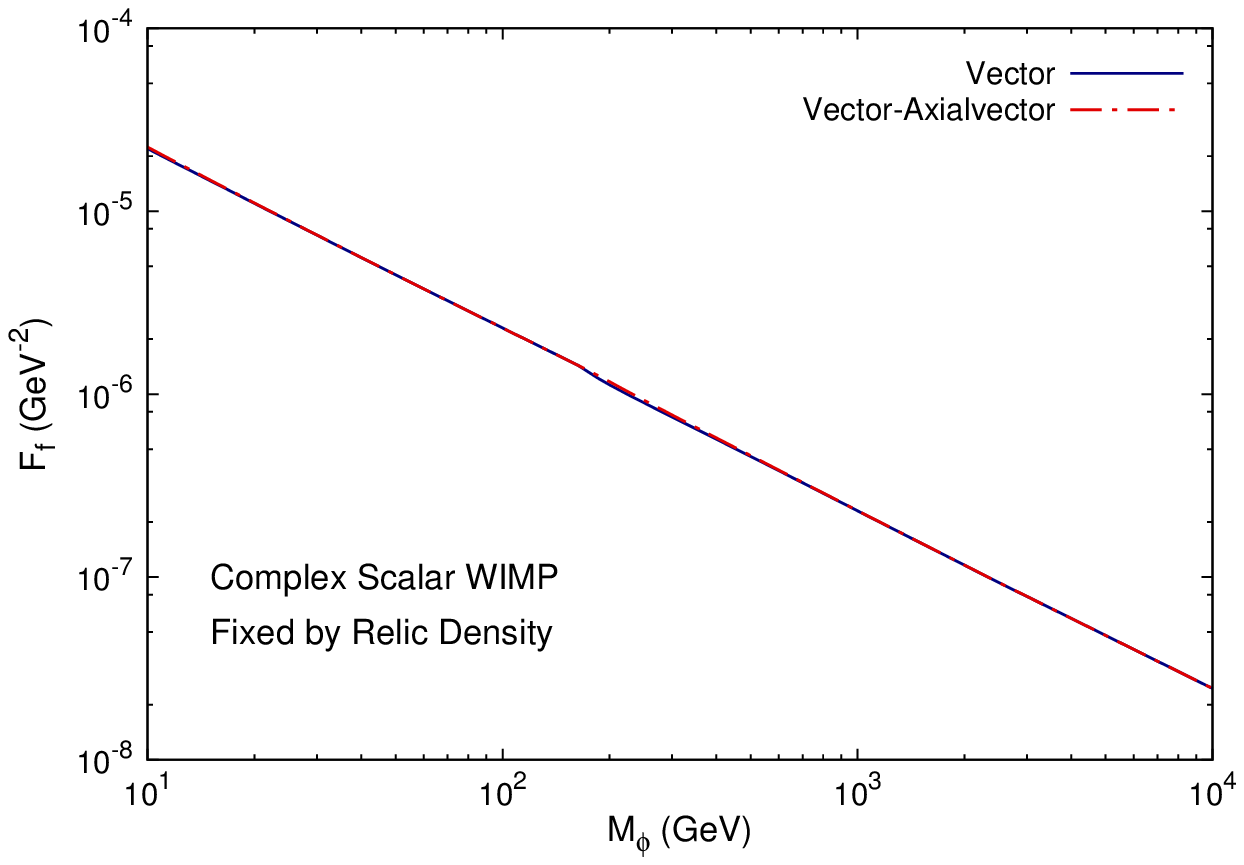}%
\hspace{0.01\textwidth}%
\includegraphics[width=0.44\textwidth]{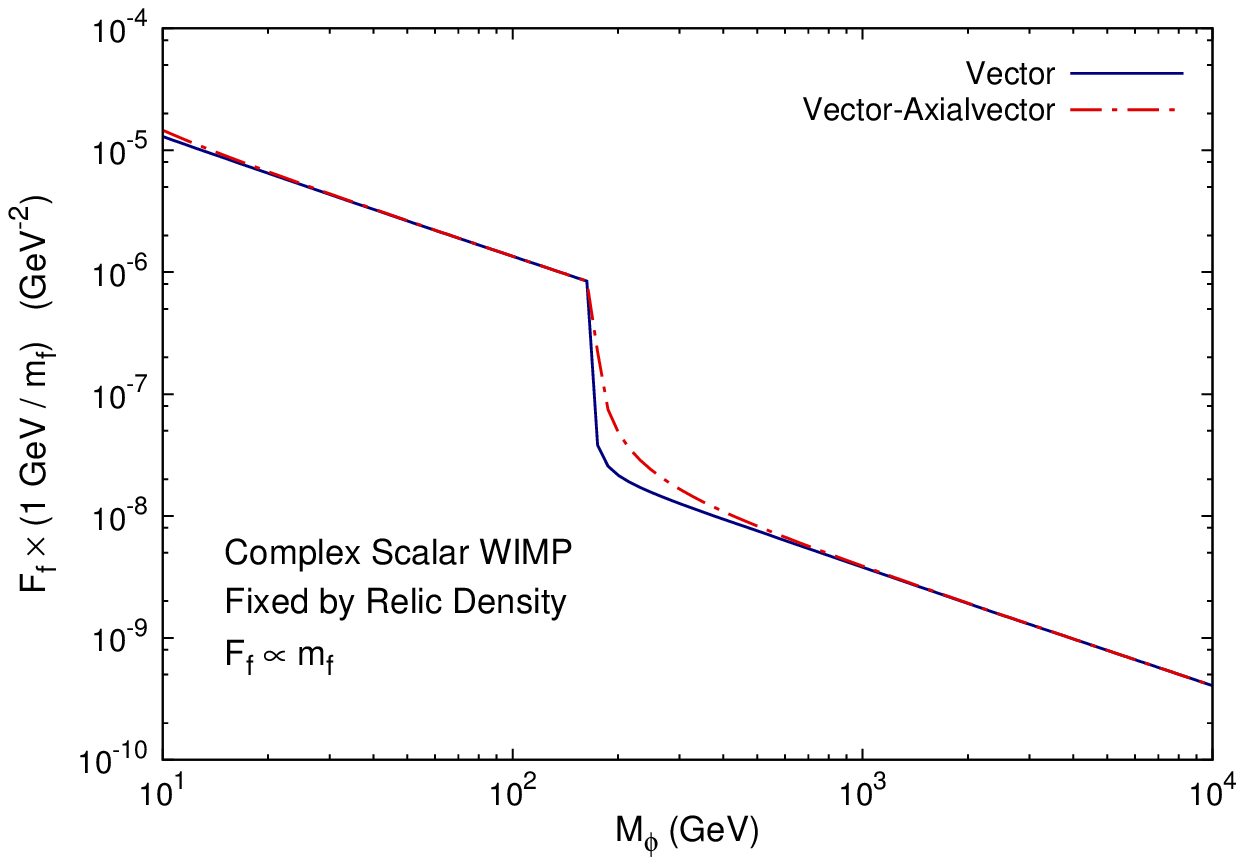}
\caption{The predicted coupling constants $F_f$ as functions of the scalar WIMP
mass $M_\phi$, fixed by the observed relic density, $\Omega_{\mathrm{DM}}
h^2=0.1109\pm0.0056$~\cite{Komatsu:2010fb}, in each effective model.
Frames in the left (right) column are results given for the case of
universal couplings ($F_f\propto m_f$).}
\label{fig:scalar:relic:comp_cpl}
\end{figure}
\begin{figure}[!htbp]
\centering
\includegraphics[width=0.44\textwidth]{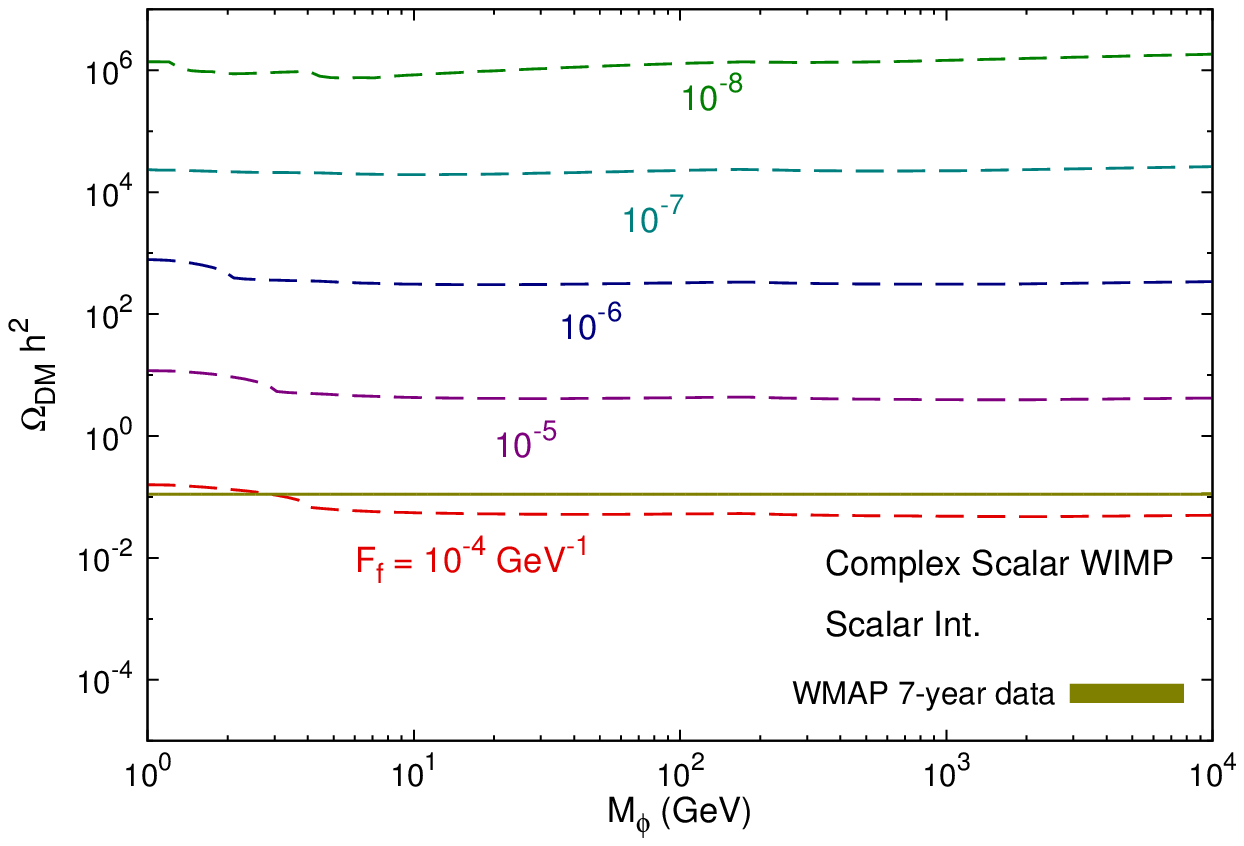}%
\hspace{0.01\textwidth}%
\includegraphics[width=0.44\textwidth]{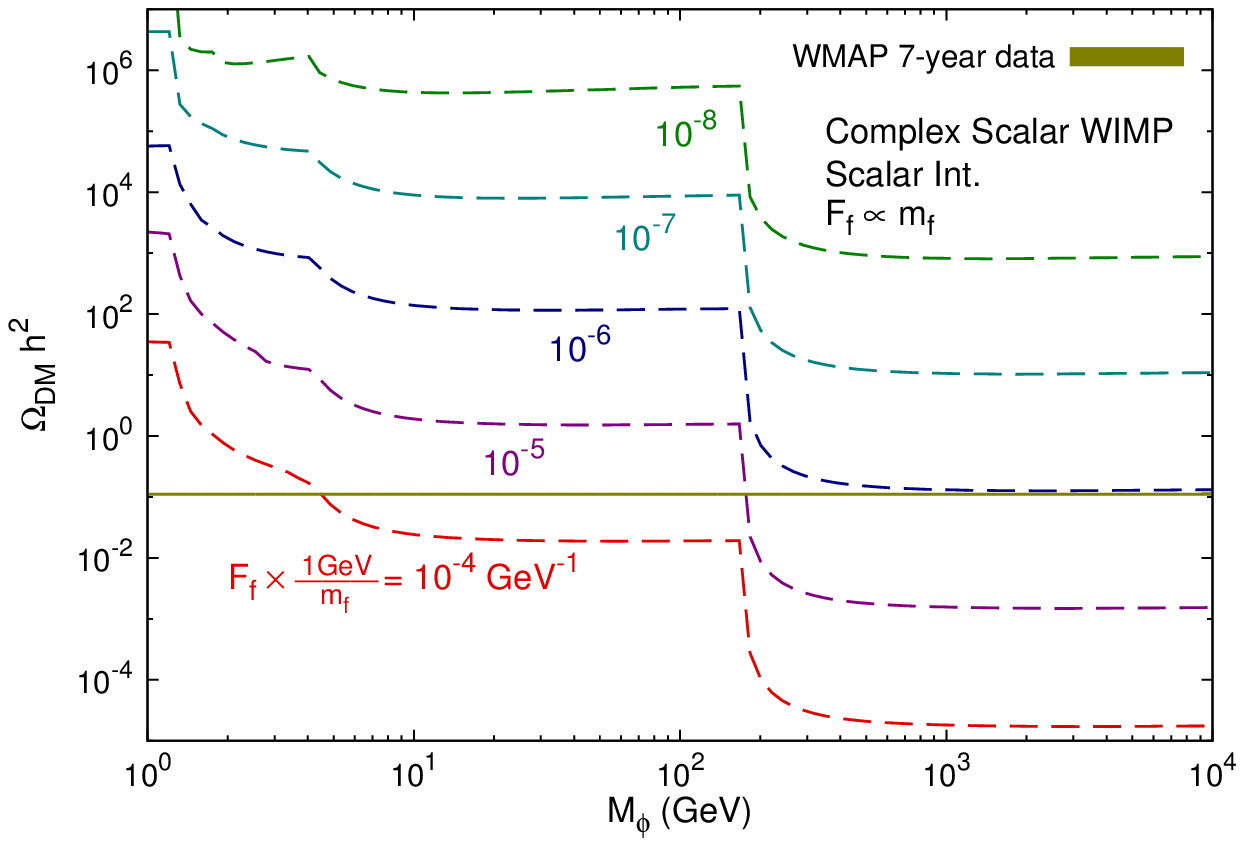}%
\\
\includegraphics[width=0.44\textwidth]{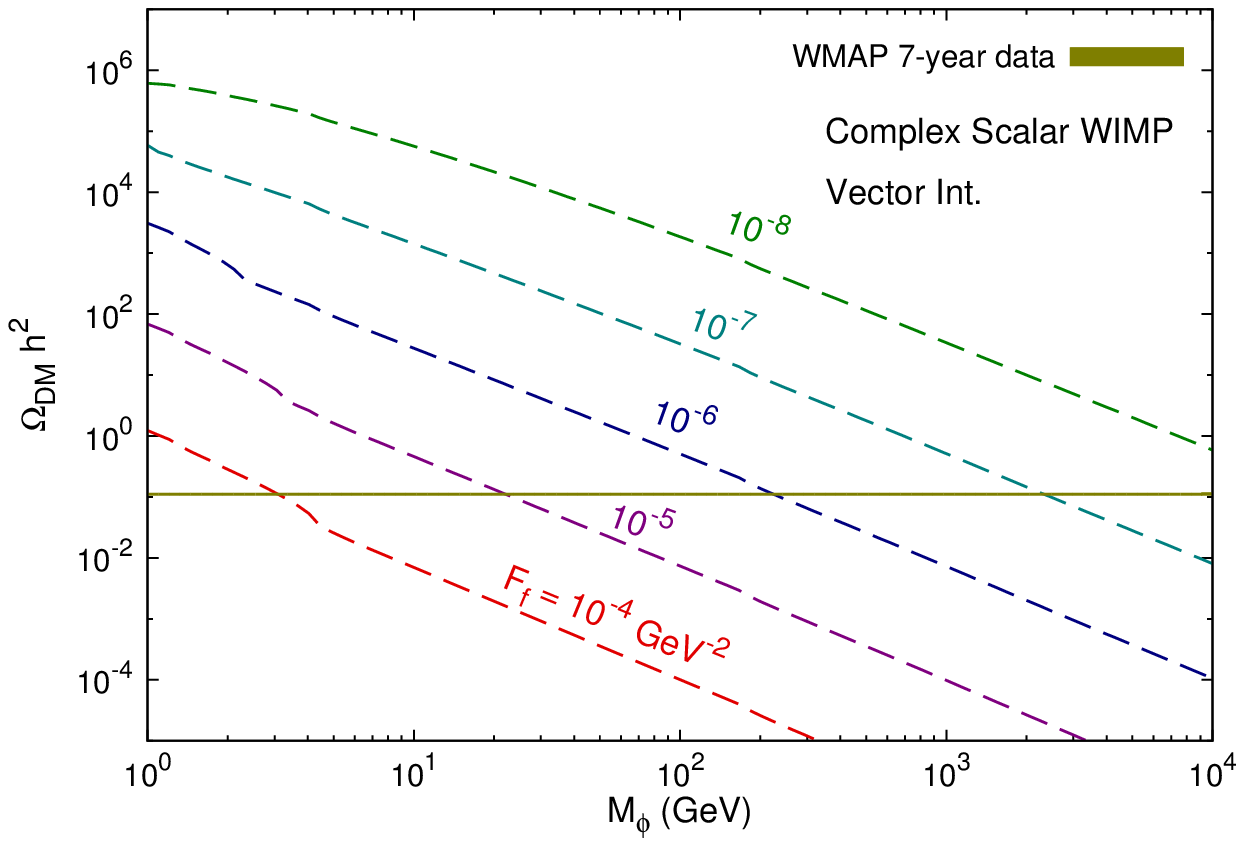}%
\hspace{0.01\textwidth}%
\includegraphics[width=0.44\textwidth]{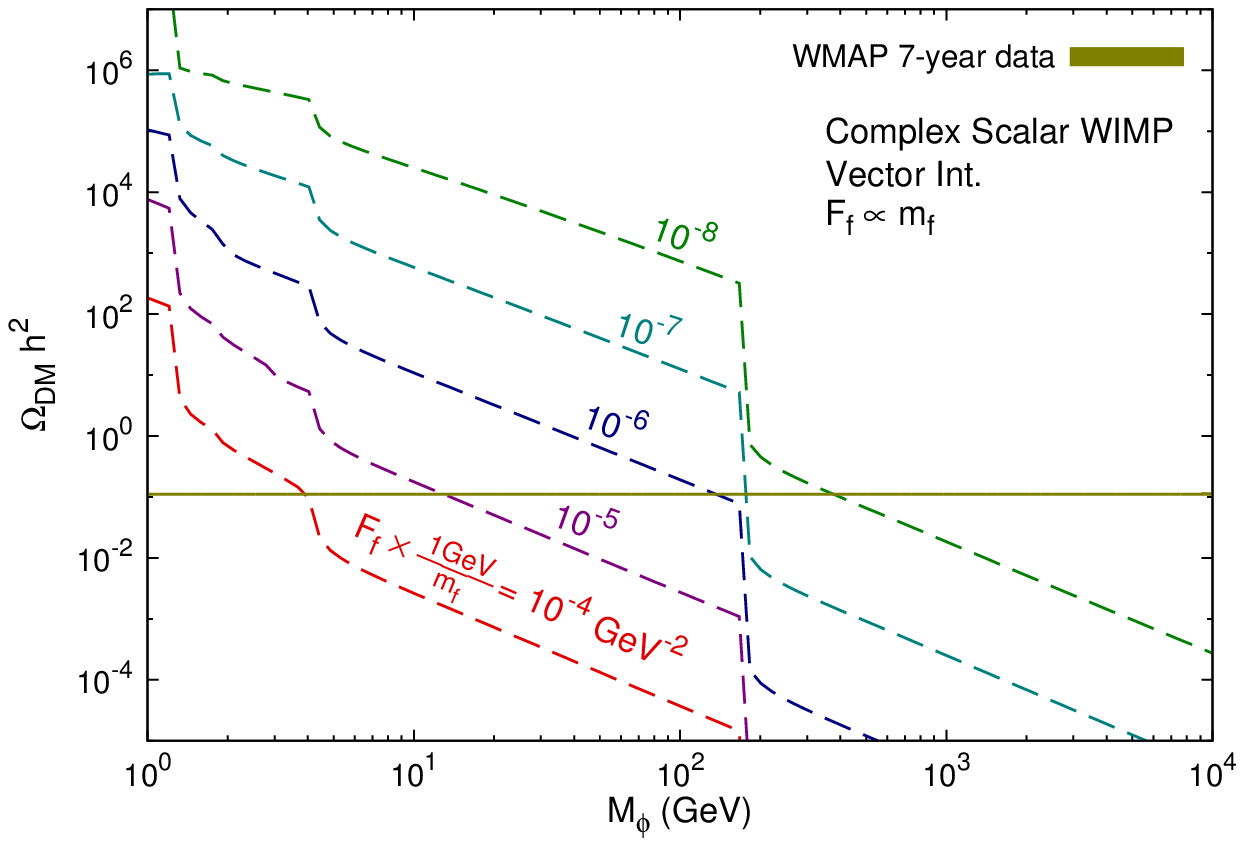}
\\
\includegraphics[width=0.44\textwidth]{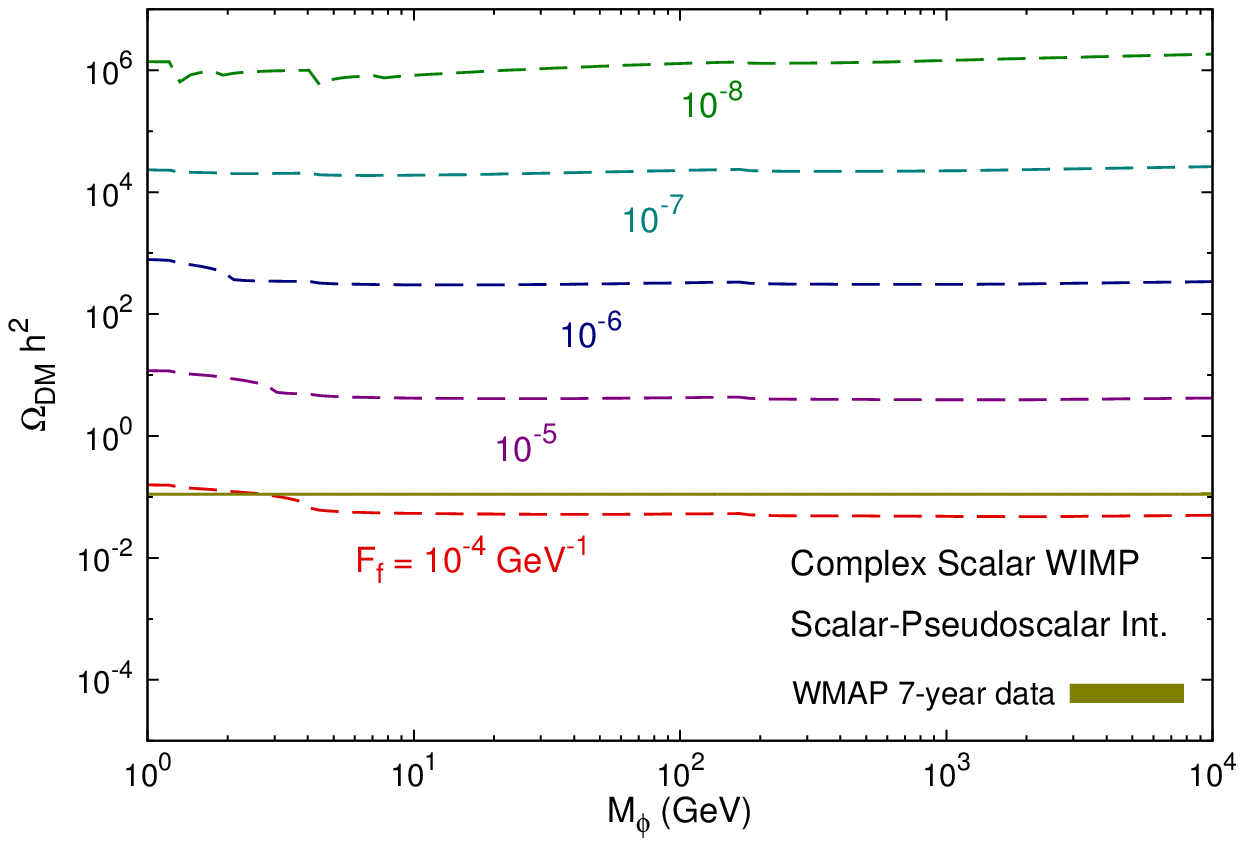}
\hspace{0.01\textwidth}%
\includegraphics[width=0.44\textwidth]{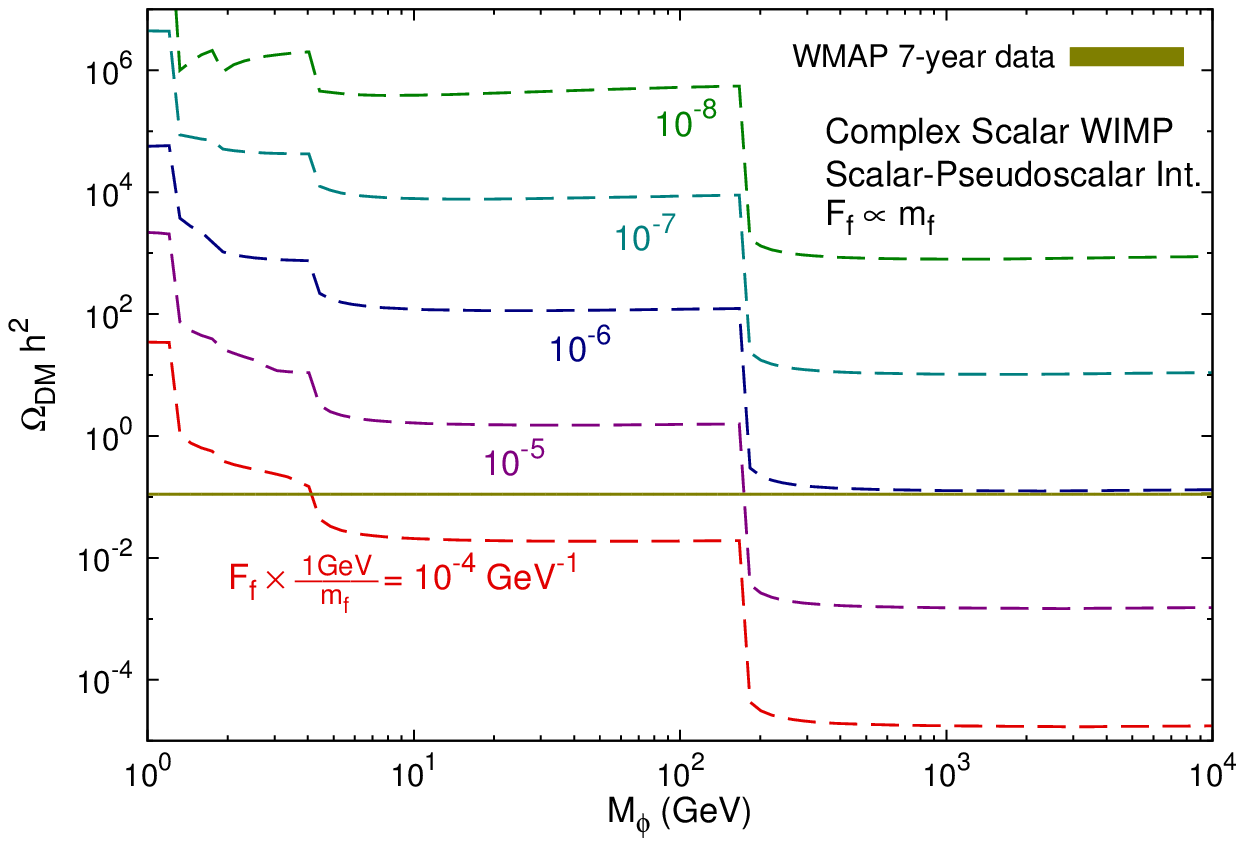}%
\\
\includegraphics[width=0.44\textwidth]{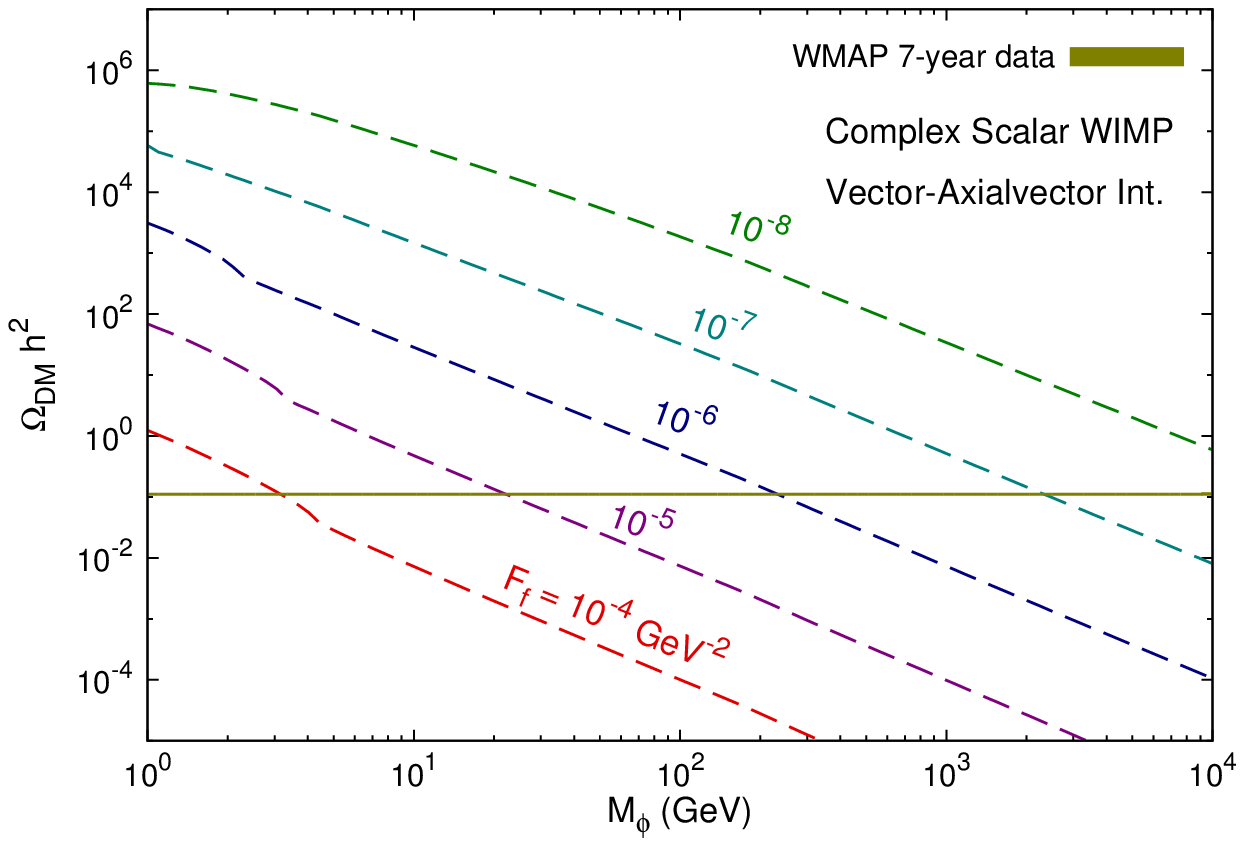}%
\hspace{0.01\textwidth}%
\includegraphics[width=0.44\textwidth]{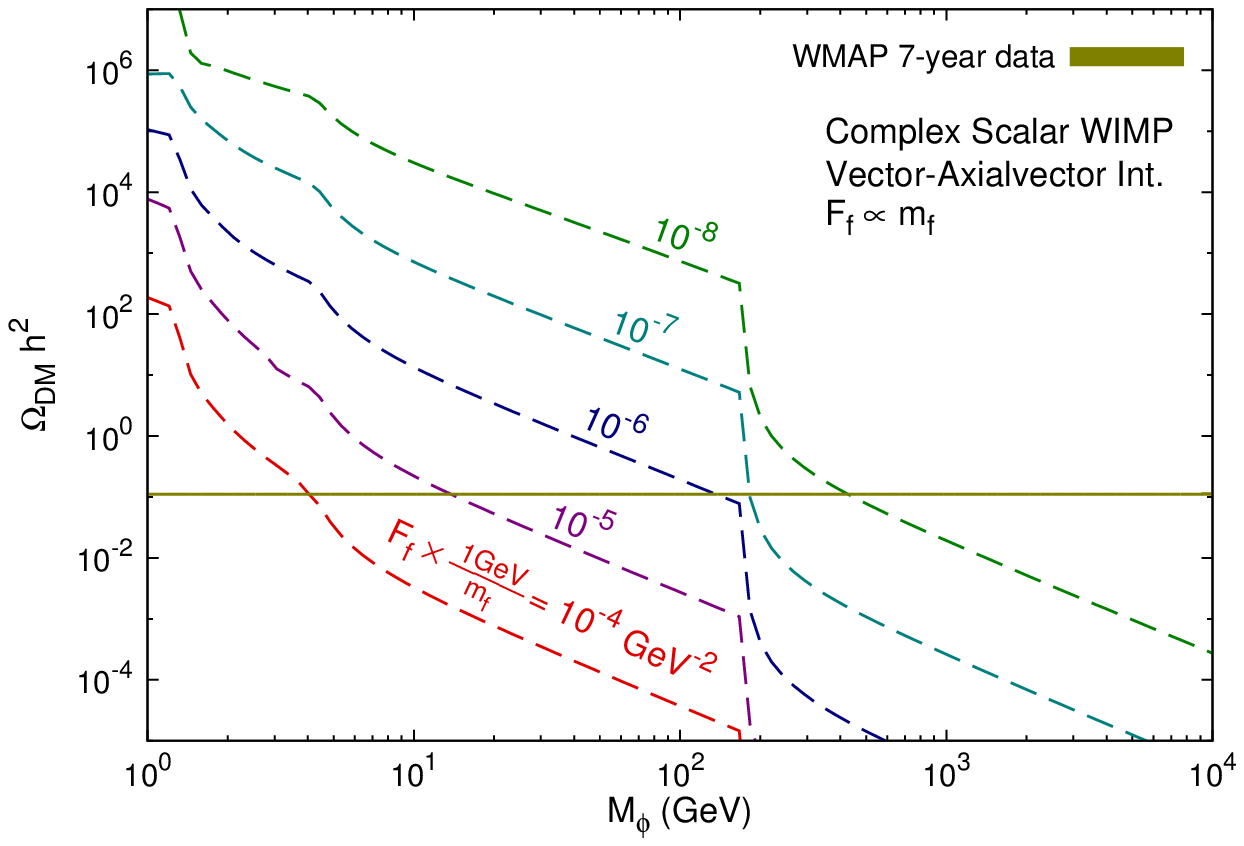}
\caption{The predicted thermal relic density (dashed lines) of
complex scalar WIMPs with S, V, SP and VA interactions respectively.
In the left (right) column, results are given for the case of
universal couplings ($F_f \propto m_f$). The narrow horizontal solid band
shows the range of the observed DM relic density, $\Omega_{\mathrm{DM}}
h^2=0.1109\pm0.0056$~\cite{Komatsu:2010fb}.}
\label{fig:scalar:rd}
\end{figure}

\subsection{Scalar WIMP annihilation and relic density\label{subsec-scal-rd}}

In order to determine the relic density of WIMPs and the source
function of cosmic-ray particles derived from WIMP annihilation in the
Galactic halo, which is relevant to the DM indirect detection, we need to
calculate the cross sections of WIMP-antiWIMP annihilation to
fermion-antifermion pairs. For each case listed above,
the result is given by
\begin{eqnarray}
\sigma_{\mathrm{S},\,\mathrm{ann}} &=& \frac{1}{8\pi}\sum\limits_f
\left( \frac{F_{\mathrm{S},f}}{\sqrt{2}} \right)^2{c_f}
\sqrt{\frac{s - 4m_f^2}{s - 4M_\phi ^2}}
\frac{(s - 4m_f^2)}{s},
\label{sigma_scalar_scal}\\
\sigma_{\mathrm{V},\,\mathrm{ann}} &=& \frac{1} {12\pi}\sum\limits_f
\left( \frac{F_{\mathrm{V},f}}{\sqrt{2}} \right)^2{c_f}
\sqrt{\frac{s - 4m_f^2}{s - 4M_\phi ^2}}
\frac{(s -4M_\phi ^2)(s + 2m_f^2)}{s},
\label{sigma_scalar_vect}\\
\sigma_{\mathrm{SP},\,\mathrm{ann}} &=& \frac{1}{8\pi}\sum\limits_f
\left( \frac{F_{\mathrm{SP},f}}{\sqrt{2}} \right)^2{c_f}
\sqrt{\frac{s - 4m_f^2}{s - 4M_\phi ^2}},
\label{sigma_scalar_s_Ps}\\
\sigma_{\mathrm{VA},\,\mathrm{ann}} &=& \frac{1}{12\pi}\sum\limits_f
\left( \frac{F_{\mathrm{VA},f}}{\sqrt{2}} \right)^2{c_f}
\sqrt{\frac{s - 4m_f^2}{s - 4M_\phi ^2}}
\frac{(s -4M_\phi ^2)(s - 4m_f^2)}{s},
\label{sigma_scalar_v_Av}
\end{eqnarray}
where $s$ is the square of the center-of-mass energy, $M_\phi$ is the WIMP mass,
and $c_f$ are the color factors, equal to 3 for quarks and 1 for leptons.
For the case of real scalar WIMPs, due to the identity of WIMP and antiWIMP,
the S and SP annihilation cross sections should include an
additional factor of 4 if the Lagrangians are taken to be
Eqs.~\eqref{lag:scalar:scal} and \eqref{lag:scalar:s_Ps}
with $\phi^\dag = \phi$.
Eqs.~\eqref{sigma_scalar_scal} -- \eqref{sigma_scalar_v_Av}
are also obtained in Ref.~\cite{Beltran:2008xg}.
Taking the energy distributions of the initial particles into account,
we need to calculate $\left<\sigma_\mathrm{ann}v\right>$, the thermal
average of cross section multiplied by relative velocity (or M{\o}ller
velocity, more exactly~\cite{Gondolo:1990dk}).

In our previous work~\cite{Zheng:2010js}, following Ref.~\cite{Gondolo:1990dk}
we calculated $\left<\sigma_\mathrm{ann}v\right>$ in the ``lab'' frame
(in which one of the two initial particles is at rest), because it can be
proved that the calculation in this frame is equivalent to that obtained in the
practical reference frame and more convenient.
Thus we can expand $s$ in the lab frame to be $s\simeq 4 M_\phi^2 +
M_\phi^2 v^2 + \frac{3}{4} M_\phi^2 v^4 + \mathcal{O}(v^6)$,
and substitute it into Eqs.~\eqref{sigma_scalar_scal}
-- \eqref{sigma_scalar_v_Av}. Then we have the form $\sigma_\mathrm{ann}v
\simeq a_0 + a_1 v^2 + \mathcal{O}(v^4)$, which leads to $\left<\sigma_
\mathrm{ann}v\right> \simeq a_0 + 6 a_1 x^{-1} + \mathcal{O}(x^{-2})$ where
$x \equiv M_\phi /T$ and $T$ is the temperature of DM.
On the other hand, we can also directly calculate
$\left<\sigma_\mathrm{ann}v\right>$ in the practical reference frame
following the method described in Ref.~\cite{Srednicki:1988ce}.
These two methods lead to the same results, i.e.,
\begin{eqnarray}
  \left<\sigma_{\mathrm{S},\,\mathrm{ann}}v\right> &\simeq& \frac{1}{4\pi}
\sum\limits_f \left(\frac{F_{\mathrm{S},f}}{\sqrt 2}\right)^2 c_f
\sqrt{1-\frac{m_f^2}{M_\phi^2}} \left[\left(1-\frac{m_f^2}{M_\phi^2}\right)
+ \frac{3}{4}\left(-2+5\frac{m_f^2}{M_\phi^2}\right) \frac{T}{M_\phi}\right],
\label{sv_scalar_scal}\\
  \left<\sigma_{\mathrm{V},\,\mathrm{ann}}v\right> &\simeq& \frac{1}{2\pi}
\sum\limits_f \left(\frac{F_{\mathrm{V},f}}{\sqrt 2}\right)^2 c_f
\sqrt{1-\frac{m_f^2}{M_\phi^2}} M_\phi^2 \left(2+\frac{m_f^2}{M_\phi^2}
\right)\frac{T}{M_\phi},
\label{sv_scalar_vect}\\
  \left<\sigma_{\mathrm{SP},\,\mathrm{ann}}v\right> &\simeq& \frac{1}{4\pi}
\sum\limits_f \left(\frac{F_{\mathrm{SP},f}}{\sqrt 2}\right)^2 c_f
\sqrt{1-\frac{m_f^2}{M_\phi^2}} \left[1+\frac{-6+9m_f^2/M_\phi^2}
{4(1-m_f^2/M_\phi^2)}\frac{T}{M_\phi}\right],
\label{sv_scalar_s_Ps}\\
  \left<\sigma_{\mathrm{VA},\,\mathrm{ann}}v\right> &\simeq& \frac{1}{\pi}
\sum\limits_f \left(\frac{F_{\mathrm{VA},f}}{\sqrt 2}\right)^2 c_f
\left(1-\frac{m_f^2}{M_\phi^2} \right)^{3/2} M_\phi^2 \frac{T}{M_\phi}.
\label{sv_scalar_v_Av}
\end{eqnarray}
Note that the above results disagree with those in Ref.~\cite{Beltran:2008xg}.

The evolution of the DM abundance is described by the
Boltzmann equation
\begin{equation}
\frac{dn_\phi}{dt}+3Hn_\phi =-\langle {\sigma
_{\mathrm{ann}}v}\rangle\left[n_\phi n_{\phi^\dag} -
n_\phi^\mathrm{eq}n_{\phi^\dag}^\mathrm{eq}\right]=-\langle {\sigma
_{\mathrm{ann}}v}\rangle\left[(n_\phi)^2 -
(n_\phi^\mathrm{eq})^2\right],
\label{boltz-eq}
\end{equation}
where $H\equiv\dot{a}/a=\sqrt{8\pi\rho/(3 M_\mathrm{Pl}^2)}$ is
the Hubble rate with $M_\mathrm{Pl}$ denoting the Planck mass,
$n_\phi$ ($n_{\phi^\dag}$) is the number density of WIMPs
(antiWIMPs), and $n_\phi^\mathrm{eq}$ ($n_{\phi^\dag}^\mathrm{eq}$) is the
corresponding equilibrium number density. For complex scalar WIMPs
without particle-antiparticle asymmetry, we have $n_\phi=n_{\phi^\dag}$.
Thus the total DM particle number density is $n_{\mathrm{DM}}=2n_\phi$
\cite{Gondolo:1990dk,Jungman:1995df}. Using the standard procedure
\cite{kolb-turner,Jungman:1995df} to approximately solve the
Boltzmann equation \eqref{boltz-eq}, we obtain a relic density of
DM particles as
\begin{equation}
\Omega_{\mathrm{DM}} h^2=2\Omega_\phi h^2\simeq 2.08
\times10^9~\mathrm{GeV}^{-1}\left(\frac{T_0}{2.725~\mathrm{K}}\right)^3
\frac{x_f}{M_{\mathrm{pl}}\sqrt{g_\ast(T_f)}(a_0+3 a_1 x_f^{-1})},
\label{relic-d}
\end{equation}
where $x_f\equiv M_\phi/T_f$ with $T_f$ being the freeze-out temperature,
$g_\ast(T_f)$ is effectively relativistic degrees of
freedom at freeze-out, $T_0=2.725\pm0.002$~K~\cite{Mather:1998gm} is the
present CMB temperature, and $a_0$ and $a_1$ are coefficients in the expansion
$\sigma _{\mathrm{ann}}v \simeq a_0 + a_1 v^2+\mathcal{O}(v^4)$.
The freeze-out temperature  parameter $x_f$ is evaluated numerically
by solving the equation
\begin{equation}
x_f=\ln\left[c(c + 2)\sqrt{\frac{45}{8}}
\frac{gM_\phi M_{\mathrm{Pl}}(a_0 + 6 a_1 x_f^{-1})}{2\pi^3\sqrt{g_\ast(x_f)}
\,x_f^{1/2}}\right],
\label{x-f}
\end{equation}
where the order one parameter $c$ is taken to be $1/2$ as usual. A notable
difference between this calculation and our previous work~\cite{Zheng:2010js}
is that the degree of freedom of a scalar WIMP or antiWIMP is $g=1$, while
that of a fermionic WIMP is $g=2$. The numerical result
of $g_\ast(T)$ is taken from Ref.~\cite{Coleman:2003hs}.  For the case of
real scalar WIMPs, $n_\mathrm{DM}=n_\phi$, and $n_{\phi^\dag}$ in
Eq.~\eqref{boltz-eq} should be replaced by $n_\phi$ so that the
relic density will be one half of that for the complex scalar case
given the same annihilation cross section.

The 7-year observation of WMAP~\cite{Komatsu:2010fb} gives the
DM relic density $\Omega_{\mathrm{DM}} h^2=0.1109\pm0.0056$.
With this result we estimate the
relation between the effective coupling constants $F_f$ and the
WIMP mass $M_\phi$ in each effective model, as shown in
Fig.~\ref{fig:scalar:relic:comp_cpl}. Two kinds of $F_f$
are considered here. In the left column of
Fig.~\ref{fig:scalar:relic:comp_cpl}, we show the results for the case
when the effective couplings $F_f$ to all the SM fermions
are equal (universal couplings). In the right column of
Fig.~\ref{fig:scalar:relic:comp_cpl}, we show the results for the case
of $F_f \propto m_f$. This kind of proportionality may come from Yukawa
couplings of a Higgs mediated interaction or some other unknown underlying
mechanisms. In both cases, $F_f$ decreases as $M_\phi$
increases for fixed $\Omega_{\mathrm{DM}} h^2$ in each effective model.
Besides, there are several interesting features
in Fig.~\ref{fig:scalar:relic:comp_cpl}:
\begin{itemize}
  \item In the case of $F_f \propto m_f$, the curves of $F_f$ vs. $M_\phi$
have obvious downward bends at about $M_\phi \sim m_t = 171$~GeV. The reason
is that when $M_\phi > m_t$, the annihilation channel $\phi \phi^\dag \to
t \bar t$ is opened. Therefore because the couplings are proportional to $m_f$,
this channel gives a tremendous contribution to the total
$\left<\sigma_\mathrm{ann}v\right>$ due to the huge mass gaps
between $t$ quark and other SM fermions.
  \item In the case of universal couplings, there are 2 pairs of nearly
identical curves, that is, S~$\simeq$~SP and V~$\simeq$~VA if we denote
this approximate identity of the two curves by a notation ``$\simeq$''
for short here and henceforth. From Eq.~\eqref{sv_scalar_scal} --
\eqref{sv_scalar_v_Av} we can see that in each pair their corresponding
$\left<\sigma_\mathrm{ann}v\right>$ differ only by terms of $\mathcal{O}
(T/M_\phi)$ and/or terms of $m_f^2/M_\phi^2$. These differences are not
important when we calculate the cold relic density.
  \item In the case of $F_f \propto m_f$, S~$\simeq$~SP and V~$\simeq$~VA
are still almost true except that in some small regions the deviations of
the two nearly identical curves in each pair become larger.
\end{itemize}

In Fig.~\ref{fig:scalar:rd} the curves of $\Omega_\mathrm{DM}h^2$ vs.
$M_\phi$ for fixed coupling constants in the complex scalar WIMP models
of S, V, SP and VA interaction operators are shown.
In the left (right) column, results are given for the case of
universal couplings ($F_f \propto m_f$). The values of the couplings
corresponding to the curves are denoted in each frames. All the curves
in Fig.~\ref{fig:scalar:rd} bend more or less at about $M_\phi \sim
1.3$, 4.2 and 171~GeV corresponding to the masses of $c$, $b$ and $t$
quarks respectively, due to the opens of these heavy quark
annihilation channels. An important qualitative difference between
the curves of effective operators with different mass dimension in both
Figs.~\ref{fig:scalar:relic:comp_cpl} and \ref{fig:scalar:rd}
is that the curves for dimension-6 operators (V and VA) slope downward,
while those for dimension-5 operators (S and SP) remain horizontal
in most mass range. The reason has been showed explicitly in
Eqs.~\eqref{sv_scalar_scal} -- \eqref{sv_scalar_v_Av}.
The annihilation cross section for dimension-6 operators have
additional factors of $M_\phi^2$ which suppress the relic density
of large WIMP mass.

It is worth noting that the results in Figs.~\ref{fig:scalar:relic:comp_cpl}
and \ref{fig:scalar:rd} are based on the assumptions
presented at the beginning of this section.
Otherwise if resonances, coannihilations or annihilations to final states
other than fermion-antifermion pairs are significant, the actual
curves in Figs.~\ref{fig:scalar:relic:comp_cpl} and
\ref{fig:scalar:rd} will be much lower than those
shown there, as pointed out in Ref.~\cite{Beltran:2008xg}.

\begin{figure}[!htbp]
\centering
\includegraphics[width=0.44\textwidth]{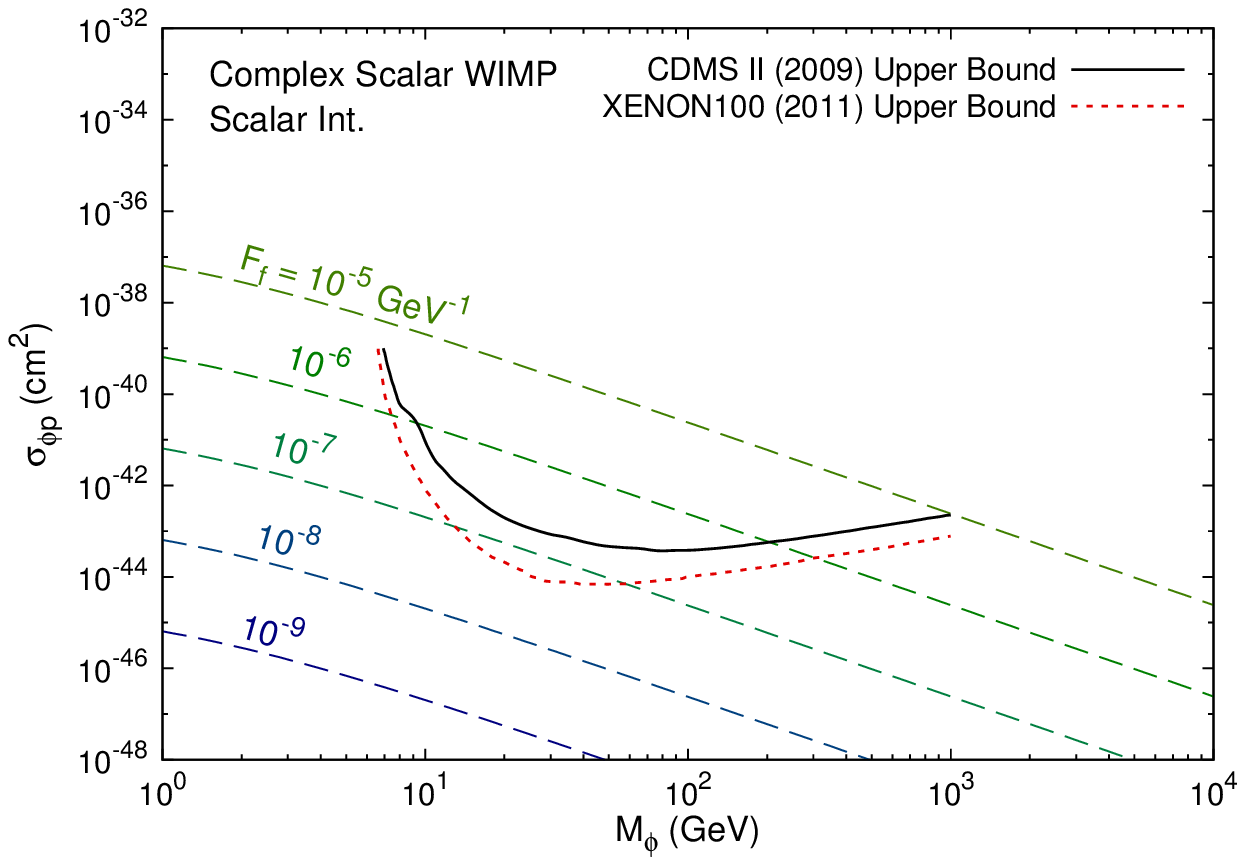}%
\hspace{0.01\textwidth}%
\includegraphics[width=0.44\textwidth]{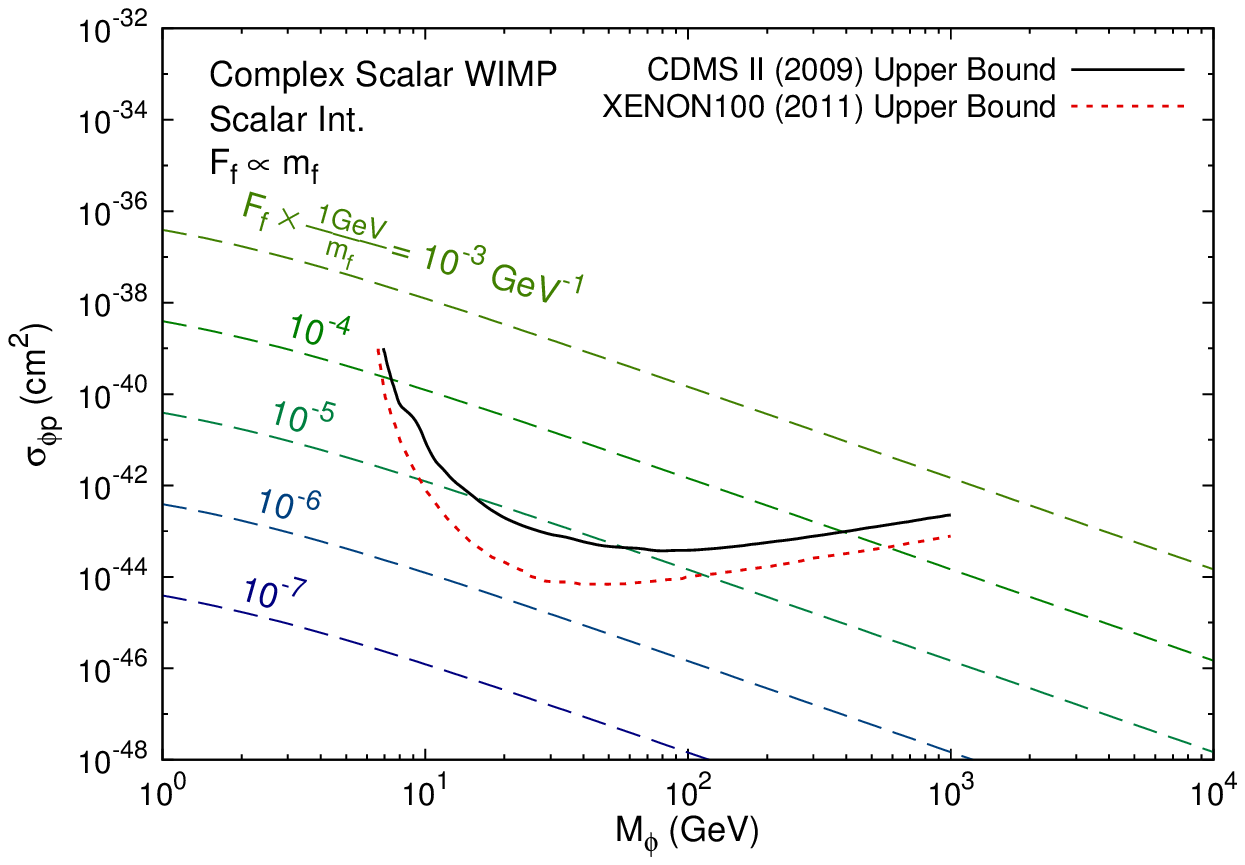}%
\\
\includegraphics[width=0.44\textwidth]{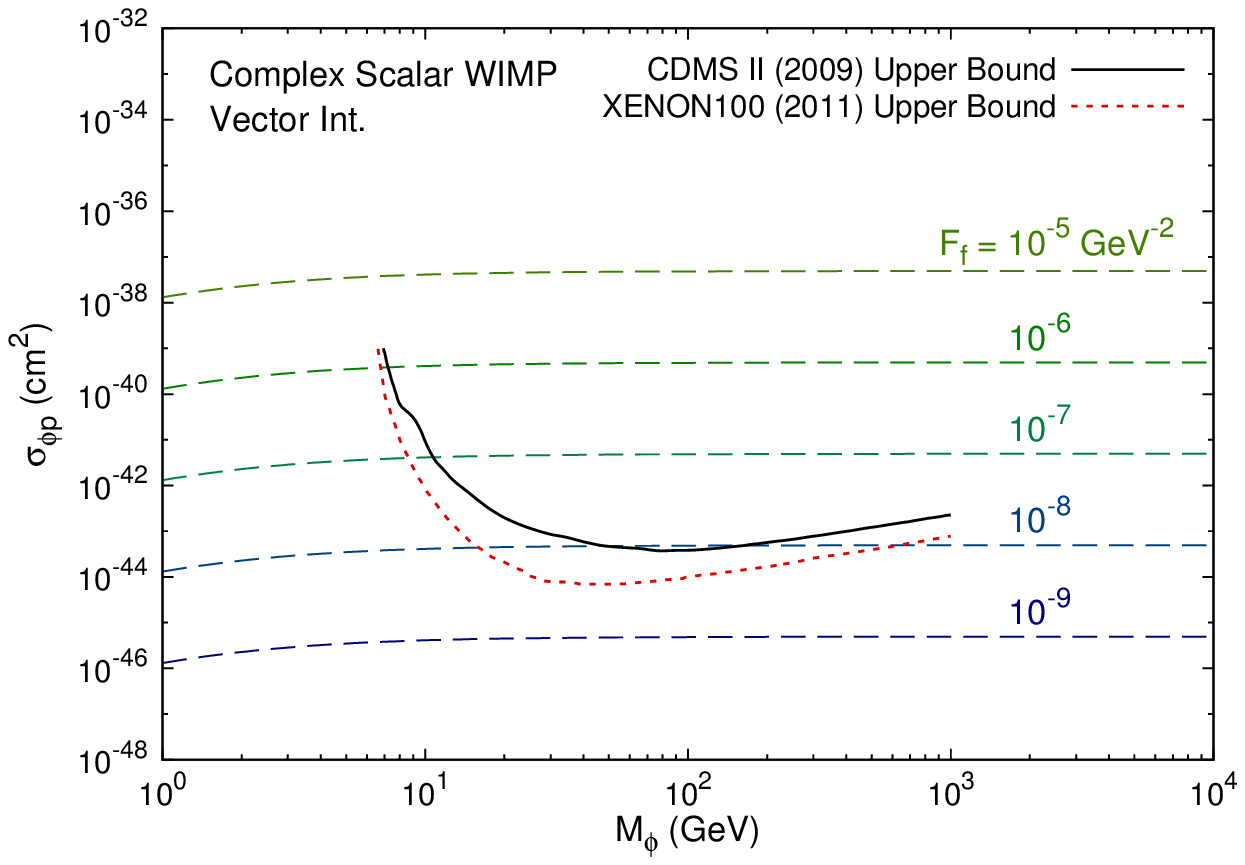}%
\hspace{0.01\textwidth}%
\includegraphics[width=0.44\textwidth]{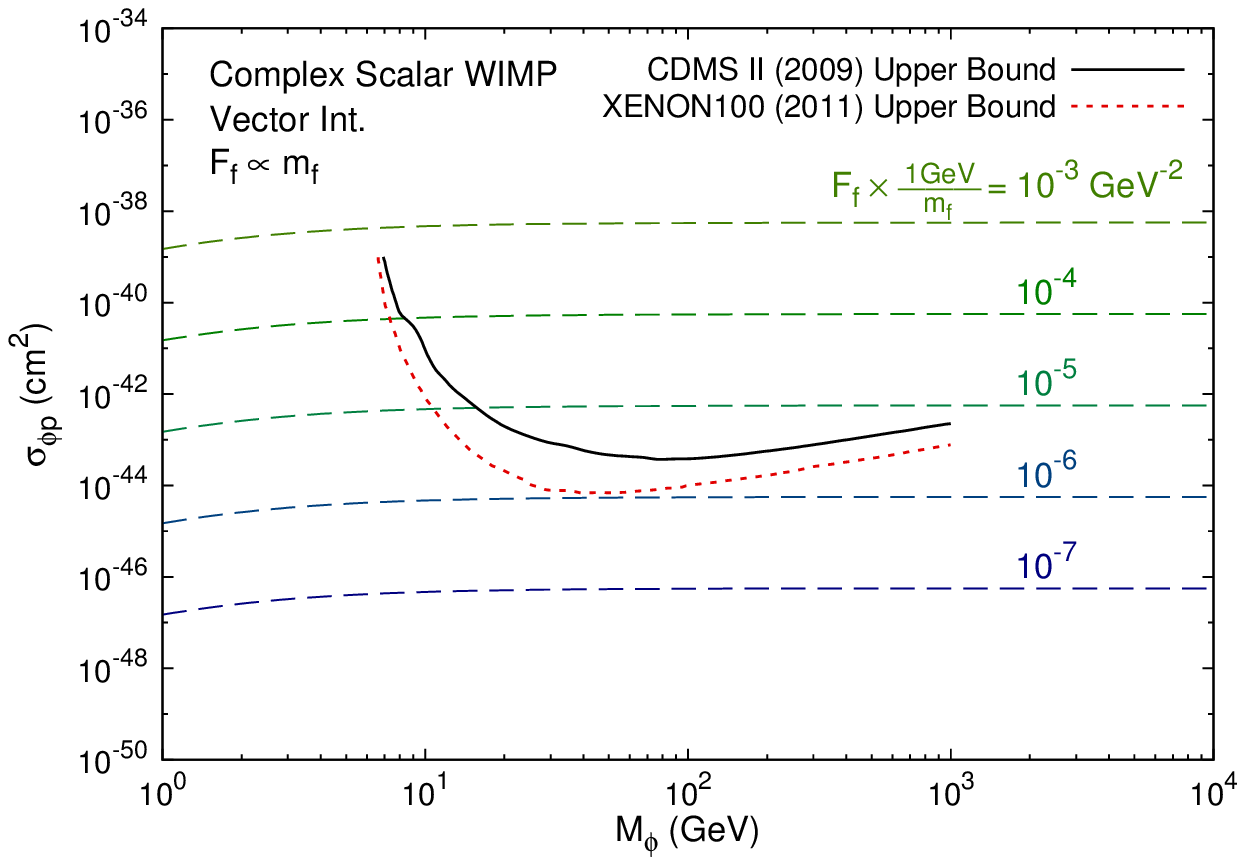}%
\caption{The spin-independent (SI) WIMP-proton cross sections (dashed lines)
for complex scalar WIMPs with S and V interactions.
In the left (right) column, results are given for the case of
universal couplings ($F_f \propto m_f$). The upper bounds set by
CDMS II~\cite{Ahmed:2009zw} and XENON100~\cite{Aprile:2011hi} are
also given in the frames for comparison.}
\label{fig:scalar:scat}
\end{figure}

\subsection{Direct detection\label{subsec-scal-dd}}

In this subsection we discuss the direct detection constraints on the
effective scalar WIMP models described by Eqs.~\eqref{lag:scalar:scal} --
\eqref{lag:scalar:v_Av}. Direct detection experiments measure the
recoil energy of the atomic nuclei when the WIMPs elastically scatter
off them. Because of the complex structure of nucleons we have to study
the WIMP-nucleon interactions with another set of effective interactions
in terms of the nucleon fields, by substituting the quark field operators
in Eqs.~\eqref{lag:scalar:scal} -- \eqref{lag:scalar:v_Av} with the nucleon
field operators $N$ ($N=p,n$) and associating the coupling constants of
these two sets of effective interactions by the form factors of nucleons.

The velocity of the WIMP near the Earth is considered to be of the
same order as the orbital velocity of the Sun, $v \simeq 0.001c$,
and we can safely calculate the scattering cross
sections in the low velocity limit.
In this limit, only the time component of $\phi^\dagger
i\overleftrightarrow{\partial_\mu }\phi$ and the spacial components
of $\bar{N}\gamma^\mu\gamma_5N$ survive, thus the VA interaction leads
to an elastically scattering cross section with velocity suppression.
On the other hand, $\bar{N}\gamma_5 N$ vanishes in this limit so that
the SP interaction is not constrained by direct detection either.
Therefore there is no measurable spin-dependent (SD) WIMP-nucleon
scattering. This situation can be also understood by realizing
that scalar WIMPs have no spin structure.
The scattering cross sections induced by the remaining
S and V interactions are both spin-independent (SI).

The calculation of scattering cross sections is much similar to our
previous calculation for fermionic DM~\cite{Zheng:2010js}, so
we will just outline the result. To compare with the result of
CDMS~\cite{Ahmed:2009zw} and XENON~\cite{Aprile:2011hi},
the WIMP-nucleon cross sections are given as follows:
\begin{eqnarray}
\text{Scalar int. :}\qquad && \sigma_{\mathrm{S},\phi N}=\frac{m^2_N}
{4\pi(M_\phi+m_N)^2}\left(\frac{F_{\mathrm{S},N}}{\sqrt{2}}\right)^2,
\label{sigmaN_scat_scalar_scal}\\
\text{Vector int. :}\qquad && \sigma_{\mathrm{V},\phi N}=\frac{m^2_N M^2_{\phi}}
{\pi(M_\phi+m_N)^2}\left(\frac{F_{\mathrm{V},N}}{\sqrt{2}}\right)^2,
\label{sigmaN_scat_scalar_vect}
\end{eqnarray}
where $F_N$ are the induced coupling constants of effective WIMP-nucleon
interactions. $F_N$ are related to
the couplings to quarks $F_q$ by form factors. For scalar interaction,
\begin{equation}
F_{\mathrm{S},N} =\sum_{q=u,d,s}F_{\mathrm{S},q}f^N_q\frac{m_N}{m_q}
+\sum_{q=c,b,t}F_{\mathrm{S},q}f^N_Q\frac{m_N}{m_q},
\label{form-S}
\end{equation}
where the nucleon form factors are $f^p_u=0.020\pm0.004$,
$f^p_d=0.026\pm0.005$, $f^p_s=0.118\pm0.062$, $f^n_u=0.014\pm0.003$,
$f^n_d=0.036\pm0.008$, $f^n_s=0.118\pm0.062$~\cite{Ellis:2000ds,Alarcon:2011zs},
and $f^N_Q=\frac{2}{27}(1-f^N_u -f^N_d -f^N_s)$ for heavy quarks.
For vector interaction,
\begin{equation}
F_{\mathrm{V},p}=2F_{\mathrm{V},u}+F_{\mathrm{V},d}\;,\quad
F_{\mathrm{V},n}=F_{\mathrm{V},u}+2F_{\mathrm{V},d}\;,
\label{form-V}
\end{equation}
which reflect the valence quark numbers in the nucleons. For the case of
real scalar WIMPs, the vector interaction vanishes, and the scattering cross
section of scalar interaction should include an additional factor 4
comparing to the case of complex scalar WIMPs.

In Fig.~\ref{fig:scalar:scat}, we show the predicted elastic
scattering cross sections between complex scalar WIMP and nucleon by
different kinds of effective interactions. Then given the
experimental bounds set by CDMS II~\cite{Ahmed:2009zw} and XENON
100~\cite{Aprile:2011hi}, we can derive the upper bounds of
coupling constants for different WIMP masses. The SP and VA
interactions are not constrained by direct detection experiments, as
stated above. Additionally, due to a difference of factor $M_\phi^2$ in
Eqs.~\eqref{sigmaN_scat_scalar_scal} and \eqref{sigmaN_scat_scalar_vect},
the predicted scattering cross section curves slope downward for scalar
interactions, while they are nearly horizontal for vector
interactions. This difference is a reflection of the different
dimensions of the two effective operators.

\begin{figure}[!htbp]
\centering
\includegraphics[width=0.44\textwidth]{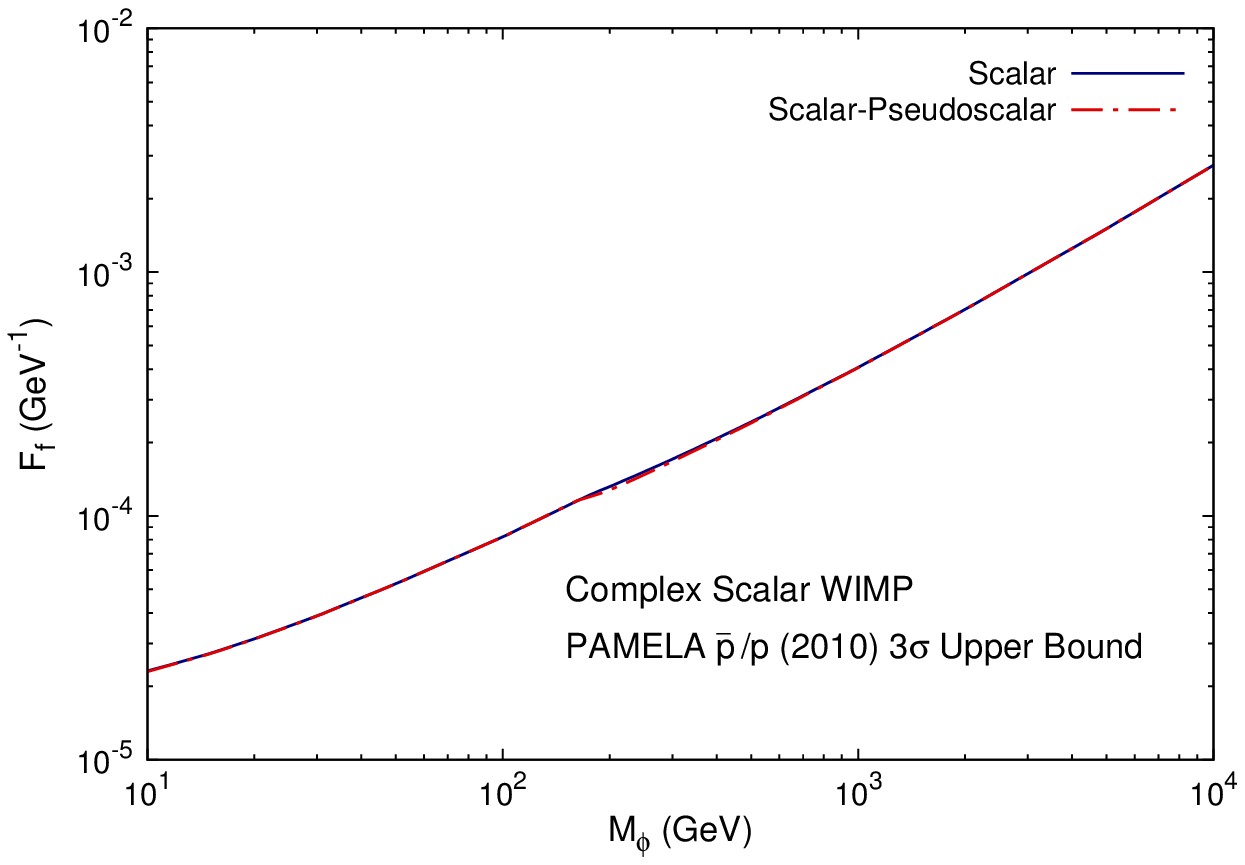}%
\hspace{0.01\textwidth}%
\includegraphics[width=0.44\textwidth]{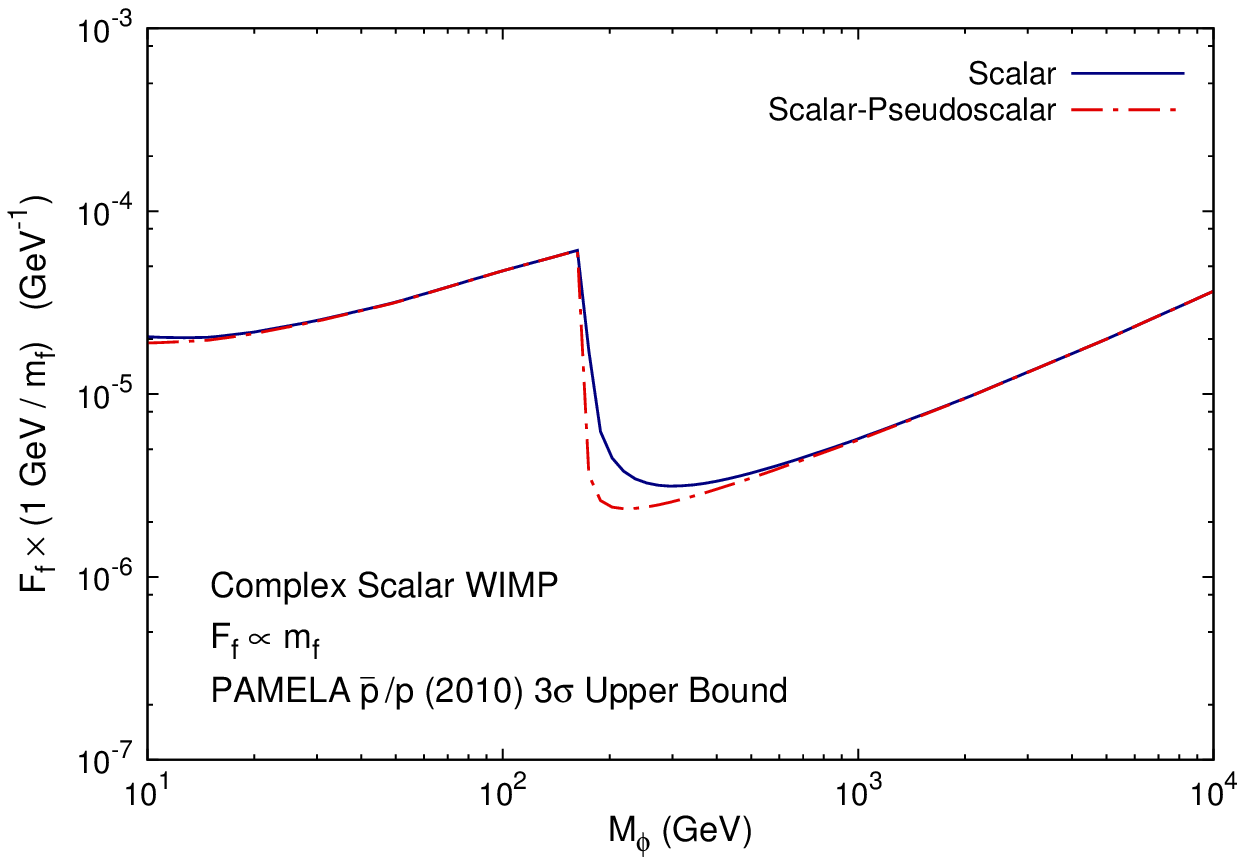}%
\\
\includegraphics[width=0.44\textwidth]{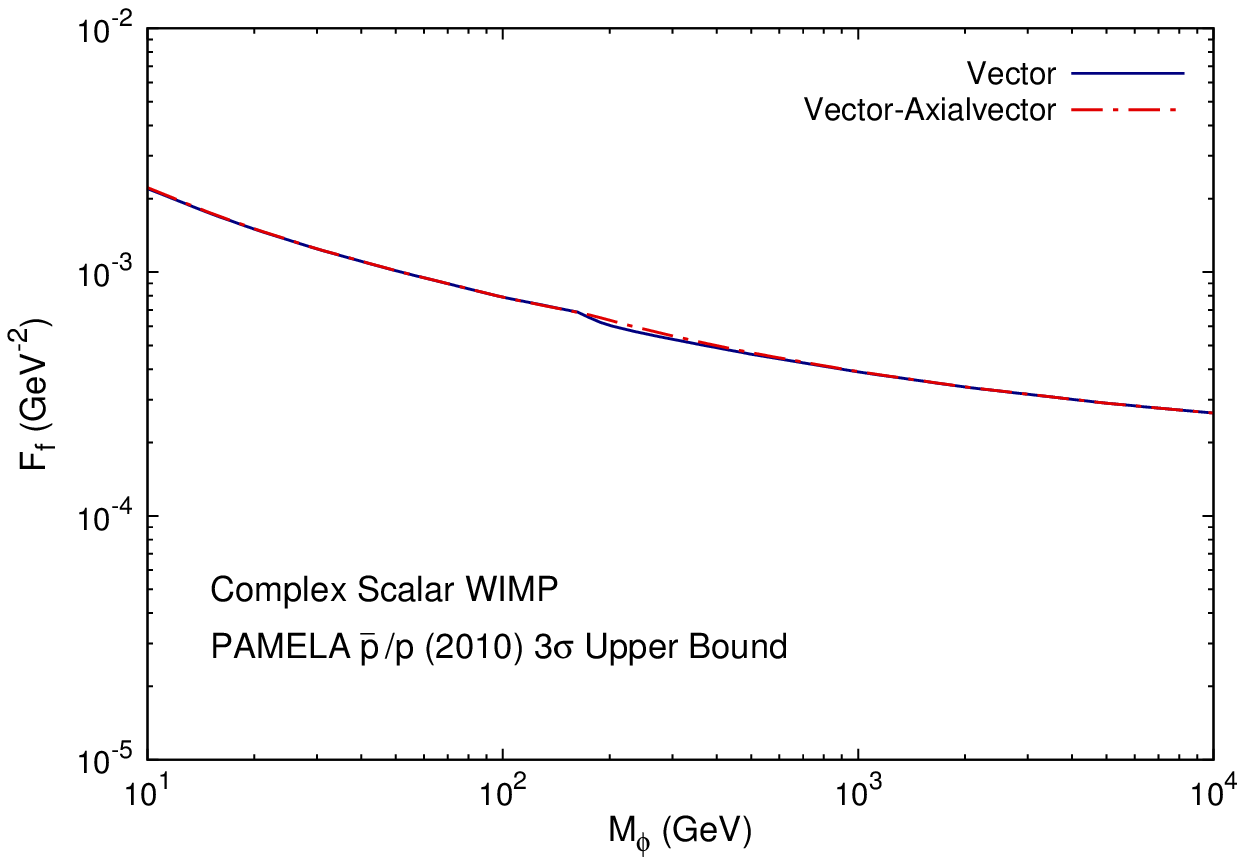}%
\hspace{0.01\textwidth}%
\includegraphics[width=0.44\textwidth]{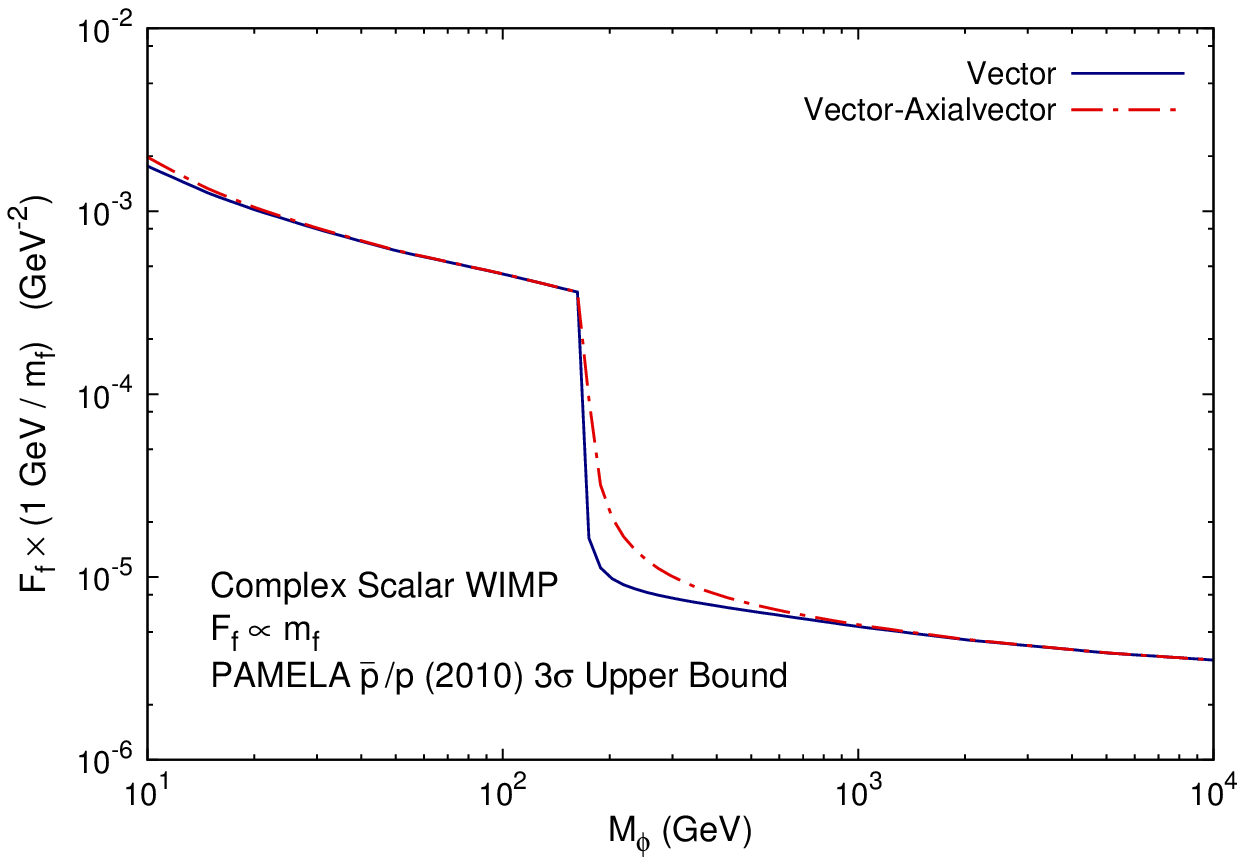}%
\caption{The $3\sigma$ upper bounds on the coupling constants $F_f$ from the
PAMELA $\bar p / p$ ratio~\cite{Adriani:2010rc} in each complex scalar
WIMP effective model. In the left (right) column, results are given
for the case of universal couplings ($F_f \propto m_f$).}
\label{fig:scalar:constr_3s}
\end{figure}

\subsection{Indirect detection\label{subsec-scal-id}}

Compared with the direct search method for the WIMP DM at
underground laboratories on Earth, an indirect detection method
concerning astrophysical effects of DM is relatively independent
and complementary. This method is used to look for the DM annihilation
or decay products which include gamma rays, neutrinos, positrons,
antiprotons and so on. These particles can be detected by satellite,
ground and underground cosmic ray experiments. When propagating in the Galaxy,
the charged products are deflected by Galactic
magnetic fields and interact with interstellar medium.
Therefore we have to consider the propagation process of the
charged particles in order to compare predictions with observations.

The propagation equation of cosmic rays in the Galaxy can be written
down as~\cite{Strong:1998pw}
\begin{eqnarray}
\frac{\partial \psi}{\partial t}= Q(\mathbf{r},p)
+\nabla\cdot(D_{xx}\nabla\psi-\mathbf{V}_c\,\psi)
+\frac{\partial}{\partial p}\left[p^2 D_{pp} \frac{\partial}{\partial p}
\left(\frac{1}{p^2}\psi\right)\right]
-\frac{\partial}{\partial p}\left[\dot p \psi  - \frac{p}{3} \left(\nabla \cdot
\mathbf{V}_c\right)\psi \right] - \frac{1}{\tau_f}\psi
- \frac{1}{\tau_r}\psi, \label{propeq}
\end{eqnarray}
where $\psi = \psi ( {\mathbf{r}},p,t) $ is the number density of
cosmic-ray particles per unit momentum interval, $Q(\mathbf{r},p)$ is
the source term, $D_{xx}$ is the spatial diffusion coefficient,
$\mathbf{V}_c$ is the convection velocity, $D_{pp}$ is the diffusion
coefficient in momentum space describing the reacceleration
process, $\dot p \equiv dp/dt$ is the momentum loss rate, $\tau_f$
and $\tau_r$ are the time scales for fragmentation and the
radioactive decay, respectively.
Following our previous work~\cite{Zheng:2010js}
we consider the antiproton-to-proton flux ratio $\bar p /
p$ measured by the satellite-borne experiment PAMELA
\cite{Adriani:2010rc} to constrain the effective models. To proceed
we solve the propagation equation~\eqref{propeq} by the numerical code GALPROP
\cite{Strong:1998pw}
and compare the solution to the PAMELA $\bar{p}/p$ data.

The source term of $\bar p$ induced by the annihilation of
complex scalar WIMPs is given by
\begin{equation}
Q_{\mathrm{ann}}\left( {{\mathbf{r}},E} \right) =
\frac{\left<\sigma_{\mathrm{ann}} v\right>_\mathrm{tot}}{4M_\phi^2}
\left[\sum\limits_q B_q\left(\frac{dN_{\bar p}}{dE_{\bar p}}\right)_q\right]
\rho^2(\mathbf{r}), \qquad q=u,d,s,c,b,t,
\label{dm_anni_pbar}
\end{equation}
where $\left\langle {\sigma_{\mathrm{ann}} v}
\right\rangle_\text{tot}$ is the total annihilation rate of WIMPs
to quarks, $B_q$ is the branching ratio of each $\bar{q}q$ channel,
$\rho(\mathbf{r})$ is the mass density distribution of the Galactic
DM halo, $(dN_{\bar p}/dE_{\bar p})_q$ is the number per unit
energy interval of the antiprotons produced by a WIMP pair-annihilation
in the $q \bar q$ channel. In particular, $(dN_{\bar p}/dE_{\bar p})_q$
is calculated by the Monte Carlo code
PYTHIA~\cite{Sjostrand:2006za}. For real scalar WIMPs, the factor
$\left<\sigma_{\mathrm{ann}} v\right>_\mathrm{tot} / 4M_\phi^2$
in Eq.~\eqref{dm_anni_pbar} is modified to
$\left<\sigma_{\mathrm{ann}} v\right>_\mathrm{tot} / 2M_\phi^2$
since the two annihilating WIMP are indistinguishable.

The NFW profile~\cite{Navarro:1996gj} is taken to describe
$\rho(\mathbf{r})$ with the characteristic density
$\rho_s = 0.334~\mathrm{GeV~cm^3}$ and the scale radius $r_s = 20$~kpc,
which leads to a local DM density $\rho(8.33~\mathrm{kpc})=
0.4~\mathrm{GeV~cm^{-3}}$. The DM particles in the Galactic halo
obey the Maxwell-Boltzmann velocity distribution $f(v_0) =
(M_\phi/2\pi k_\mathrm{B} T)^{3/2} \exp (-M_\phi v_0^2/2k_\mathrm{B} T)$.
Their velocity dispersion $\bar v_0 \equiv \sqrt{\left<v_0^2
\right>} = \sqrt{3T/M_\phi}$ is taken to be the canonical value
$270~{\mathrm{km~s^{-1}}}$~\cite{Jungman:1995df}.
In the calculation of the $\bar p / p$ spectrum with GALPROP,
the Galaxy propagation model with diffusion and convection is adopted
and the half-height of the Galaxy propagation halo $z_h$ is set to be
4~kpc. The PAMELA $\bar p / p$ spectrum is best fitted by astrophysical
background with a solar modulation potential of $\Phi = 335~\mathrm{MV}$,
which gives the minimal $\chi^2$.
Under such potential if the contribution of DM is added in,
the resulted spectrum deviates from the background and gives a
larger $\chi ^2$. Thus for the fixed $M_\phi$ in both the cases of
universal couplings and $F_f\propto m_f$ for each type of
effective interactions, we can derive the 3$\sigma$ upper bounds on
$\left<\sigma_{\mathrm{ann}}v\right>_\mathrm{tot}$,
which eventually set the bounds on effective coupling constants $F_f$,
as shown in Fig.~\ref{fig:scalar:constr_3s}.

In Fig.~\ref{fig:scalar:constr_3s}, the upper bounds set for S and SP
interactions and those for V and VA interactions are nearly identical
respectively, with slight differences at $M_\phi \sim m_t$ in the case
of $F_f\propto m_f$. The reason is the same as that explained in
Subsection \ref{subsec-scal-rd}.
It is also worth noting that the upper bound curves for the operators
with different dimensions have different trends of slope,
due to the additional factor $M_\phi^2$ in the annihilation cross sections
of dimension-6 operators (V and VA). For the case of $F_f\propto m_f$,
downward bends at $M_\phi \sim m_t$ in the curves appear again, because
of the $\phi \phi^\dag \to t \bar t$ threshold effect.

\begin{figure}[!htbp]
\centering
\includegraphics[width=0.49\textwidth]{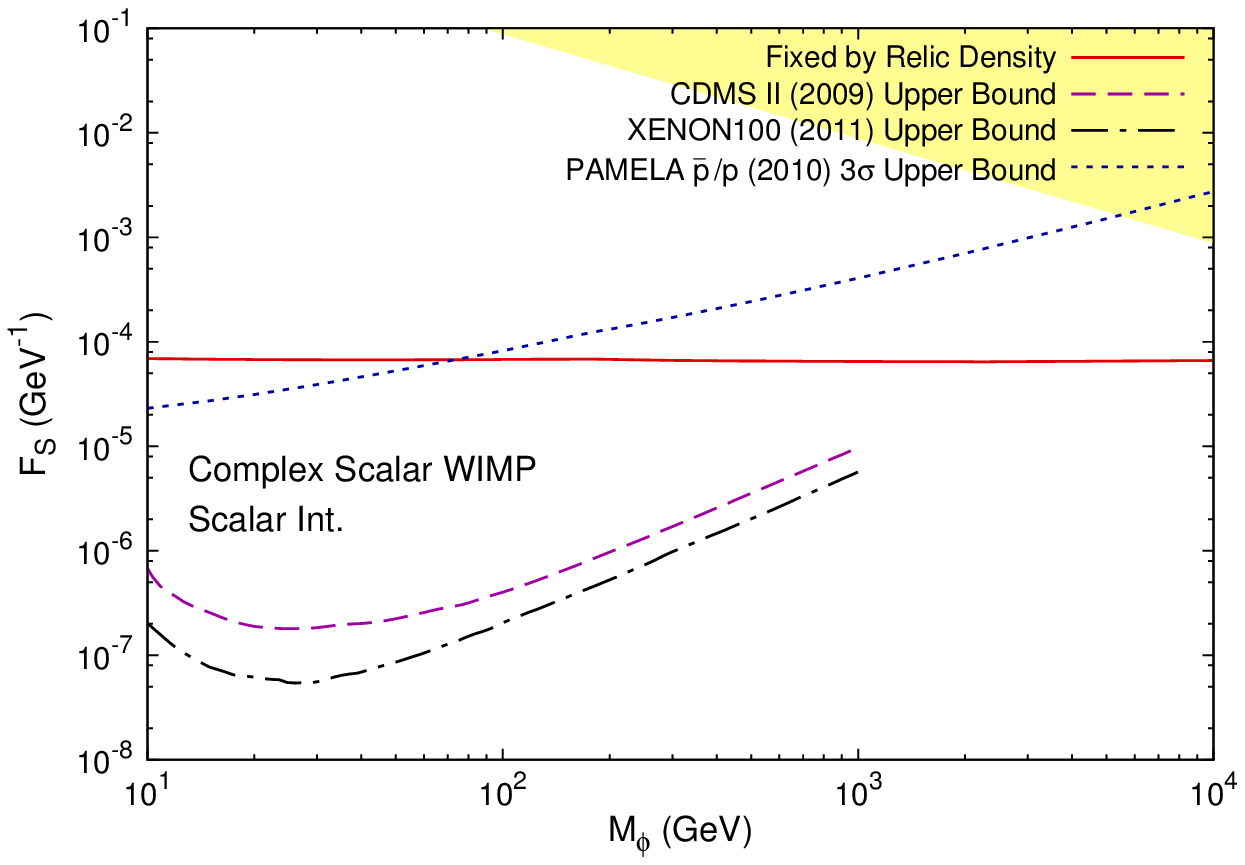}%
\hspace{0.008\textwidth}%
\includegraphics[width=0.49\textwidth]{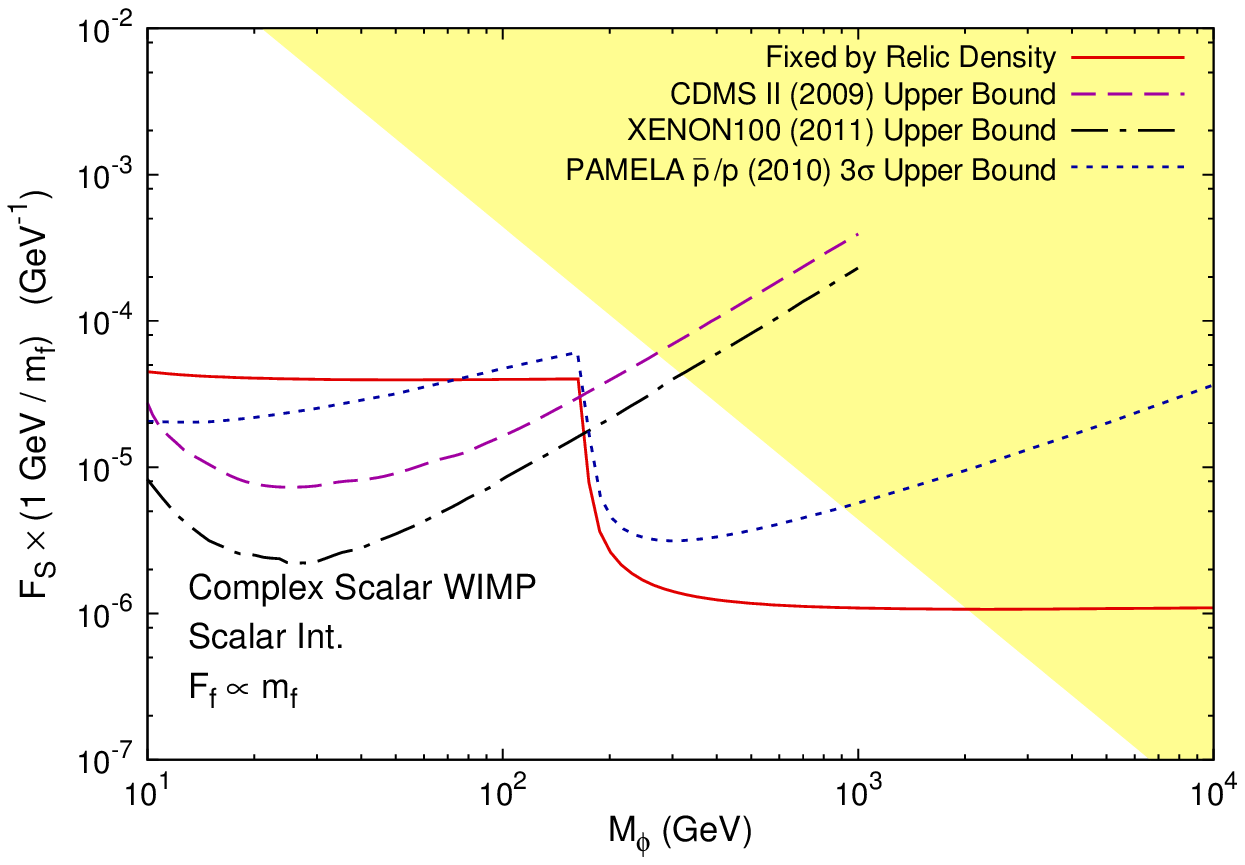}%
\caption{Combined constraints on coupling constants $F_f$ of complex scalar
WIMPs with scalar (S) interaction from relic density, direct detection
experiments of CDMS II and XENON100, PAMELA $\bar p / p$ ratio,
and validity of effective theory. The yellow region denotes the invalid
parameter space of effective field theory.
The left (right) frame is shown for the case of universal couplings
($F_f\propto m_f$).
\label{fig-scal-combined-S}}
\end{figure}
\begin{figure}[!htbp]
\centering
\includegraphics[width=0.49\textwidth]{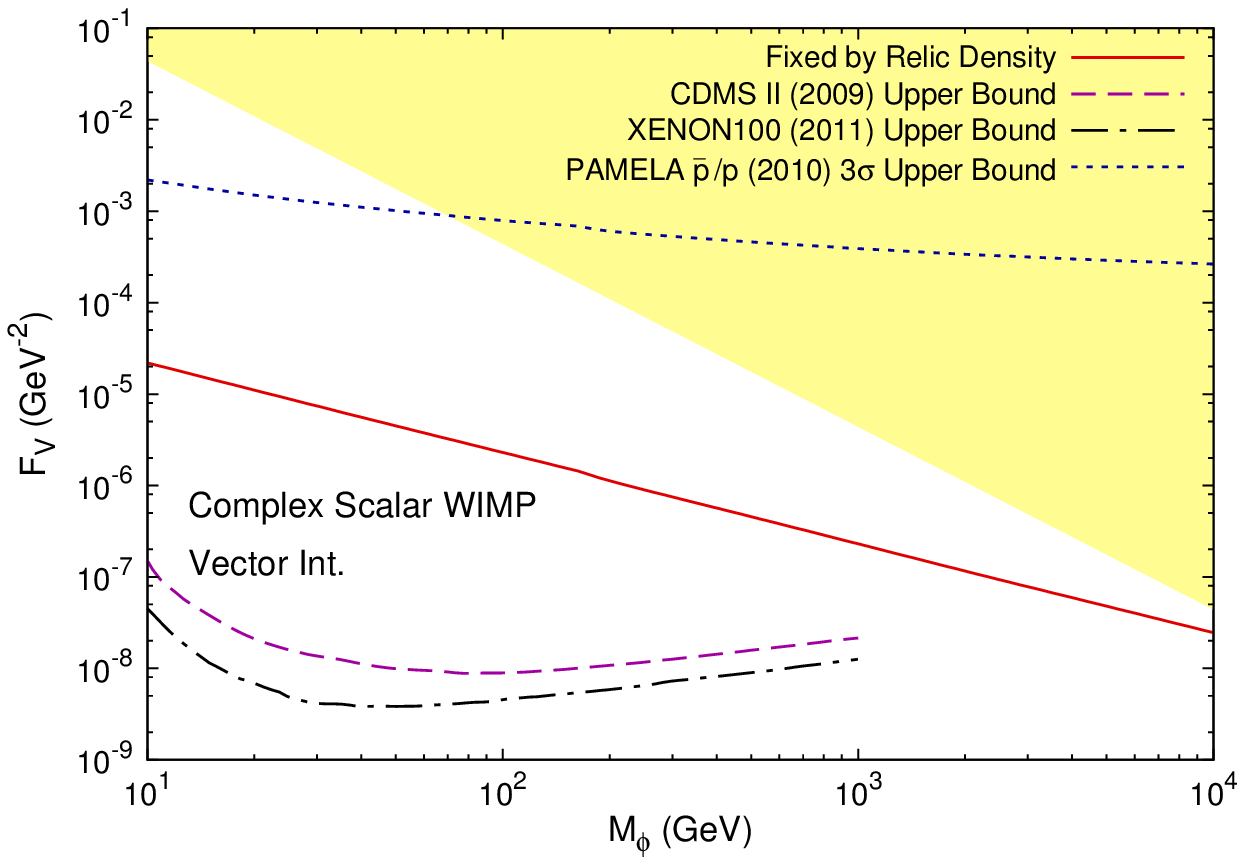}%
\hspace{0.008\textwidth}%
\includegraphics[width=0.49\textwidth]{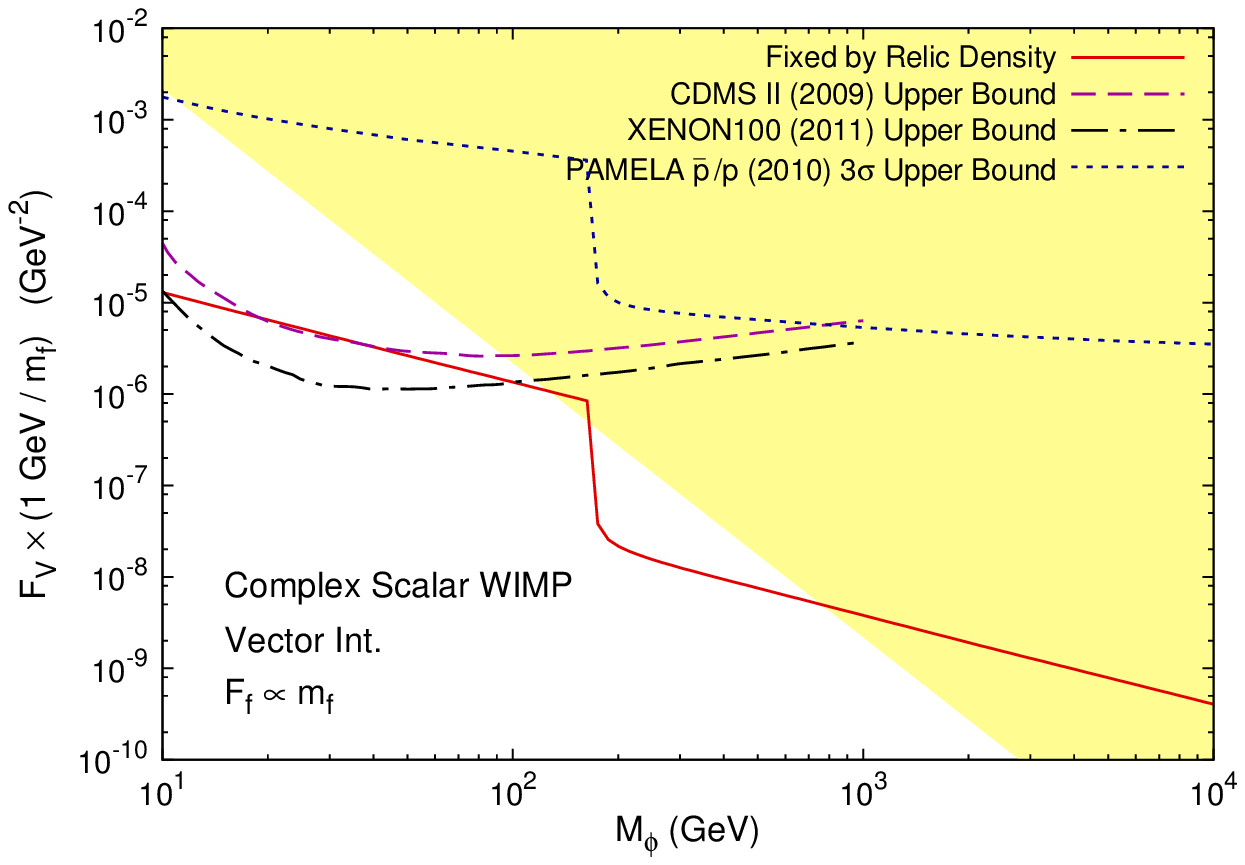}%
\caption{Combined constraints on coupling constants $F_f$ of
complex scalar WIMPs with vector (V) interaction from relic density,
direct detection experiments of CDMS II and XENON100,
PAMELA $\bar p / p$ ratio, and validity of effective theory.
\label{fig-scal-combined-V}}
\end{figure}
\begin{figure}[!htbp]
\centering
\includegraphics[width=0.49\textwidth]{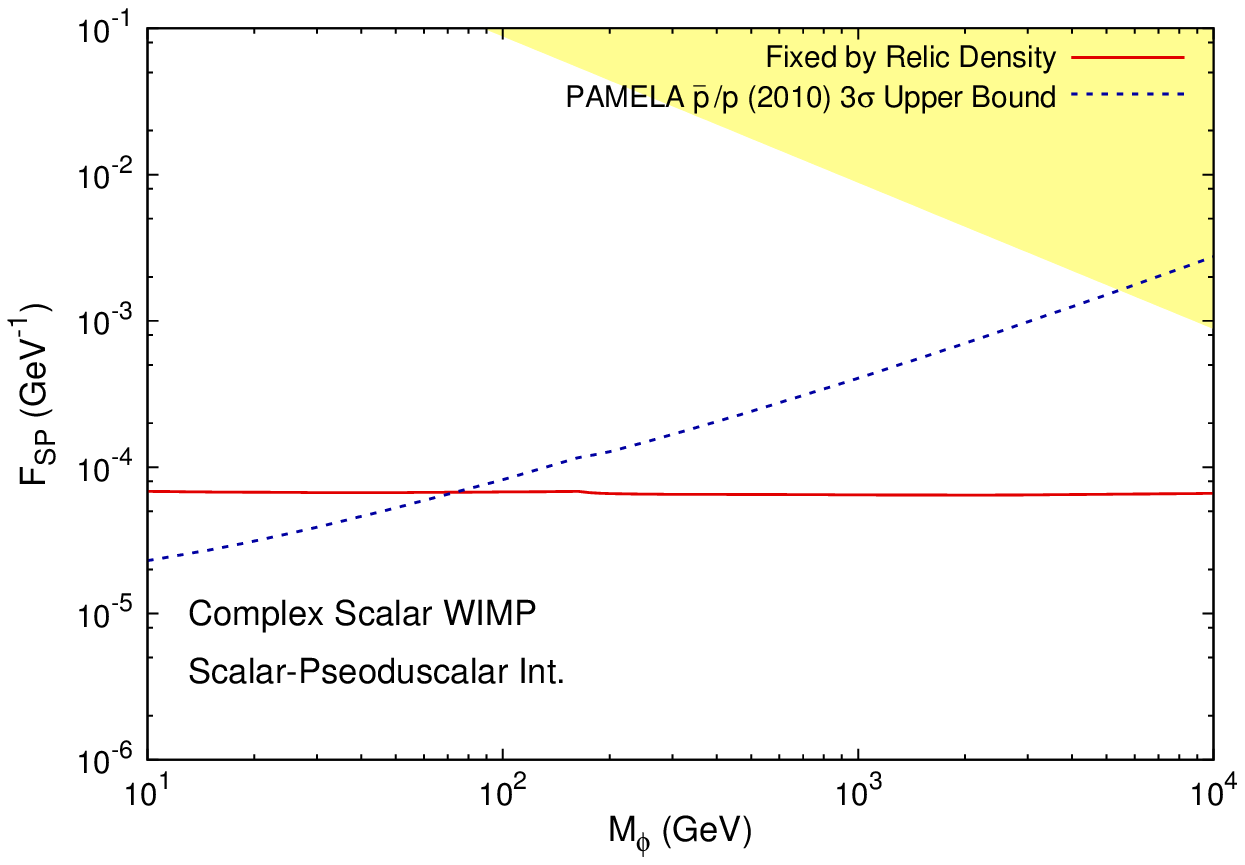}
\hspace{0.008\textwidth}%
\includegraphics[width=0.49\textwidth]{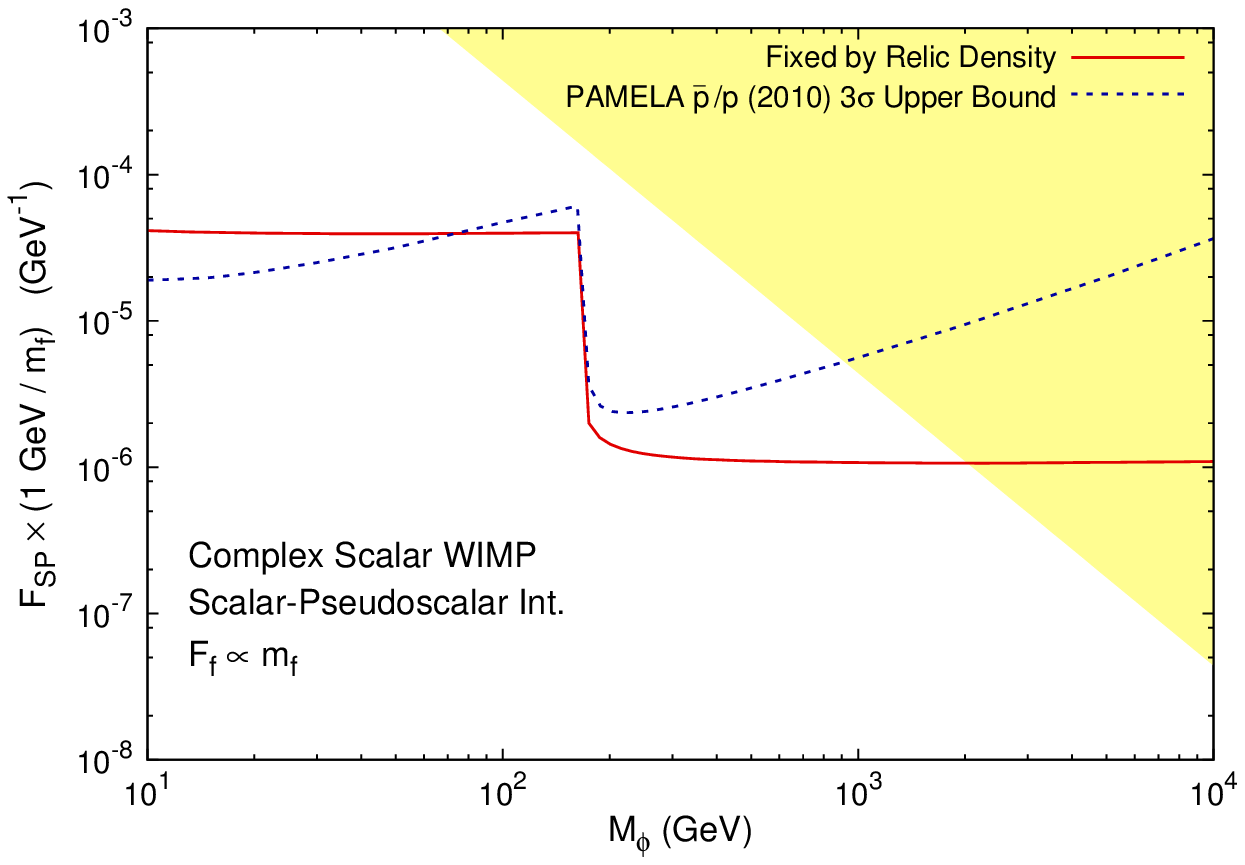}%
\caption{Combined constraints on coupling constants $F_f$ of
complex scalar WIMPs with scalar-pseudoscalar (SP) interaction
from relic density, PAMELA $\bar p / p$ ratio, and validity of
effective theory.
\label{fig-scal-combined-s_Ps}}
\end{figure}
\begin{figure}[!htbp]
\centering
\includegraphics[width=0.49\textwidth]{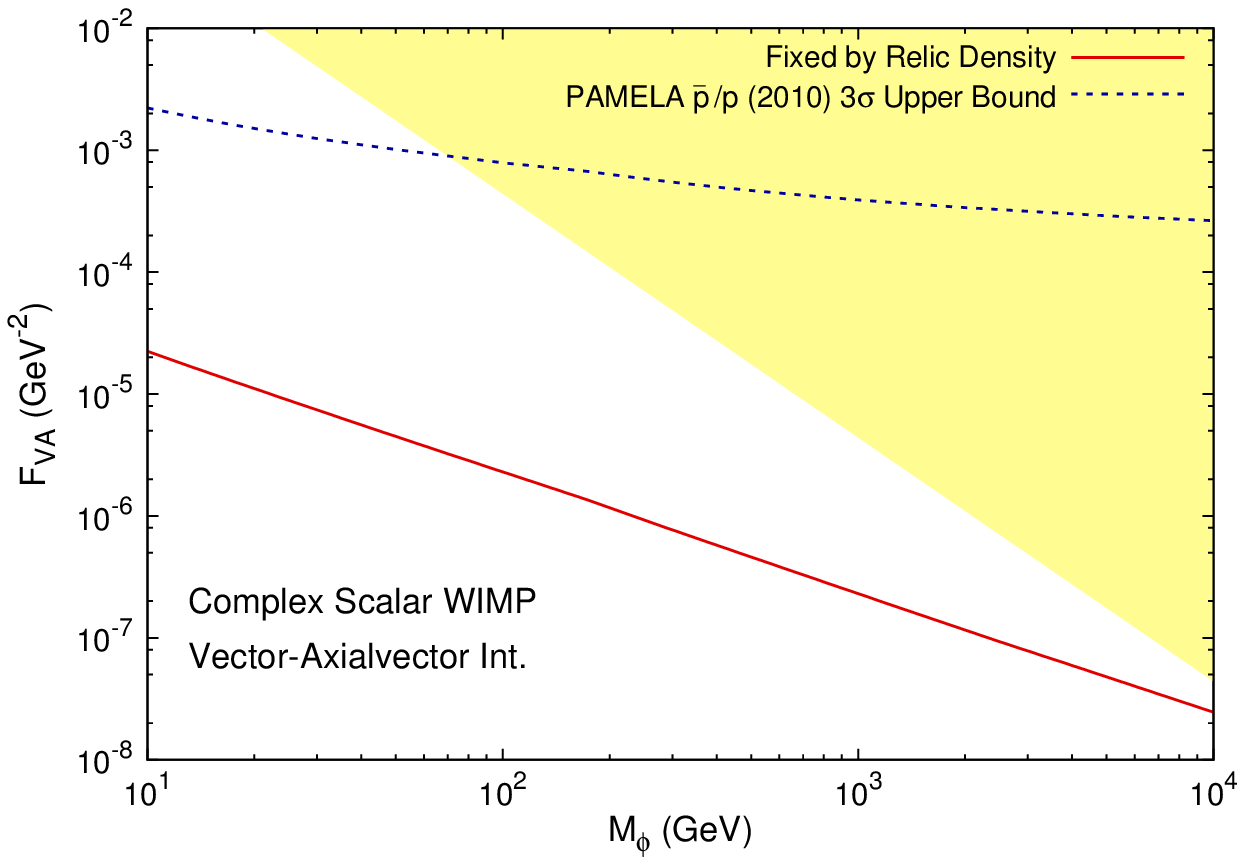}%
\hspace{0.008\textwidth}%
\includegraphics[width=0.49\textwidth]{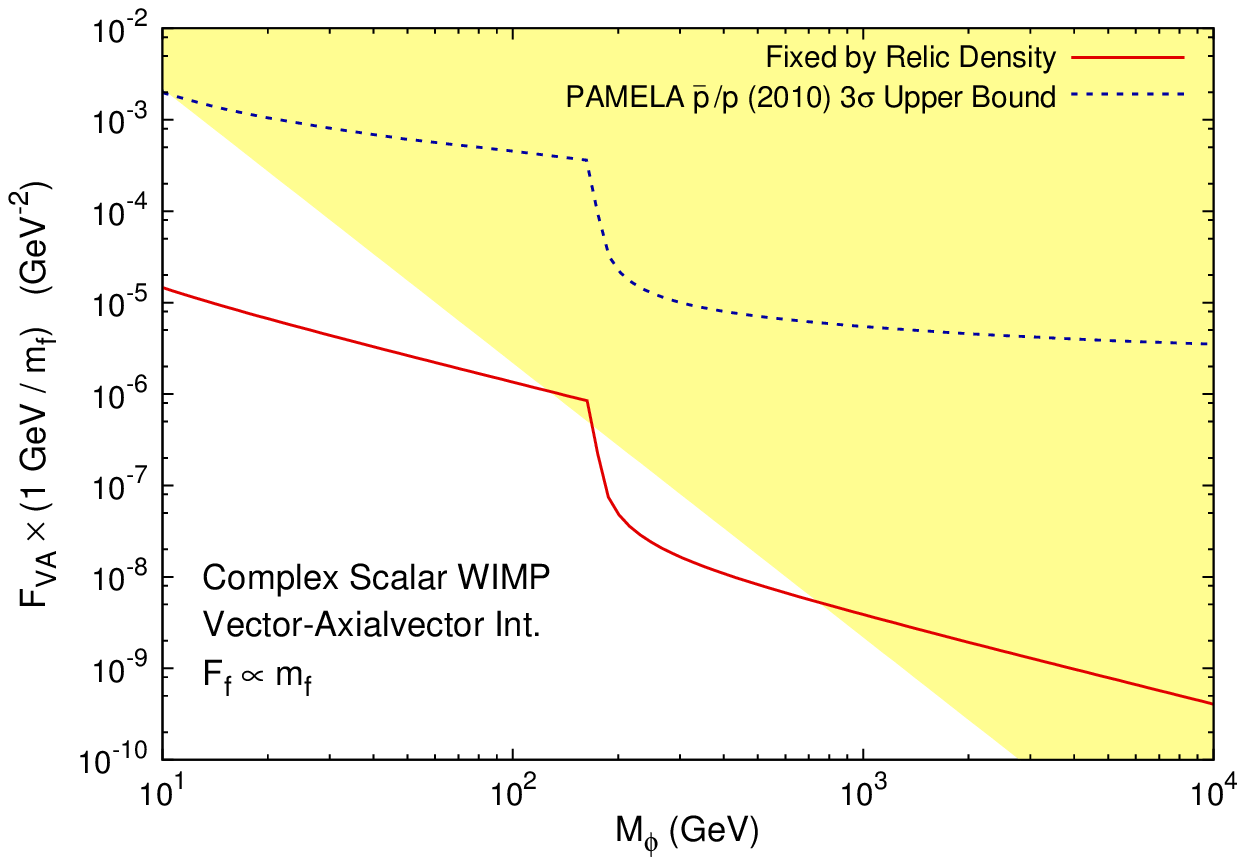}%
\caption{Combined constraints on coupling constants $F_f$ of
complex scalar WIMPs with vector-axialvector (VA) interaction
from relic density, PAMELA $\bar p / p$ ratio, and validity of
effective theory.
\label{fig-scal-combined-v_Av}}
\end{figure}

\subsection{Validity region of effective models and combined constraints
\label{subsec-scal-comb}}

In this subsection, we will discuss the validity region where the method
of effective theory can be used. Since our previous analysis is made in
both the cases of universal couplings and $F_f\propto m_f$, we consider them
case by case in the following.
\begin{itemize}
  \item In the case of universal couplings, for dimension-5 operators
S and SP (dimension-6 operators V and VA), $F_f$ can be written as
$F_f / \sqrt{2}=\alpha / \Lambda$ ($F_f / \sqrt{2} = \alpha / \Lambda^2$),
where $\Lambda$ is the cutoff energy scale and $\alpha$ is the coupling
of the fundamental theory beyond $\Lambda$, which may be of order 1.
The transfer momentum of the annihilation process $\phi \phi^\dag
\to f \bar f$ must be well below the cutoff, i.e., $2M_\phi \ll \Lambda$,
so that the effective theory can be used. On the other hand, a weakly
coupled UV completion of the effective theory usually requires
$\alpha < 4\pi$ such that the perturbative method is valid
\cite{Goodman:2010ku,Bi:2009am}. From the above 3 relations, we obtain
    \begin{eqnarray}
    F_f \ll \frac{2\sqrt{2}\pi}{M_\phi},&\quad&
        \text{for S and SP interactions}, \label{validity-condi-scal-5}\\
    F_f \ll \frac{\sqrt{2}\pi}{M_\phi^2},&\quad&
        \text{for V and VA interactions}. \label{validity-condi-scal-6}
    \end{eqnarray}
  \item In the case of $F_f\propto m_f$, for dimension-5 operators S and SP
(dimension-6 operators V and VA), $F_f$ can be written as
$F_f / \sqrt{2} = \alpha m_f / \Lambda^2$ ($F_f / \sqrt{2}
= \alpha m_f / \Lambda^3$).
Likewise, we still have $2M_\phi\ll \Lambda$
and $\alpha < 4\pi$. From the above 3 relations, we obtain
    \begin{eqnarray}
    \frac{F_f}{m_f} \ll \frac{\sqrt{2}\pi}{M_\phi^2},&\quad&
        \text{for S and SP interactions}, \label{validity-condi-scal-mf-5}\\
    \frac{F_f}{m_f} \ll \frac{\pi}{\sqrt{2}M_\phi^3},&\quad&
        \text{for V and VA interactions}. \label{validity-condi-scal-mf-6}
    \end{eqnarray}
\end{itemize}

\begin{table}[!htbp]
\begin{center}
\belowcaptionskip=0.2cm
\caption{A summary for complex scalar WIMPs
with various effective interactions. The excluded regions of
$M_\phi$ given by direct and indirect experiments are indicated.}
\label{tab:scalar_sum}
\renewcommand{\arraystretch}{1.3}
\begin{tabular}{ccc}
\hline \hline \multicolumn{3}{c}{Universal coupling} \\
Interaction & Direct detection & PAMELA $\bar p / p$ \\
\hline
S  & Excluded $M_\phi \simeq 10~\mathrm{GeV} - \mathrm{above}~1~\mathrm{TeV}$
   & Excluded $M_\phi \simeq 10 - 73~\mathrm{GeV}$ \\
V  & Excluded $M_\phi \simeq 10~\mathrm{GeV} - \mathrm{above}~1~\mathrm{TeV}$
   & Not sensitive \\
SP & Not sensitive
   & Excluded $M_\phi \simeq 10 - 73~\mathrm{GeV}$ \\
VA & Not sensitive & Not sensitive \\
\hline \hline \multicolumn{3}{c}{$F_f \propto m_f$} \\
Interaction & Direct detection & PAMELA $\bar p / p$ \\
\hline
S  & Excluded $M_\phi \simeq 10 - 168~\mathrm{GeV}$
   & Excluded $M_\phi \simeq 10 - 73~\mathrm{GeV}$ \\
V  & Excluded $M_\phi \simeq 10 - 100~\mathrm{GeV}$ & Not sensitive \\
SP & Not sensitive & Excluded $M_\phi \simeq 10 - 73~\mathrm{GeV}$ \\
VA & Not sensitive & Not sensitive \\
\hline \hline
\end{tabular}
\end{center}
\end{table}

Considering the validity conditions \eqref{validity-condi-scal-5} --
\eqref{validity-condi-scal-mf-6} altogether with the other phenomenological
constraints, we derive the combined constraints on the effective models.
In Figs.~\ref{fig-scal-combined-S} and \ref{fig-scal-combined-V} we show
the combined constraints on coupling constants $F_f$ of complex scalar WIMPs
with S and V interactions from relic density, direct detection experiments
of CDMS II and XENON100, PAMELA $\bar p / p$ ratio and validity of
effective theory, while in Figs.~\ref{fig-scal-combined-s_Ps} and
\ref{fig-scal-combined-v_Av}, we show the combined constraints on
$F_f$  with SP and VA interactions from
relic density, PAMELA $\bar p / p$ ratio and validity of effective theory.
The invalid parameter space of effective field theory is denoted by
the yellow region in each figure.
The constraints are much similar between S and SP interactions, and so are
those between V and VA interactions, except that the S and V interactions
are constrained by direct detection experiments, while the SP and VA
interactions are not.

There are some interesting features in
Figs.~\ref{fig-scal-combined-S} -- \ref{fig-scal-combined-v_Av}.
In the case of universal coupling for S and V interactions,
the WIMP with $M_\phi < 1$~TeV are strongly excluded by direct detection
experiments, while in the case of $F_f \propto m_f$, the direct
detection exclude a much smaller range of $M_\phi$.
The indirect detection constraints also rule out the WIMP with $M_\phi \lesssim
70$~GeV for S and SP interactions, while the constraints becomes much weaker
for V and VA interactions. The reason is that for V and VA interactions
the annihilation rate \eqref{sv_scalar_vect} and \eqref{sv_scalar_v_Av} are
of order $\mathcal{O}(T/M_\phi)$, which lead to looser bounds by
indirect detection because of the much lower WIMP temperature today
comparing with that of the WIMP freeze-out epoch.
Additionally, in the case of $F_f\propto m_f$ the effective
interactions are valid only up to $M_\phi \sim$~TeV.
A larger WIMP mass in these cases may implies violation to our
previous assumptions, such as the existence of a light intermediate
state, or the effect of resonances or coannihilations.

If the effective coupling is much weaker than that derived from the
observed relic density, the thermal DM production may overclose
the Universe. Therefore the DM models in which the direct or indirect
detection constraints are stronger than that of relic density should
be excluded, or else some exotic entropy generation processes should
occur after DM froze out.
As a summary of the study on the scalar WIMP,
in Table~\ref{tab:scalar_sum},
we indicate the excluded regions of $M_\phi$ given by direct and indirect
experiments for complex scalar WIMPs with various effective interactions.

\section{Vector dark matter\label{sec-vect}}

In this section, we will discuss the case that DM consists of complex vector
WIMPs ($X^\mu$ and its antiparticle $X^{\mu*}$). Notable differences between
this case and the case of real vector WIMPs will be remarked below.
Following the procedure in which we discuss scalar WIMPs
in Section \ref{sec-scal}, we construct effective interaction operators
of $X^{\mu\ast}$-$X^{\nu}$-$\bar f$-$f$ type between
the WIMPs and the SM fermions, which are limited only by the requirements
of Hermiticity, Lorentz invariance and CPT invariance.

The assumptions listed in Section \ref{sec-scal} are also adopted. Thus the
effective interaction term of Lagrangian between two complex vector WIMPs
($X^\mu$ and $X^{\mu*}$) and two SM fermions ($f$ and $\bar f$) is given by
only one of the following expressions:
\begin{eqnarray}
\text{Scalar int. (S)}:&&\qquad{} \mathcal{L}_\mathrm{S} =
\sum\limits_f \frac{K_{\mathrm{S},f}}{\sqrt{2}}X_\mu^* X^\mu \bar f f,
\label{lag:vector:scal} \\
\text{Vector int. (V)}:&&\qquad{} \mathcal{L}_\mathrm{V} =
\sum\limits_f \frac{K_{\mathrm{V},f}}{\sqrt{2}} (X_\nu^*i
\overleftrightarrow{\partial_\mu} X^\nu)\bar f \gamma^\mu f,
\label{lag:vector:vect} \\
\text{Tensor int. (T)}:&&\qquad{} \mathcal{L}_\mathrm{T} =
\sum\limits_f \frac{K_{\mathrm{T},f}}{\sqrt{2}}i (X_\mu^* X_\nu
- X_\nu^* X_\mu)\bar f  \sigma^{\mu\nu} f,
 \label{lag:vector:tens} \\
\text{Scalar-Pseudoscalar int. (SP)}:&&\qquad{}
\mathcal{L}_\mathrm{SP} = \sum\limits_f \frac{K_{\mathrm{SP},f}}
{\sqrt{2}}X_\mu^* X^\mu\bar f i{\gamma_5}f,
\label{lag:vector:s_Ps} \\
\text{Vector-Axialvector int. (VA)}:&&\qquad{}
\mathcal{L}_\mathrm{VA} = \sum\limits_f \frac{K_{\mathrm{VA},f}}
{\sqrt{2}}(X_\nu^* i\overleftrightarrow{\partial_\mu}
X^\nu)\bar f \gamma^\mu \gamma_5 f,
\label{lag:vector:v_Av} \\
\text{Alternative Vector int. ($\widetilde{\mathrm{V}}$)}:&&\qquad{}
\mathcal{L}_{\widetilde{\mathrm{V}}} = \sum\limits_f
\frac{\tilde{K}_{\mathrm{V},f}}{\sqrt{2}} \varepsilon^{\mu\nu\rho\sigma}
(X_\mu^* \overleftrightarrow{\partial_\nu} X_\rho)\bar f \gamma_\sigma f,
\label{lag:vector:alt_vect} \\
\text{Alternative Vector-Axialvector int. ($\widetilde{\mathrm{VA}}$)}:
&&\qquad{}\mathcal{L}_{\widetilde{\mathrm{VA}}} = \sum\limits_f
\frac{\tilde{K}_{\mathrm{VA},f}}{\sqrt{2}}\varepsilon^{\mu\nu\rho\sigma}
(X_\mu^* \overleftrightarrow{\partial_\nu}X_\rho)\bar f
\gamma_\sigma \gamma_5 f,
\label{lag:vector:alt_v_Av} \\
\text{Alternative Tensor int. ($\widetilde{\mathrm{T}}$)}:&&\qquad{}
\mathcal{L}_{\widetilde{\mathrm{T}}} = \sum\limits_f
\frac{\tilde{K}_{\mathrm{T},f}}{\sqrt{2}} \varepsilon^{\mu\nu\rho\sigma}
i(X_\mu^* X_\nu  - X_\nu^* X_\mu)\bar f \sigma_{\rho\sigma}f,
\label{lag:vector:alt_tens}
\end{eqnarray}
where the sum of $f$ is over all the SM fermions,
and the effective coupling constants $K_f$ are real numbers.
For the dimension-5 operators (S, SP, T and $\widetilde{\mathrm{T}}$),
the coupling constants $K_f$ have mass dimension of $-1$,
while for the dimension-6 operators (V, VA, $\widetilde{\mathrm{V}}$ and
$\widetilde{\mathrm{VA}}$), $K_f$ have mass dimension of $-2$.
Operators with more derivatives on the fields are not considered,
since such operators are more suppressed at low energy scale due to
their higher mass dimensions.
Note that for the case of real vector WIMPs, the V, T, VA and
$\widetilde{\mathrm{T}}$ interactions vanish.
In the following, we will discuss the phenomenological constraints
on $K_f$ given by the recent results of the DM relic density,
direct and indirect detection experiments.

\begin{figure}[!htbp]
\centering
\includegraphics[width=0.44\textwidth]{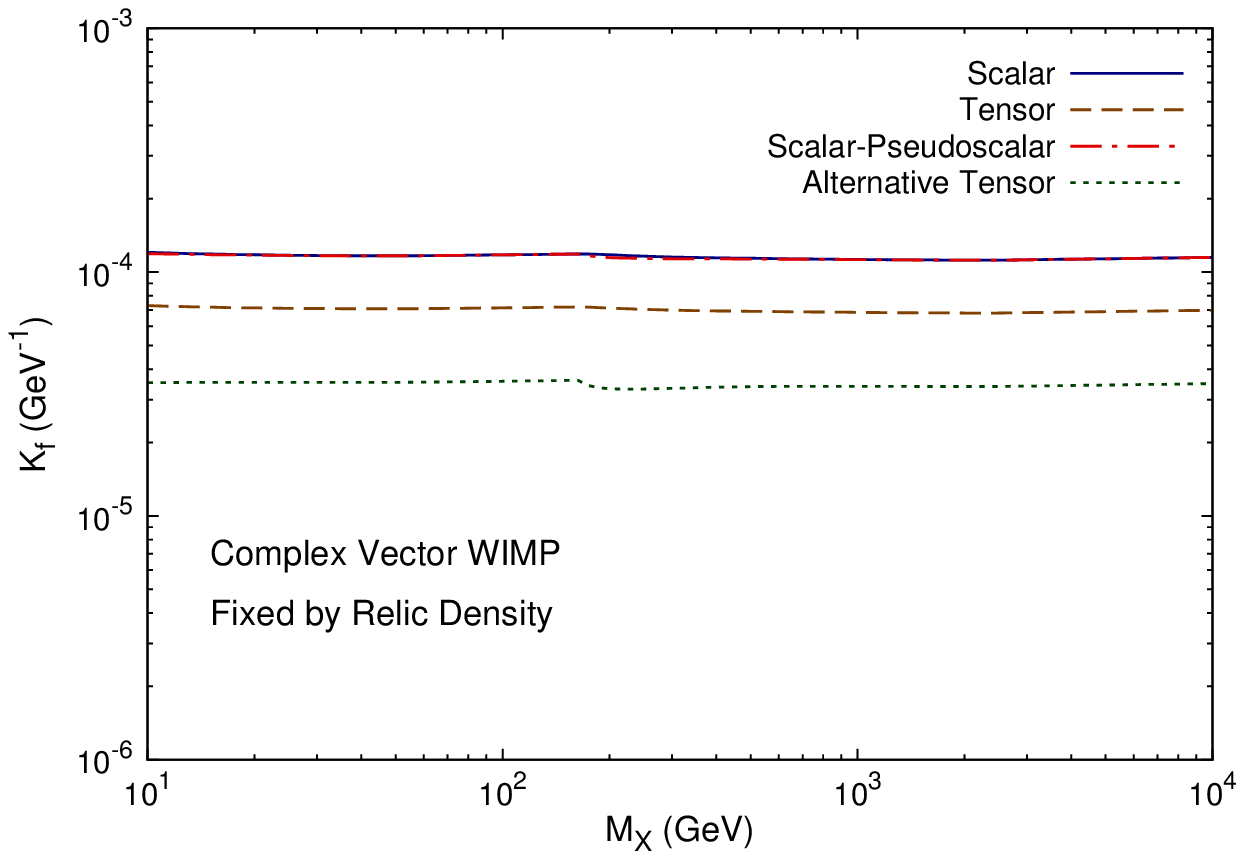}%
\hspace{0.01\textwidth}%
\includegraphics[width=0.44\textwidth]{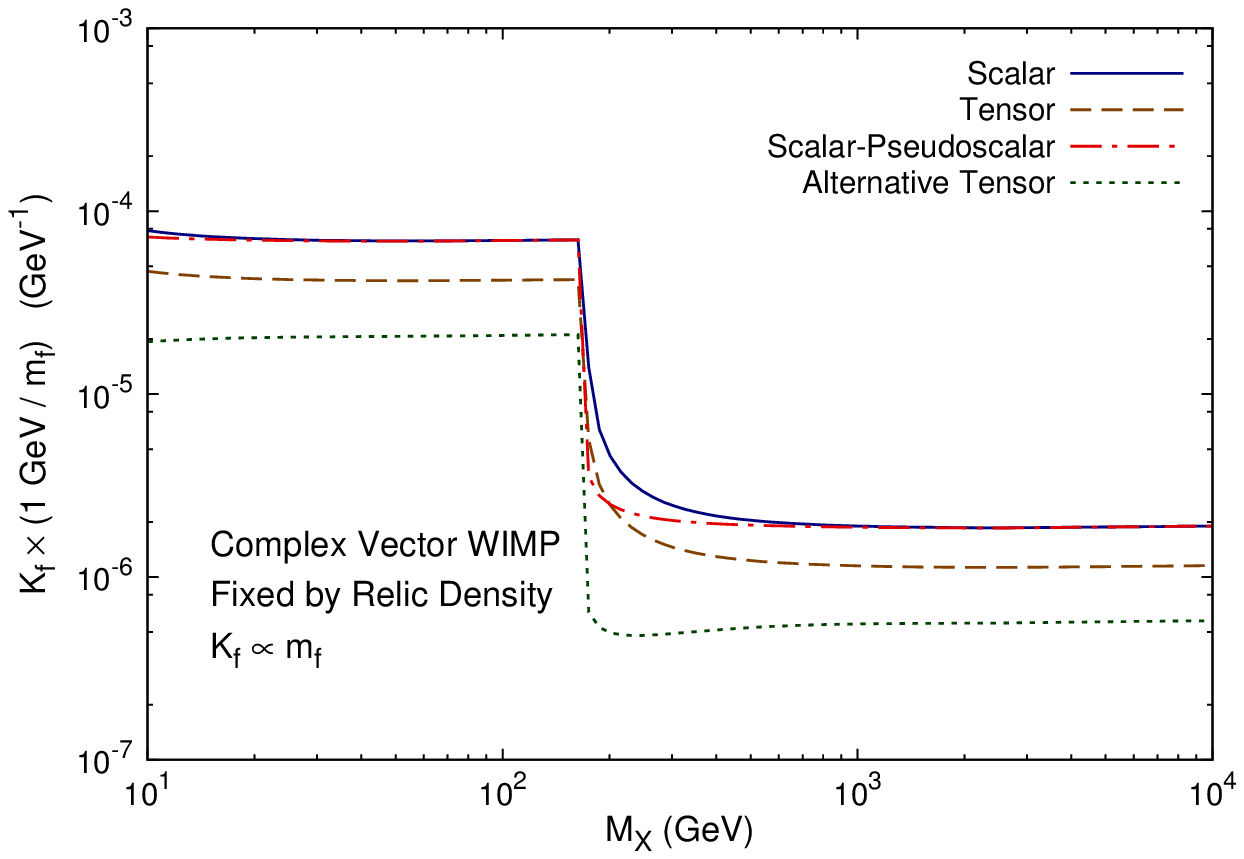}
\\
\includegraphics[width=0.44\textwidth]{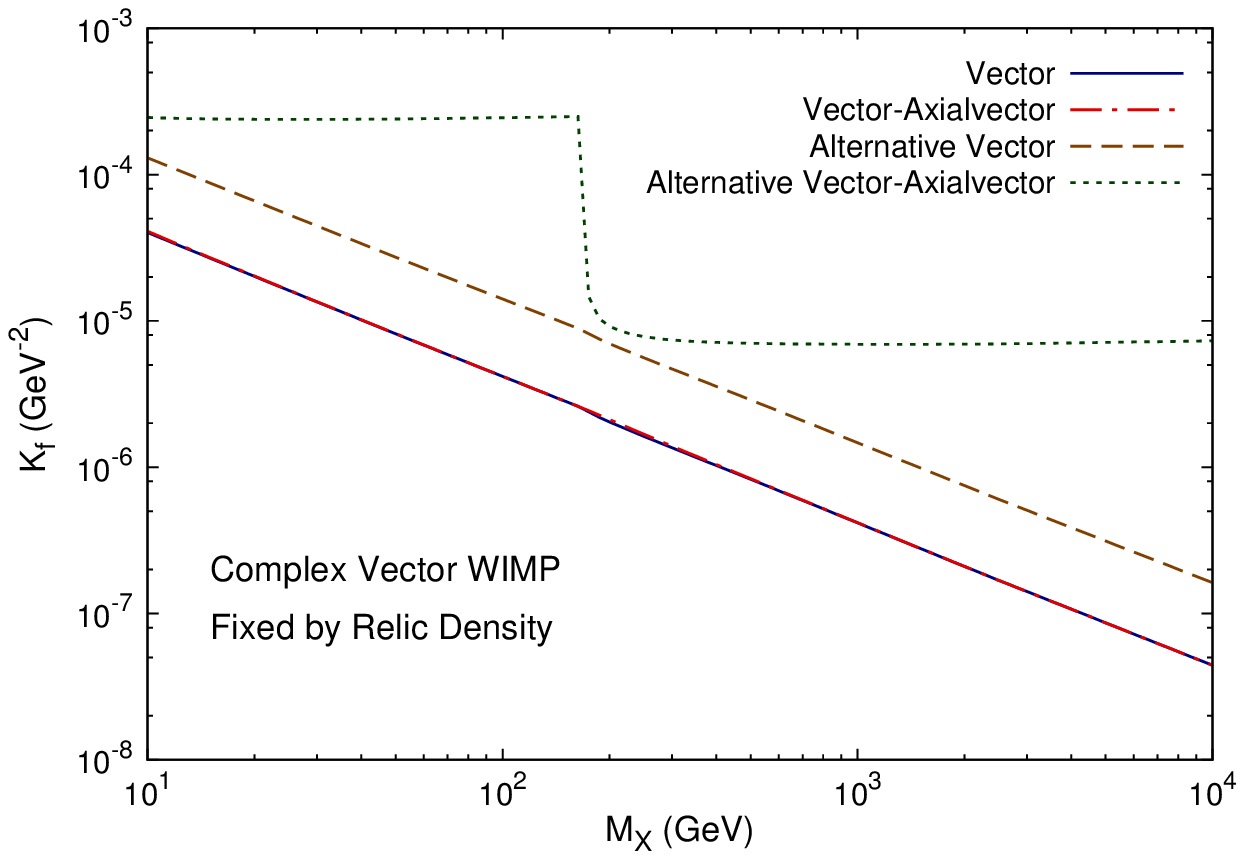}%
\hspace{0.01\textwidth}%
\includegraphics[width=0.44\textwidth]{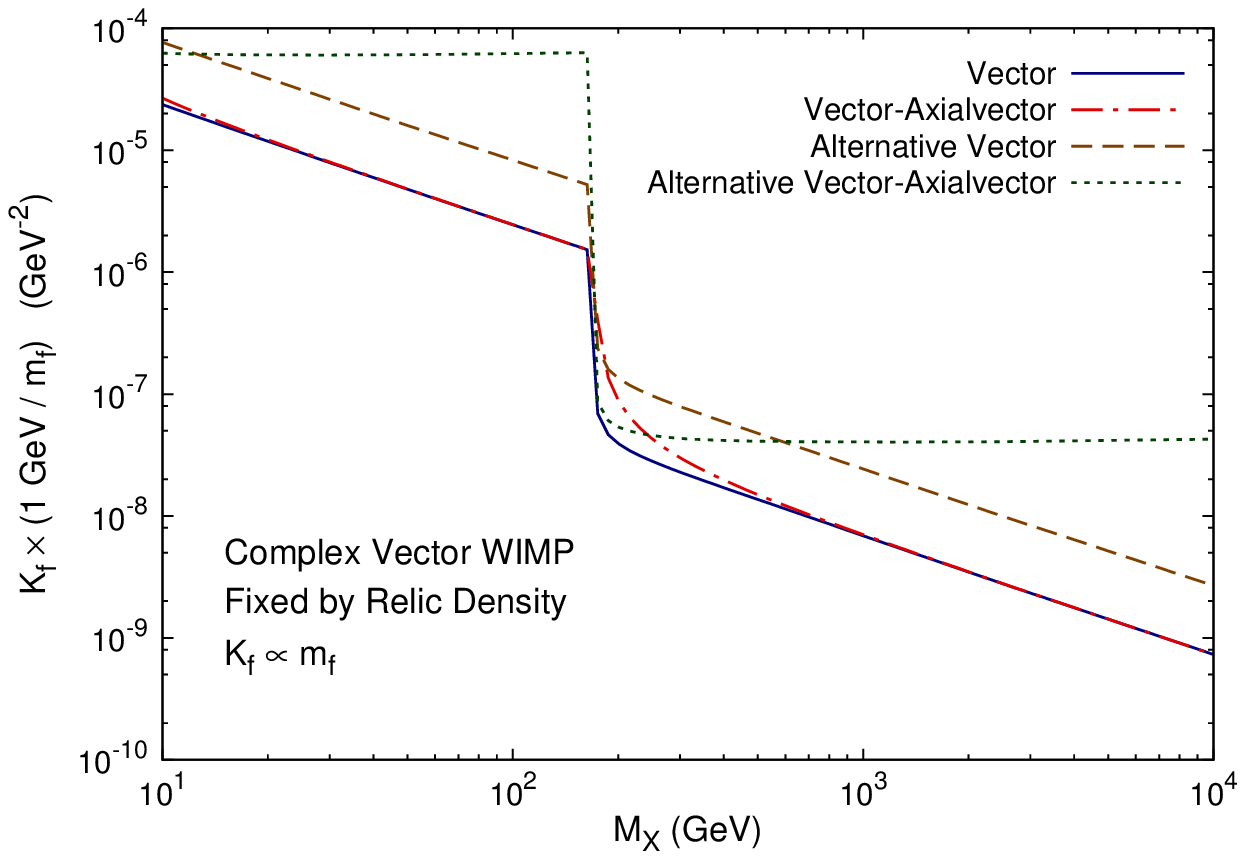}
\caption{The predicted coupling constants $K_f$ as functions of the vector
WIMP mass $M_X$, fixed by the observed relic density,
$\Omega_{\mathrm{DM}} h^2=0.1109\pm0.0056$~\cite{Komatsu:2010fb}.
Frames in the left (right) column are results given for the case of
universal couplings ($K_f\propto m_f$).}
\label{fig:vector:relic:comp_cpl}
\end{figure}
\begin{figure}[!htbp]
\centering
\includegraphics[width=0.44\textwidth]{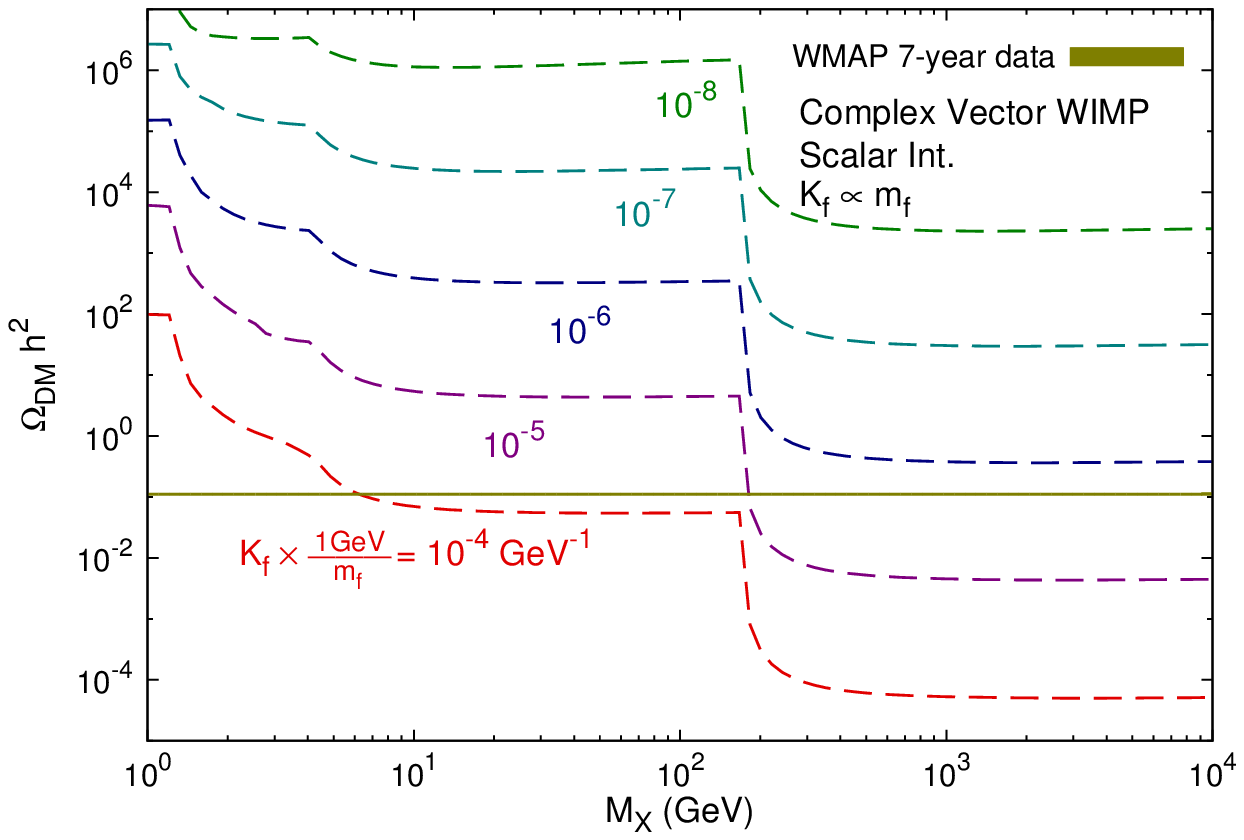}%
\hspace{0.01\textwidth}%
\includegraphics[width=0.44\textwidth]{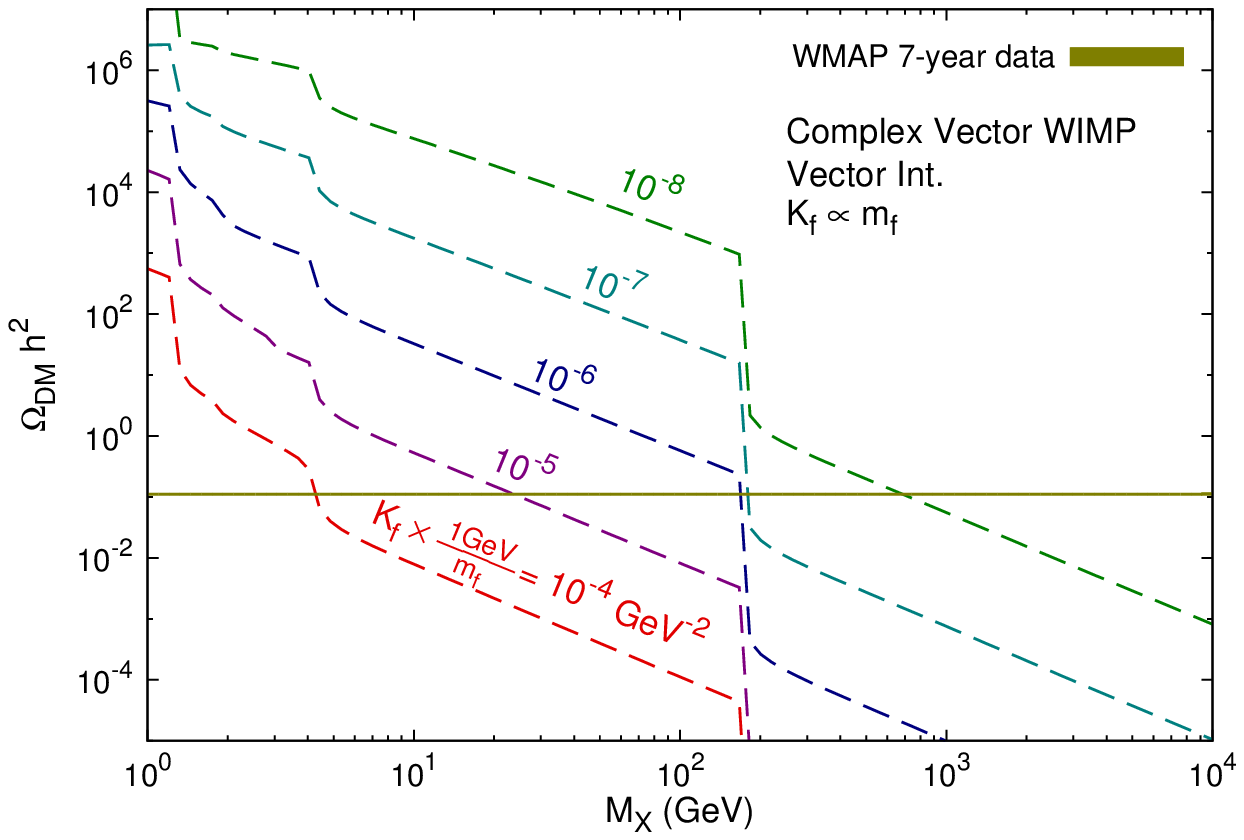}%
\\
\includegraphics[width=0.44\textwidth]{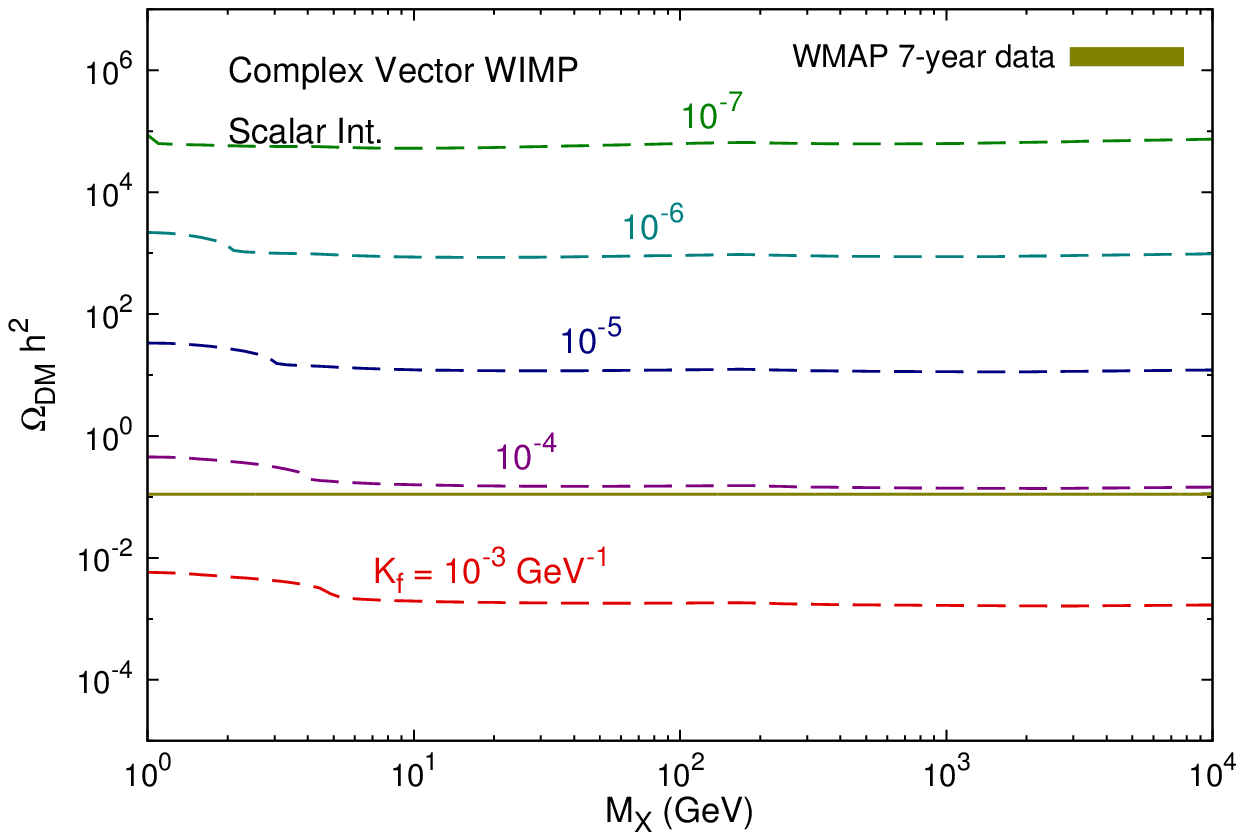}%
\hspace{0.01\textwidth}%
\includegraphics[width=0.44\textwidth]{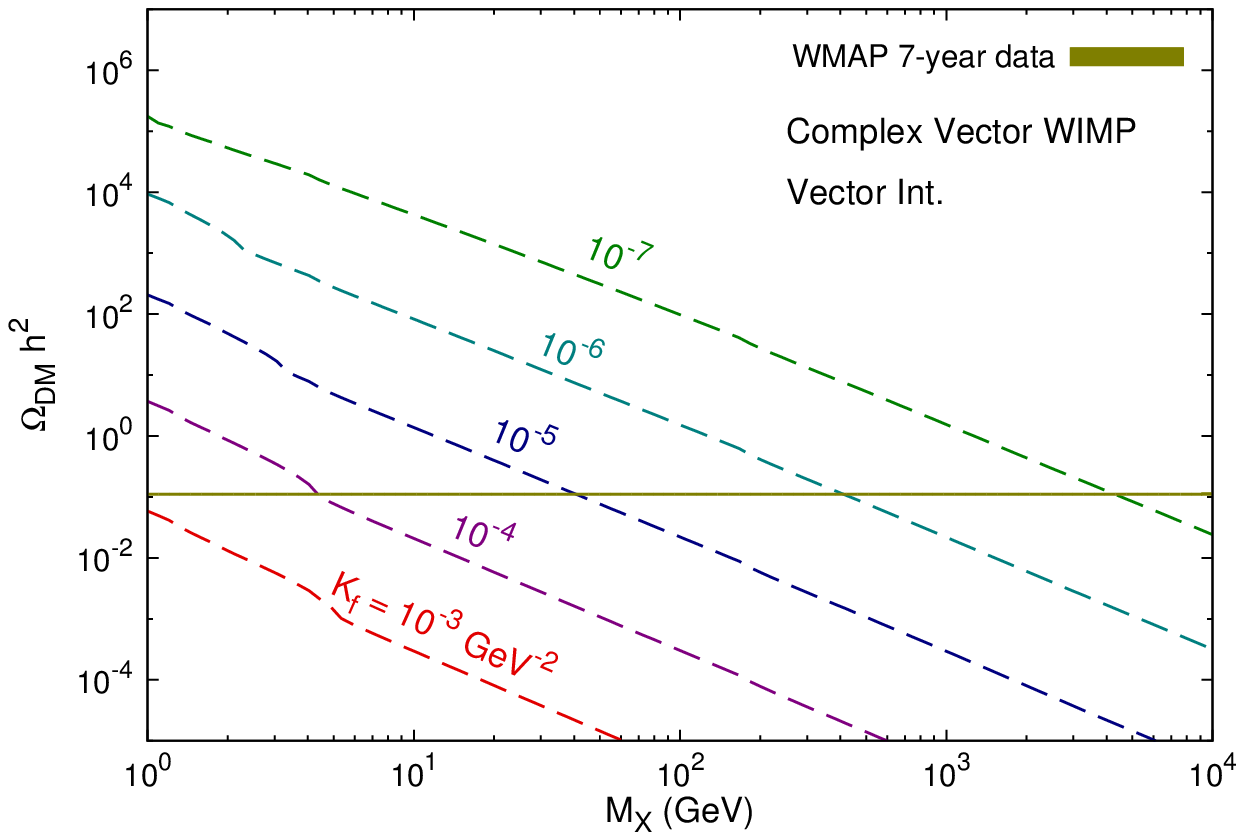}%
\\
\includegraphics[width=0.44\textwidth]{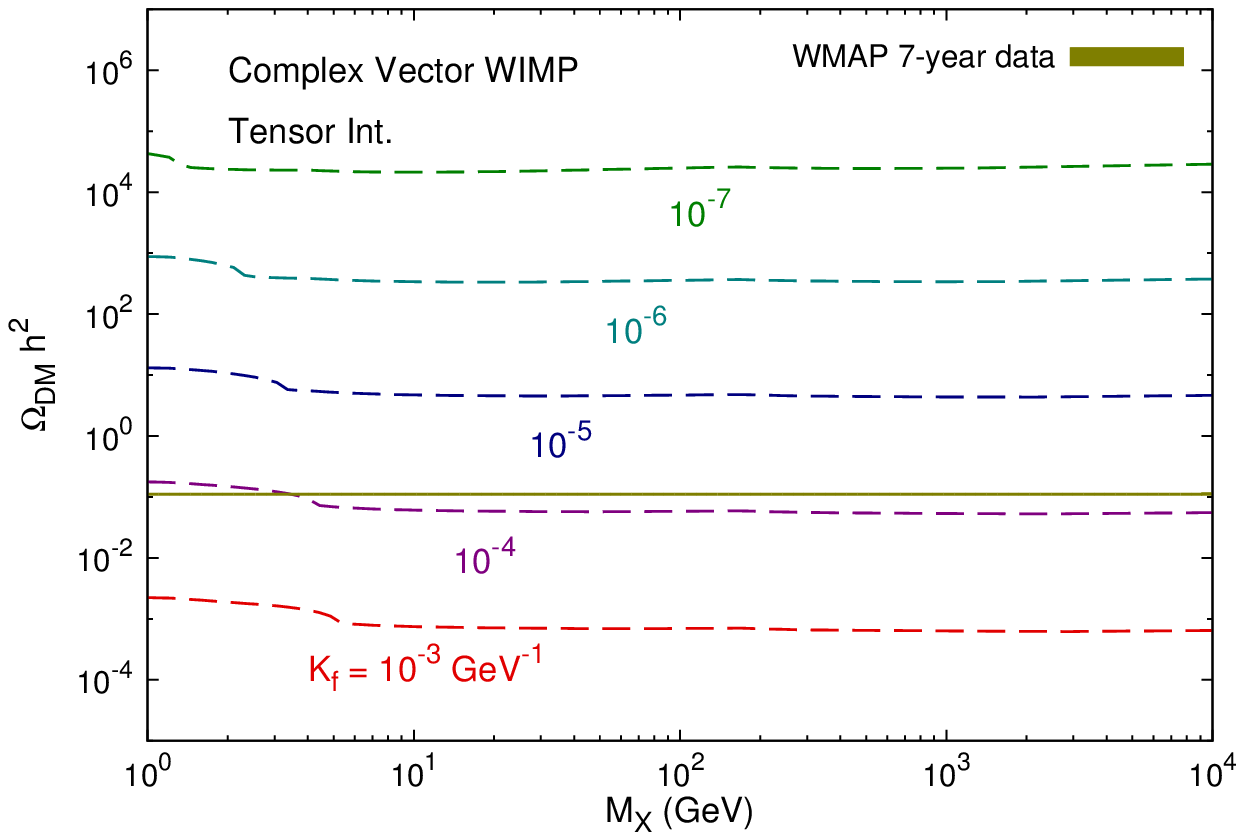}%
\hspace{0.01\textwidth}%
\includegraphics[width=0.44\textwidth]{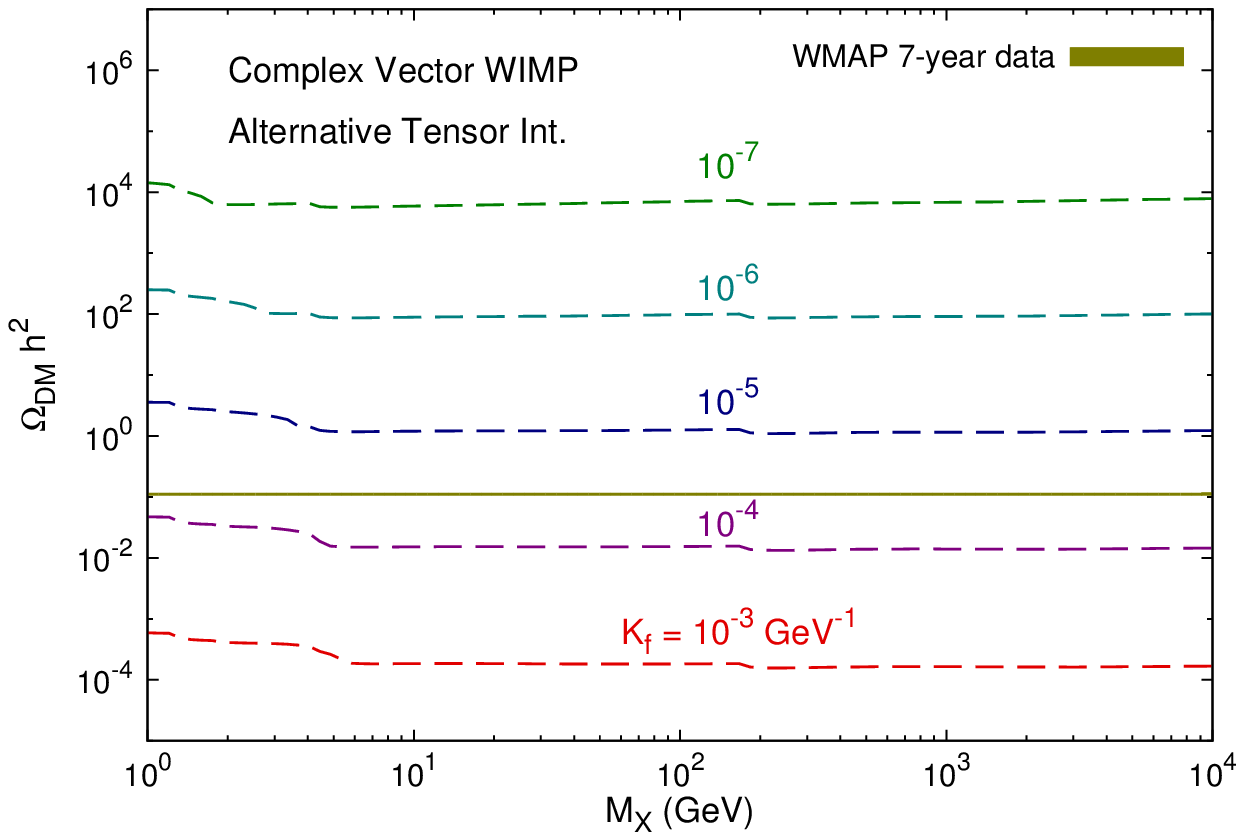}%
\\
\includegraphics[width=0.44\textwidth]{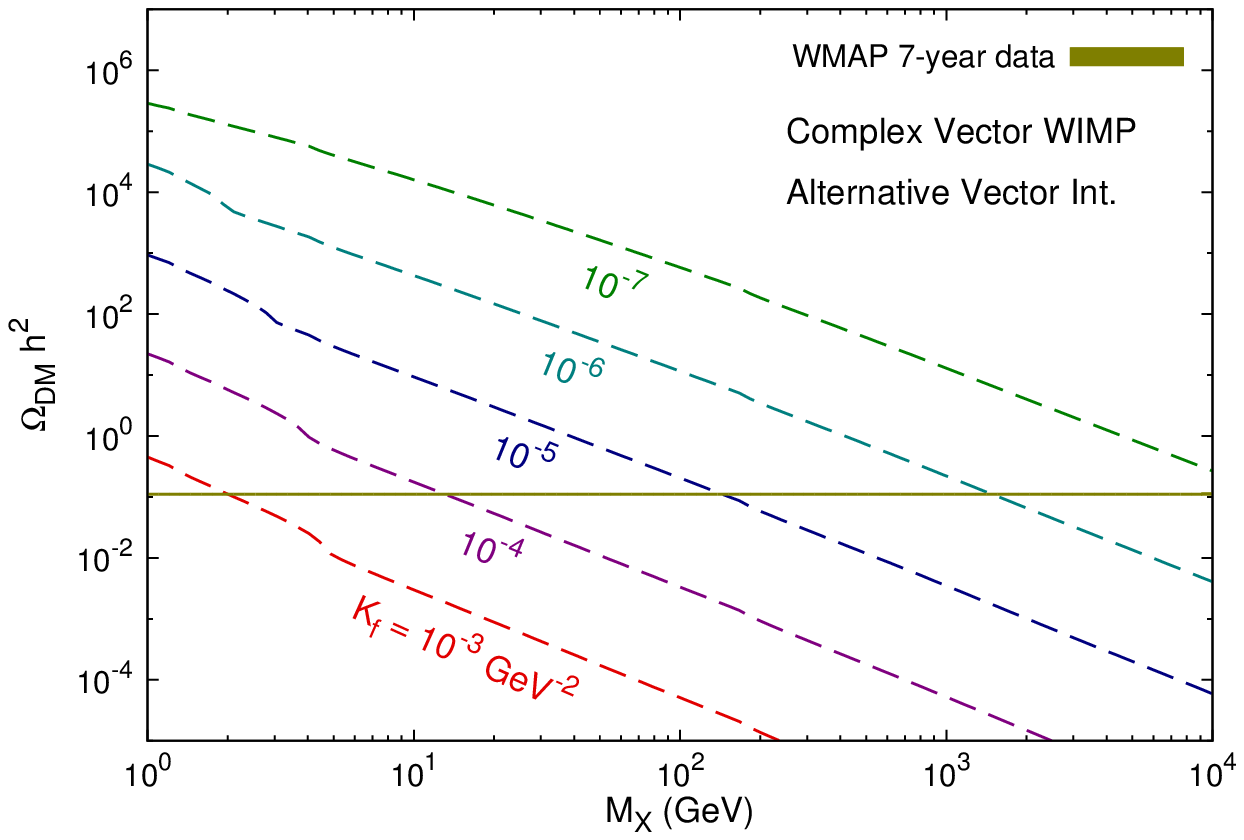}%
\hspace{0.01\textwidth}%
\includegraphics[width=0.44\textwidth]{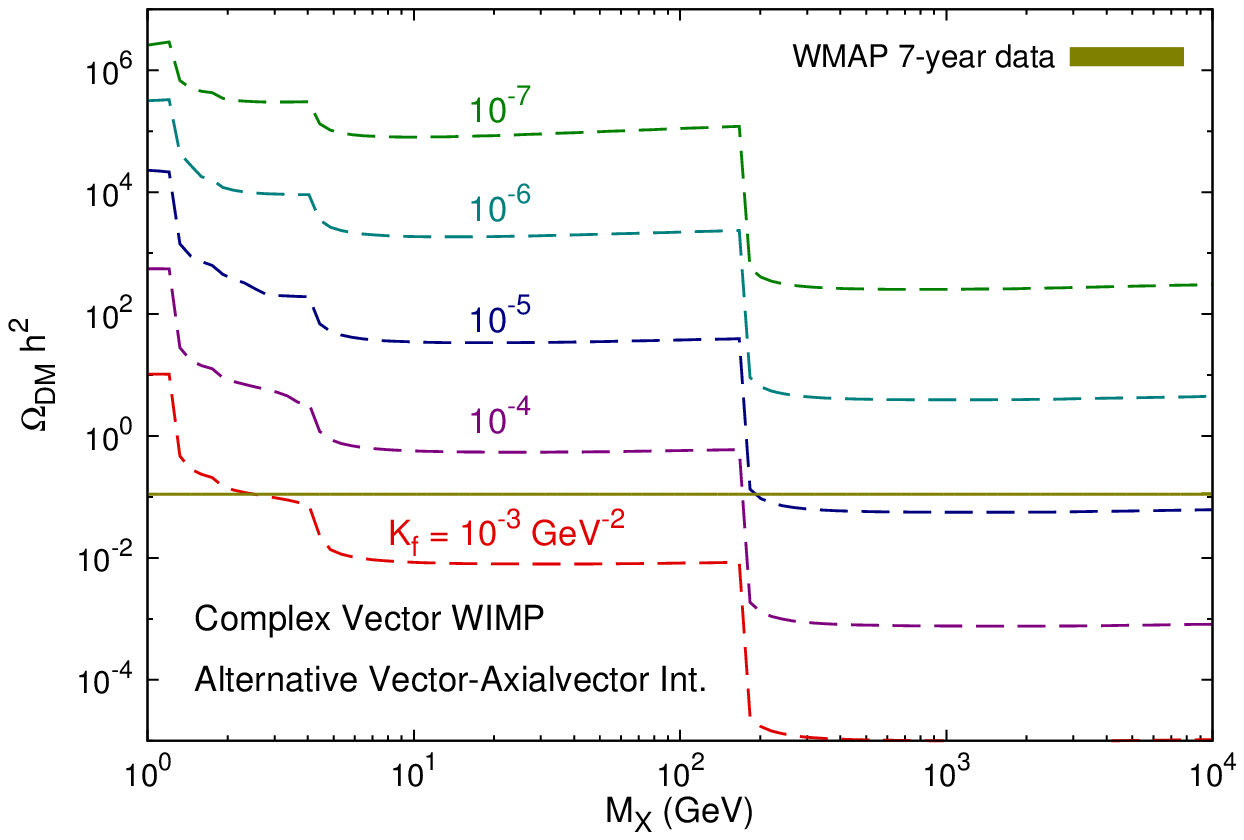}%
\caption{The predicted thermal relic density (dashed lines)
of complex vector WIMPs with S, V, T, $\widetilde{\mathrm{T}}$,
$\widetilde{\mathrm{V}}$ and $\widetilde{\mathrm{VA}}$ interactions,
respectively. In the upper two frames, results are shown for the case of
$K_f\propto m_f$ with S and V interactions. In the remaining frames, results
are given for the case of universal couplings. The narrow horizontal solid band
shows the range of the observed DM relic density, $\Omega_{\mathrm{DM}}
h^2=0.1109\pm0.0056$~\cite{Komatsu:2010fb}.}
\label{fig:vector:rd}
\end{figure}
\subsection{Vector WIMP annihilation and relic density\label{subsec-vect-rd}}

In order to calculate relic density and DM contribution in cosmic ray,
the annihilation cross sections of WIMP-antiWIMP to SM fermion-antifermion
pairs are needed. For each effective model listed above, the result
is given by
\begin{eqnarray}
  \sigma_{\mathrm{S},\,\mathrm{ann}} &=& \frac{1}{288\pi M_X^4}\sum\limits_f
\left( {\frac{K_{\mathrm{S},f}}{\sqrt 2}} \right)^2 c_f \sqrt {\frac{s
- 4m_f^2}{s - 4M_X^2}}\frac{(s-4m_f^2)(s^2 - 4M_X^2s + 12M_X^4)}{s},
\label{sigma_vector_scal}\\
  \sigma_{\mathrm{V},\,\mathrm{ann}} &=& \frac{1}{432\pi M_X^4}\sum\limits_f
\left(\frac{K_{\mathrm{V},f}}{\sqrt 2} \right)^2 c_f \sqrt{(s - 4m_f^2)
(s - 4M_X^2)}\frac{(s + 2m_f^2)(s^2 - 4M_X^2s + 12M_X^4)}{s},
\label{sigma_vector_vect}\\
  \sigma _{\mathrm{T},\,\mathrm{ann}} &=& \frac{1}{216\pi M_X^4}\sum\limits_f
\left(\frac{K_{\mathrm{T},f}}{\sqrt 2} \right)^2 c_f \sqrt
{\frac{s - 4m_f^2}{s - 4M_X^2}} \left[s^2 + (8m_f^2 + 4M_X^2)s
- 16m_f^2M_X^2 - 8M_X^4 - 160\frac{m_f^2M_X^4} {s}\right],
\label{sigma_vector_tens}\\
  \sigma _{\mathrm{SP},\,\mathrm{ann}} &=& \frac{1}{288\pi M_X^4}\sum\limits_f
\left(\frac{K_{\mathrm{SP},f}}{\sqrt 2}\right)^2 c_f \sqrt{\frac{s - 4m_f^2}
{s - 4M_X^2}} (s^2 - 4M_X^2s + 12M_X^4),
\label{sigma_vector_s_Ps}\\
  \sigma _{\mathrm{VA},\,\mathrm{ann}} &=& \frac{1}{432\pi M_X^4}\sum\limits_f
\left(\frac{K_{\mathrm{VA},f}}{\sqrt 2}\right)^2 c_f \sqrt{(s - 4m_f^2)
(s - 4M_X^2)}\frac{(s - 4m_f^2)(s^2 - 4M_X^2s + 12M_X^4)}{s},
\label{sigma_vector_v_Av}\\
  \sigma _{\widetilde{\mathrm{V}},\,\mathrm{ann}} &=& \frac{1}{108\pi M_X^2}
\sum\limits_f \left(\frac{\tilde K_{\mathrm{V},f}}{\sqrt 2} \right)^2 c_f
\sqrt{(s - 4m_f^2)(s - 4M_X^2)} \frac{(s-4M_X^2)(s+2m_f^2)}{s},
\label{sigma_vector_alt_vect}\\
  \sigma _{\widetilde{\mathrm{VA}},\,\mathrm{ann}} &=& \frac{1}{108\pi M_X^2}
\sum\limits_f \left(\frac{\tilde K_{\mathrm{VA},f}}{\sqrt 2}\right)^2 c_f
\sqrt{(s - 4m_f^2)(s - 4M_X^2)} \left[s - 4(M_X^2 + m_f^2)
+ 28\frac{m_f^2M_X^2}{s}\right],
\label{sigma_vector_alt_v_Av}\\
  \sigma _{\widetilde{\mathrm{T}},\,\mathrm{ann}} &=& \frac{1}{54\pi M_X^4}
\sum\limits_f \left(\frac{\tilde K_{\mathrm{T},f}}{\sqrt 2}\right)^2 c_f
\sqrt{\frac{s - 4m_f^2}{s - 4M_X^2}} \left[s^2 + 4(M_X^2 - m_f^2)s +
32 m_f^2 M_X^2 - 8 M_X^4 + 128\frac{m_f^2M_X^4}{s}\right],
\label{sigma_vector_alt_tens}
\end{eqnarray}
where $M_X$ is the WIMP mass.
For the case of real vector WIMPs, each non-vanishing annihilation
cross section should include an additional factor of 4
because the WIMP and its antipartner are identical.
After taking the thermal average, we obtain the annihilation rates, i.e.,
\begin{eqnarray}
  \left<\sigma_{\mathrm{S},\,\mathrm{ann}}v\right> &\simeq& \frac{1}{12\pi}
\sum\limits_f \left(\frac{K_{\mathrm{S},f}}{\sqrt 2}\right)^2 c_f
\sqrt{1-\frac{m_f^2}{M_X^2}} \left[\left(1-\frac{m_f^2}{M_X^2}\right)
+ \frac{1}{4}\left(2+7\frac{m_f^2}{M_X^2}\right) \frac{T}{M_X} \right],
\label{sv_vector_scal}\\
  \left<\sigma_{\mathrm{V},\,\mathrm{ann}}v\right> &\simeq& \frac{1}{6\pi}
\sum\limits_f \left(\frac{K_{\mathrm{V},f}}{\sqrt 2}\right)^2 c_f
\sqrt{1-\frac{m_f^2}{M_X^2}} M_X^2 \left(2+\frac{m_f^2}{M_X^2}\right)
\frac{T}{M_X},
\label{sv_vector_vect}\\
  \left<\sigma_{\mathrm{T},\,\mathrm{ann}}v\right> &\simeq& \frac{1}{18\pi}
\sum\limits_f \left(\frac{K_{\mathrm{T},f}}{\sqrt 2}\right)^2 c_f
\sqrt{1-\frac{m_f^2}{M_X^2}} \left[4\left(1-\frac{m_f^2}{M_X^2}\right)
+ 3\left(2+9\frac{m_f^2}{M_X^2}\right)\frac{T}{M_X}\right],
\label{sv_vector_tens}\\
  \left<\sigma_{\mathrm{SP},\,\mathrm{ann}}v\right> &\simeq& \frac{1}{12\pi}
\sum\limits_f \left(\frac{K_{\mathrm{SP},f}}{\sqrt 2}\right)^2 c_f
\sqrt{1-\frac{m_f^2}{M_X^2}} \left[1+\frac{2+m_f^2/M_X^2}{4(1-m_f^2/M_X^2)}
\frac{T}{M_X}\right],
\label{sv_vector_s_Ps}\\
  \left<\sigma_{\mathrm{VA},\,\mathrm{ann}}v\right> &\simeq& \frac{1}{3\pi}
\sum\limits_f \left(\frac{K_{\mathrm{VA},f}}{\sqrt 2}\right)^2 c_f
\left(1-\frac{m_f^2}{M_X^2}\right)^{3/2} M_X^2 \frac{T}{M_X},
\label{sv_vector_v_Av}\\
  \big<\sigma_{\widetilde{\mathrm{V}},\,\mathrm{ann}}v\big> &\simeq&
\frac{5}{9\pi} \sum\limits_f \left(\frac{\tilde K_{\mathrm{V},f}}{\sqrt 2}
\right)^2 c_f \sqrt{1- \frac{m_f^2}{M_X^2}} M_X^2 \left(2+\frac{m_f^2}
{M_X^2}\right) \frac{T^2}{M_X^2},
\label{sv_vector_alt_vect}\\
  \big<\sigma_{\widetilde{\mathrm{VA}},\,\mathrm{ann}}v\big> &\simeq&
\frac{1}{3\pi} \sum\limits_f \left(\frac{\tilde K_{\mathrm{VA},f}}{\sqrt 2}
\right)^2 c_f \sqrt{1- \frac{m_f^2}{M_X^2}} m_f^2 \frac{T}{M_X},
\label{sv_vector_alt_v_Av}\\
  \big<\sigma_{\widetilde{\mathrm{T}},\,\mathrm{ann}}v\big> &\simeq&
\frac{1}{9\pi} \sum\limits_f \left(\frac{\tilde K_{\mathrm{T},f}}{\sqrt 2}
\right)^2 c_f \sqrt{1- \frac{m_f^2}{M_X^2}} \left[8\left(1+2\frac{m_f^2}
{M_X^2}\right) +\frac{6(2-5 m_f^2/M_X^2)(1-2 m_f^2/M_X^2)}{1-m_f^2/M_X^2}
\frac{T}{M_X}\right].
\label{sv_vector_alt_tens}
\end{eqnarray}
Note that in the case of $\widetilde{\mathrm{V}}$ interaction,
$\left<\sigma_\mathrm{ann}v\right>$ vanishes in the zeroth and the first orders
of $T/M_X$, thus we retain the term of order $T^2/M_X^2$.

The procedure to obtain the relic density of vector WIMPs is similar to
that in the case of scalar WIMPs. A notable difference is that the degree
of freedom for a vector WIMP is $g=3$. Besides, since
$\left<\sigma_\mathrm{ann}v\right>$ of $\widetilde{\mathrm{V}}$ interaction
\eqref{sv_vector_alt_vect} is of order $T^2/M_X^2$, Eqs.~\eqref{relic-d}
and \eqref{x-f} need to be extended. If $\sigma_\mathrm{ann}v$
in the lab frame is expanded to be $\sigma _{\mathrm{ann}}v \simeq a_0
+ a_1 v^2 + a_2 v^4 + \mathcal{O}(v^6)$, we have
$\left<\sigma_\mathrm{ann}v\right> \simeq a_0 + 6 a_1 x^{-1}
+ (60 a_2 - 27 a_1) x^{-2}
+ \mathcal{O}(x^{-3})$ with $x \equiv M_X /T$. Then the factor
$(a_0+3a_1 x_f^{-1})$ in the denominator of the right-hand side of
Eq.~\eqref{relic-d} should be modified to be $[a_0+3a_1 x_f^{-1}
+(20a_2 - 9a_1) x_f^{-2}]$, and the factor $(a_0+6a_1 x_f^{-1})$
in the square bracket of the right-hand side of Eq.~\eqref{x-f} should
be modified to be $[a_0+6a_1 x_f^{-1}+(60a_2-27a_1) x_f^{-2}]$.
For the case of real vector WIMPs, the relic density just one half of that
for the case of complex vector WIMPs given the same annihilation cross section.

In Fig.~\ref{fig:vector:relic:comp_cpl}, the predicted coupling constants
$K_f$ fixed by the WMAP observed relic density are shown as functions of
$M_X$. The upper two frames are plotted for dimension-5 operators
(S, SP, T and $\widetilde{\mathrm{T}}$), while the remaining two are
plotted for dimension-6 operators (V, VA, $\widetilde{\mathrm{V}}$ and
$\widetilde{\mathrm{VA}}$). Two cases of coupling constants are
considered again. In the left (right) column of the frames, results
are given for universal couplings ($K_f \propto m_f$).
There are some interesting features in Fig.~\ref{fig:vector:relic:comp_cpl}:
\begin{itemize}
  \item The curves for the case of $K_f\propto m_f$ have sharp downward
bends at $M_X \sim m_t$. The reason is the same as that in the case of
scalar WIMPs. Even for the case of universal couplings, the curve of
$\widetilde{\mathrm{VA}}$ interaction also exhibits the same bending
behavior because of $\big<\sigma_{\widetilde{\mathrm{VA}},\,\mathrm{ann}}v\big>
\propto m_f^2$.
  \item There are two pairs of nearly identical curves,
S~$\simeq$~SP and V~$\simeq$ VA. This is because of the similarities
of the annihilation rates corresponding to these two pairs
of interactions, respectively. The leading terms of these two pairs of
the annihilation rates tend to be the same when $M_X$ is much
larger than that of the final state fermions.
  \item The curves of V and VA interactions are obviously lower
than the other two curves of $\widetilde{\mathrm{V}}$ and
$\widetilde{\mathrm{VA}}$ interactions.
That is due to the forms of the annihilation rates.
The leading terms of the annihilation rates for V and VA interactions
are of order $T/M_X$, while that for $\widetilde{\mathrm{V}}$ interaction
is of order $T^2/M_X^2$. The leading term of the annihilation rate for
$\widetilde{\mathrm{VA}}$ interaction is proportional to $m_f^2 T/M_X$,
which is suppressed by the SM fermion masses $m_f$ that are smaller than
$M_X$ almost in the whole region.
On the other hand, the curves of T and $\widetilde{\mathrm{T}}$
interactions are lower than those of S and SP interactions due to
the larger zeroth order terms of the annihilation rates for
T and $\widetilde{\mathrm{T}}$ interactions.
\end{itemize}

In Fig.~\ref{fig:vector:rd}, we show the predicted thermal relic density
$\Omega_\mathrm{DM}h^2$ of complex vector WIMPs for S, V, T,
$\widetilde{\mathrm{T}}$, $\widetilde{\mathrm{V}}$ and $\widetilde{\mathrm{VA}}$
interactions. The curves for SP and VA interactions are very
similar to those for S and V interactions respectively, as discussed above.
In the upper two frames, results are shown for the case of $K_f\propto m_f$
with S and V interactions. The remaining frames give the results for the case
of universal couplings. Similar to the case of scalar WIMPs, the curves for
dimension-5 operators (S, SP, T and $\widetilde{\mathrm{T}}$) are nearly
horizontal in almost the whole mass range, while those for dimension-6
operators slope downward (except for those of $\widetilde{\mathrm{VA}}$
interaction, due to $\big<\sigma_{\widetilde{\mathrm{VA}},\,\mathrm{ann}}v\big>
\propto m_f^2$).
Additionally, if resonances, coannihilations, or annihilations to final states
other than fermion-antifermion pairs are significant, the actual
curves in Figs.~\ref{fig:vector:relic:comp_cpl} and \ref{fig:vector:rd}
will be significantly lower than they are shown there.

\begin{figure}[!htbp]
\centering
\includegraphics[width=0.40\textwidth]{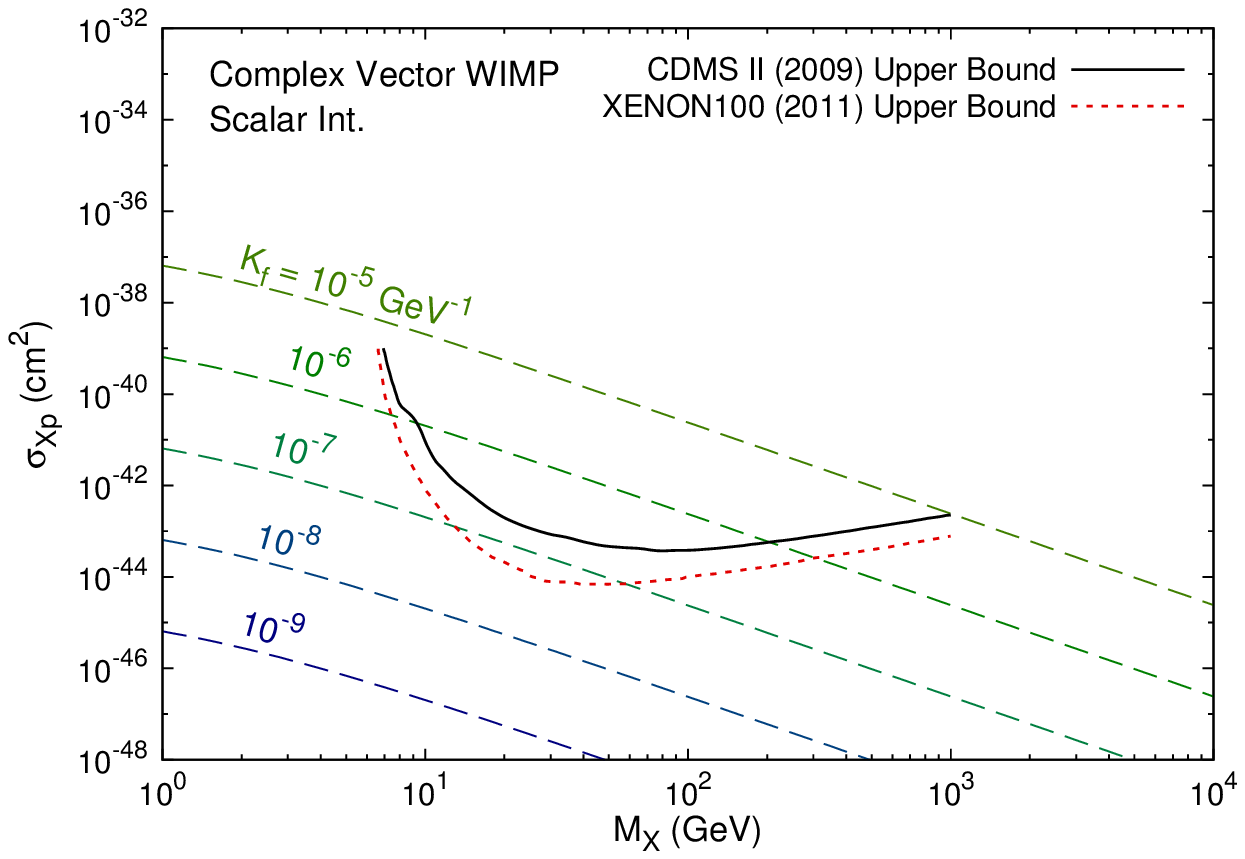}%
\hspace{0.01\textwidth}%
\includegraphics[width=0.40\textwidth]{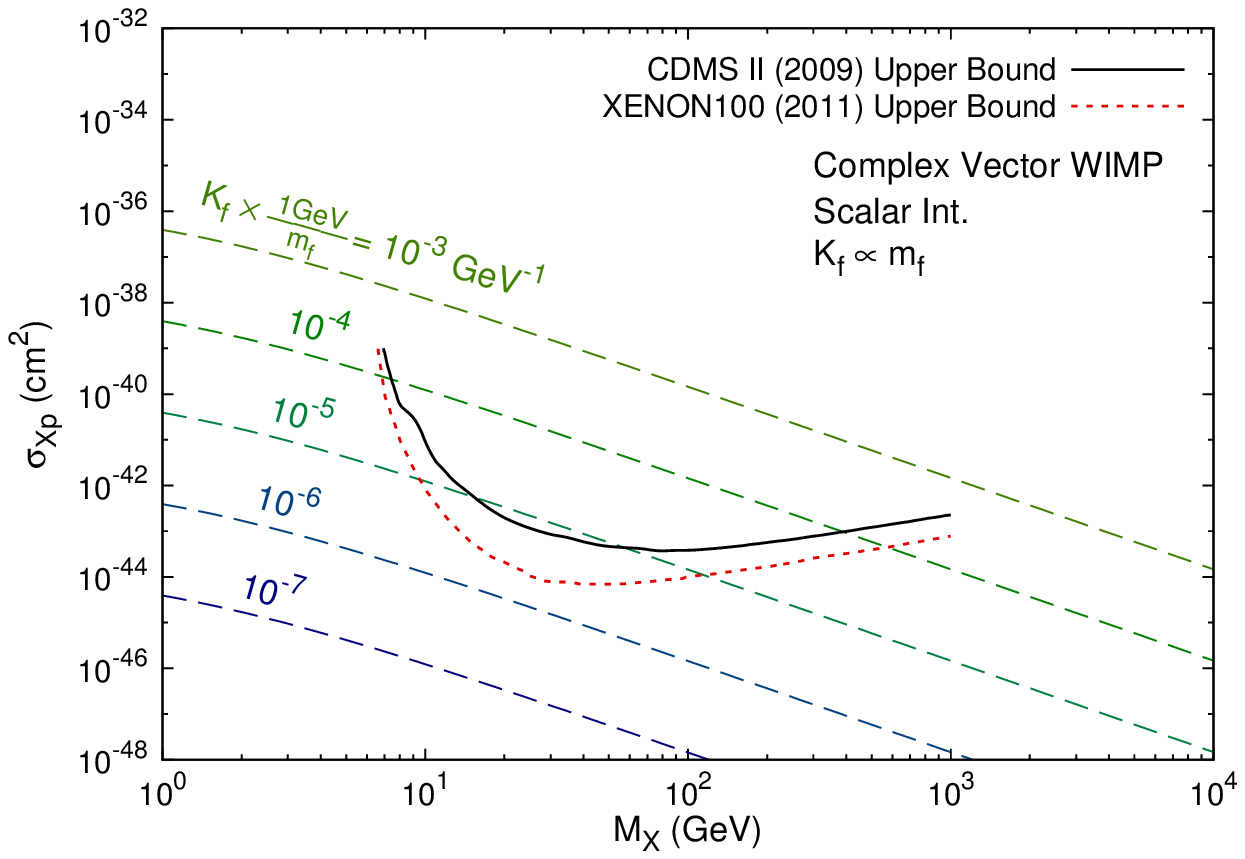}%
\\
\includegraphics[width=0.40\textwidth]{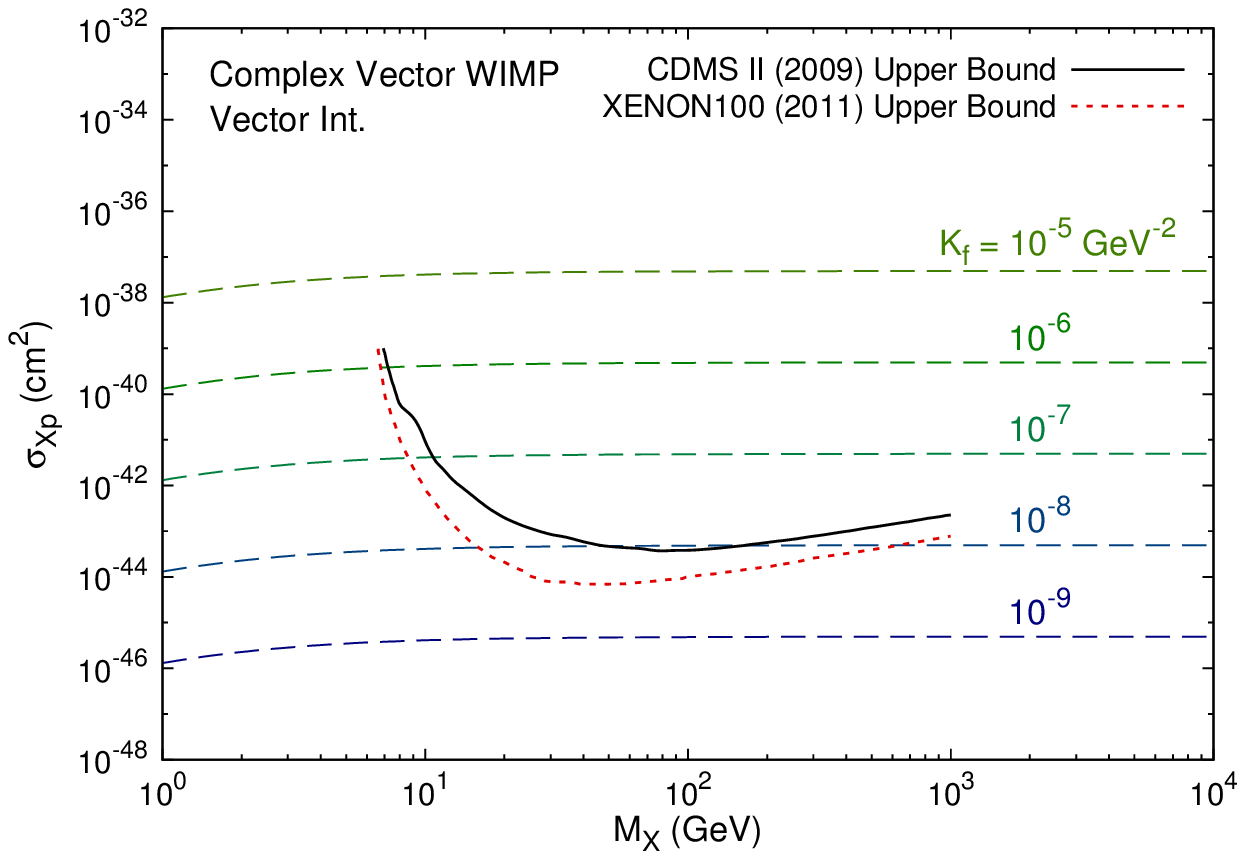}%
\hspace{0.01\textwidth}%
\includegraphics[width=0.40\textwidth]{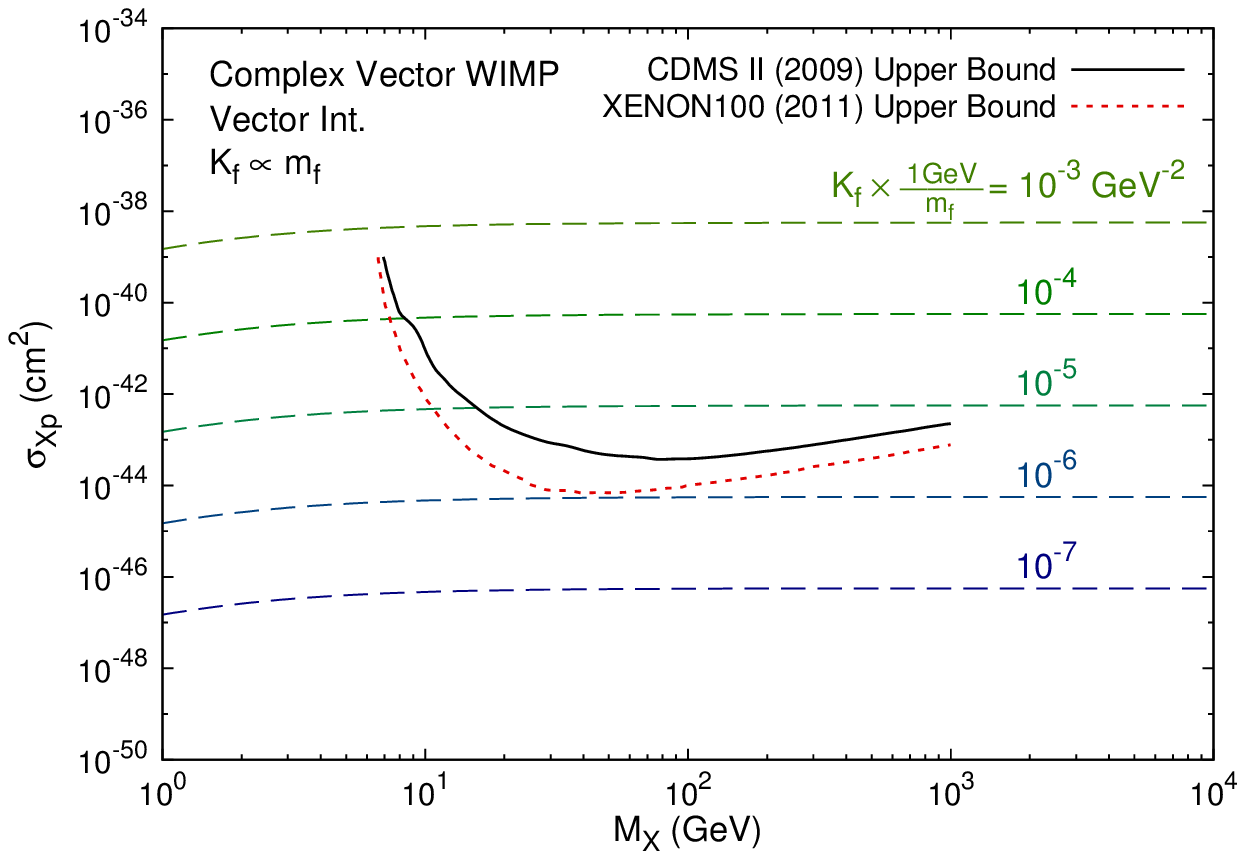}
\caption{The spin-independent (SI) WIMP-proton cross sections (dashed lines)
for complex vector WIMPs with S and V interactions.
In the left (right) column, results are given for universal couplings
($K_f \propto m_f$). The upper bounds set by CDMS II~\cite{Ahmed:2009zw}
and XENON100~\cite{Aprile:2011hi} are also given in the frames for comparison.}
\label{fig:vector:SI}
\vspace*{1.0em}
\includegraphics[width=0.40\textwidth]{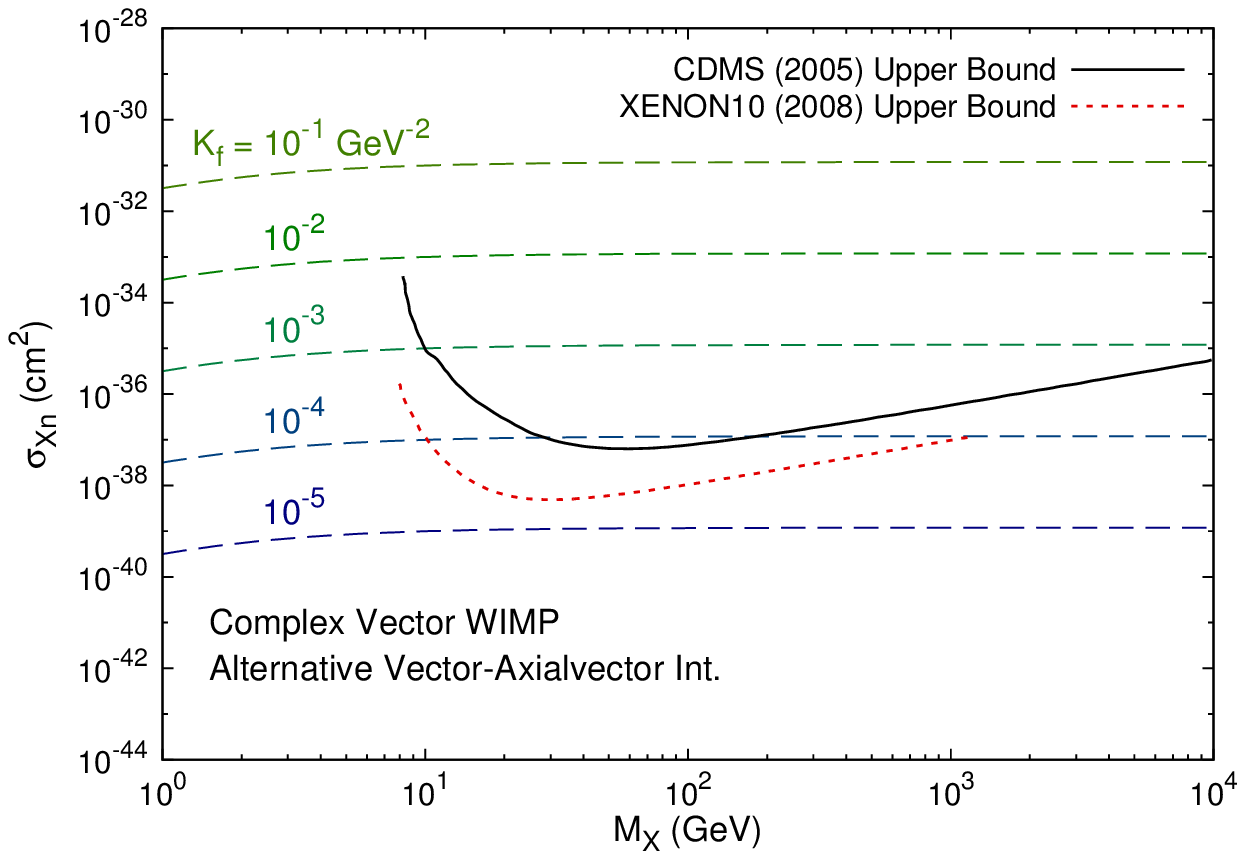}%
\hspace{0.01\textwidth}%
\includegraphics[width=0.40\textwidth]{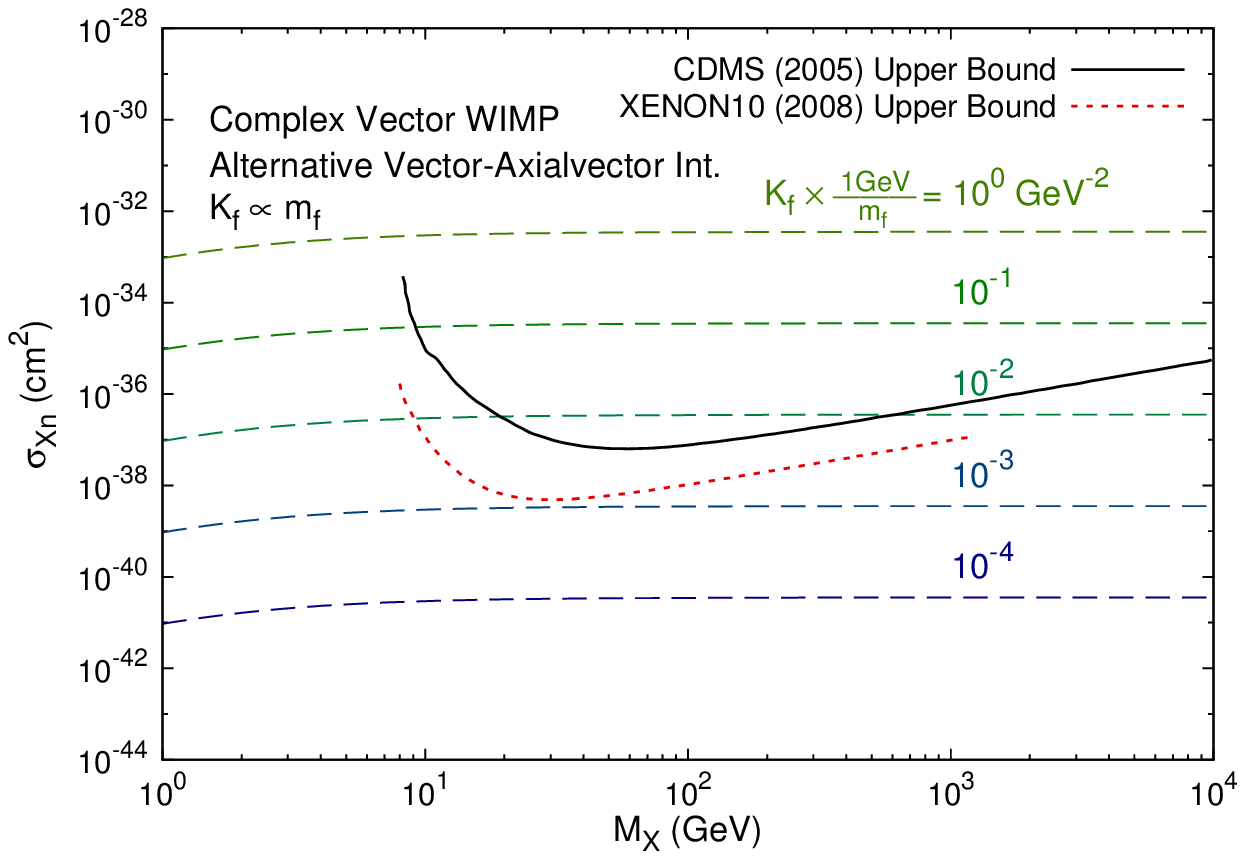}%
\\
\includegraphics[width=0.40\textwidth]{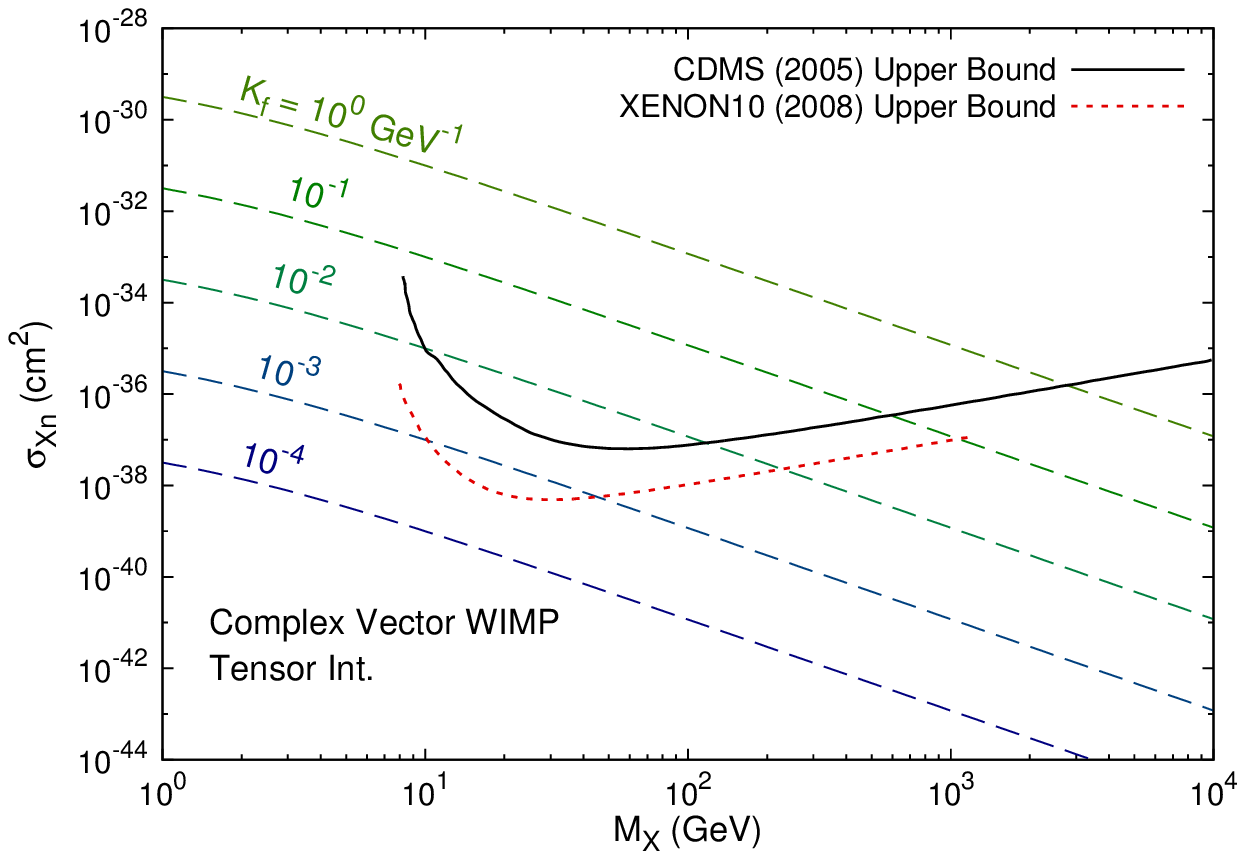}%
\hspace{0.01\textwidth}%
\includegraphics[width=0.40\textwidth]{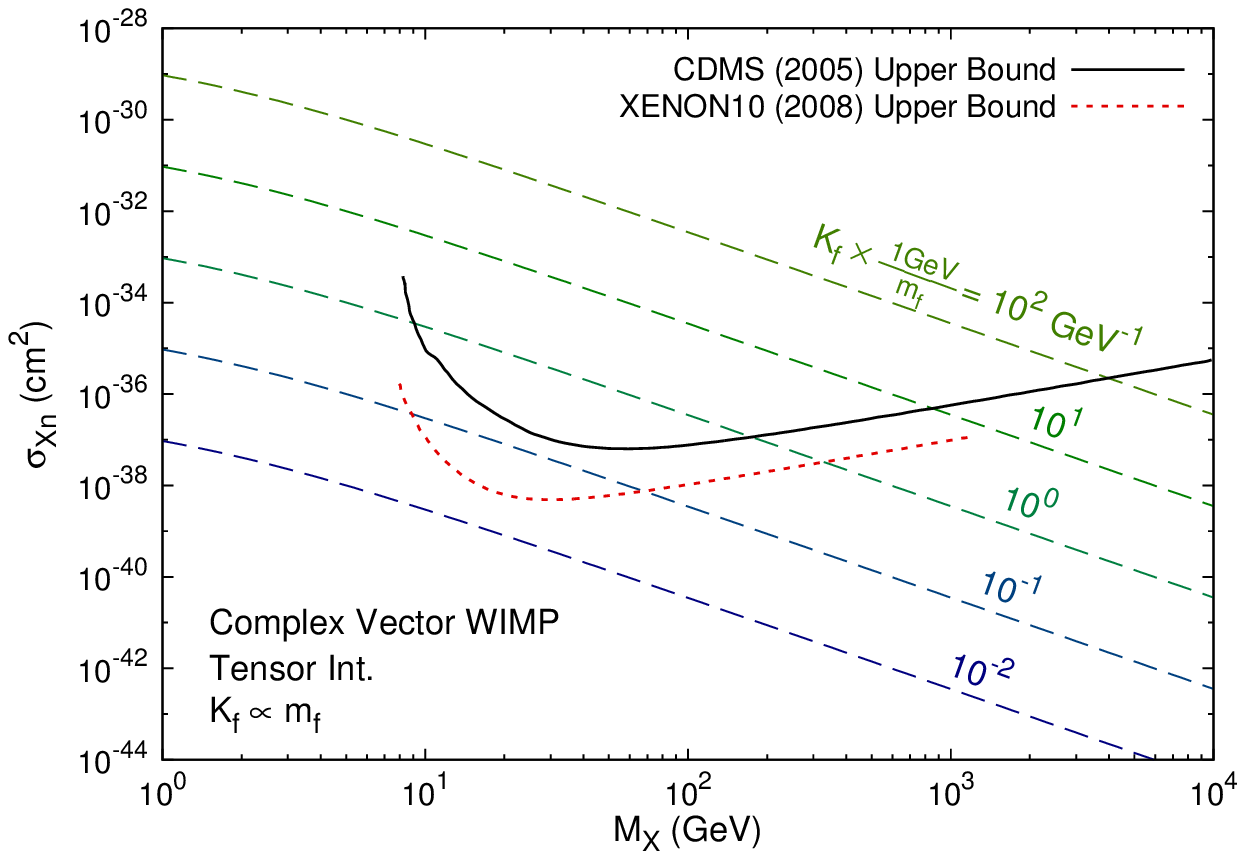}%
\caption{The spin-dependent (SD) WIMP-neutron cross sections (dashed lines)
for complex vector WIMPs with $\widetilde{\mathrm{VA}}$ and T interactions.
In the left (right) column, results are given for universal couplings
($K_f \propto m_f$). The upper bounds set by CDMS~\cite{Akerib:2005za}
and XENON10~\cite{Angle:2008we} are also given in the frames for comparison.}
\label{fig:vector:SD}
\end{figure}
\subsection{Direct detection and indirect detection\label{subsec-vect-dd-id}}
In this subsection we discuss the direct and indirect detection constraints
on the effective vector WIMP models described by Eqs.~\eqref{lag:vector:scal}
-- \eqref{lag:vector:alt_tens}. In the low velocity limit,
only the scattering cross sections of S, V, T and $\widetilde{\mathrm{VA}}$
interactions survive, and others are suppressed by the low WIMP velocity.
The S and V interactions between WIMPs and nuclei are spin-independent (SI),
while the T and $\widetilde{\mathrm{VA}}$ interactions are spin-dependent (SD).
In order to be compared with the results of CDMS
\cite{Ahmed:2009zw}\cite{Akerib:2005za} and XENON
\cite{Aprile:2011hi}\cite{Angle:2008we}, the WIMP-nucleon cross sections
are given as follows:
\begin{eqnarray}
\text{Scalar int.}:\qquad && \sigma_{\mathrm{S},\,X N}=\frac{m^2_N}
{4\pi\left(M_X+m_N\right)^2}\left(\frac{K_{\mathrm{S},N}}{\sqrt{2}}\right)^2,
\label{sigmaN_scat_vector_scal}\\
\text{Vector int.}:\qquad && \sigma_{\mathrm{V},\,X N}=\frac{m^2_N M^2_X}
{\pi\left(M_X+m_N\right)^2}\left(\frac{K_{\mathrm{V},N}}{\sqrt{2}}\right)^2,
\label{sigmaN_scat_vector_vect}\\
\text{Tensor int.}:\qquad && \sigma_{\mathrm{T},\,X N}=\frac{2m^2_N}
{\pi\left(M_X+m_N\right)^2}\left(\frac{K_{\mathrm{T},N}}{\sqrt{2}}\right)^2,
\label{sigmaN_scat_vector_tens}\\
\widetilde{\mathrm{VA}}\text{ int.}:\qquad && \sigma_{\widetilde{\mathrm{VA}},
\,X N}=\frac{2m^2_N M^2_X} {\pi\left(M_X+m_N\right)^2}\left(\frac{{\tilde
K}_{\mathrm{VA},N}}{\sqrt{2}}\right)^2,
\label{sigmaN_scat_vector_alt_v_Av}
\end{eqnarray}
where $K_N$ ($N=p,n$) are the induced coupling constants of effective
WIMP-nucleon interactions, related to the couplings of WIMP-quark $K_q$ by
form factors. For S and V interactions, the relations between
$K_N$ and $K_q$ are the same as those between $F_N$ and $F_q$
(Eqs.~\eqref{form-S} and \eqref{form-V}), respectively.
For $\widetilde{\mathrm{VA}}$ and T interactions,
\begin{eqnarray}
\tilde{K}_{\mathrm{VA},N} &=& \sum_{q=u,d,s} \tilde{K}_{\mathrm{VA},q}
\Delta^N_q,
\label{form-A}\\
K_{\mathrm{T},N} &=& \sum_{q=u,d,s} K_{\mathrm{T},q} \Delta^N_q,
\label{form-T}
\end{eqnarray}
with the form factors $\Delta^p_u=0.842\pm0.012$,
$\Delta^p_d=-0.427\pm0.013$, $\Delta^p_s=-0.085\pm0.018$
\cite{Airapetian:2007mh}, $\Delta^n_u=\Delta^p_d$,
$\Delta^n_d=\Delta^p_u$, $\Delta^n_s=\Delta^p_s$,
which reflect the contributions of quark components to the nucleon spin.
Note again for real vector WIMPs,
the V and T interactions vanish, and the scattering cross sections of
S and $\widetilde{\mathrm{VA}}$ interactions should include
an additional factor 4 comparing to the case of complex vector WIMPs.

\begin{figure}[!bp]
\centering
\includegraphics[width=0.44\textwidth]{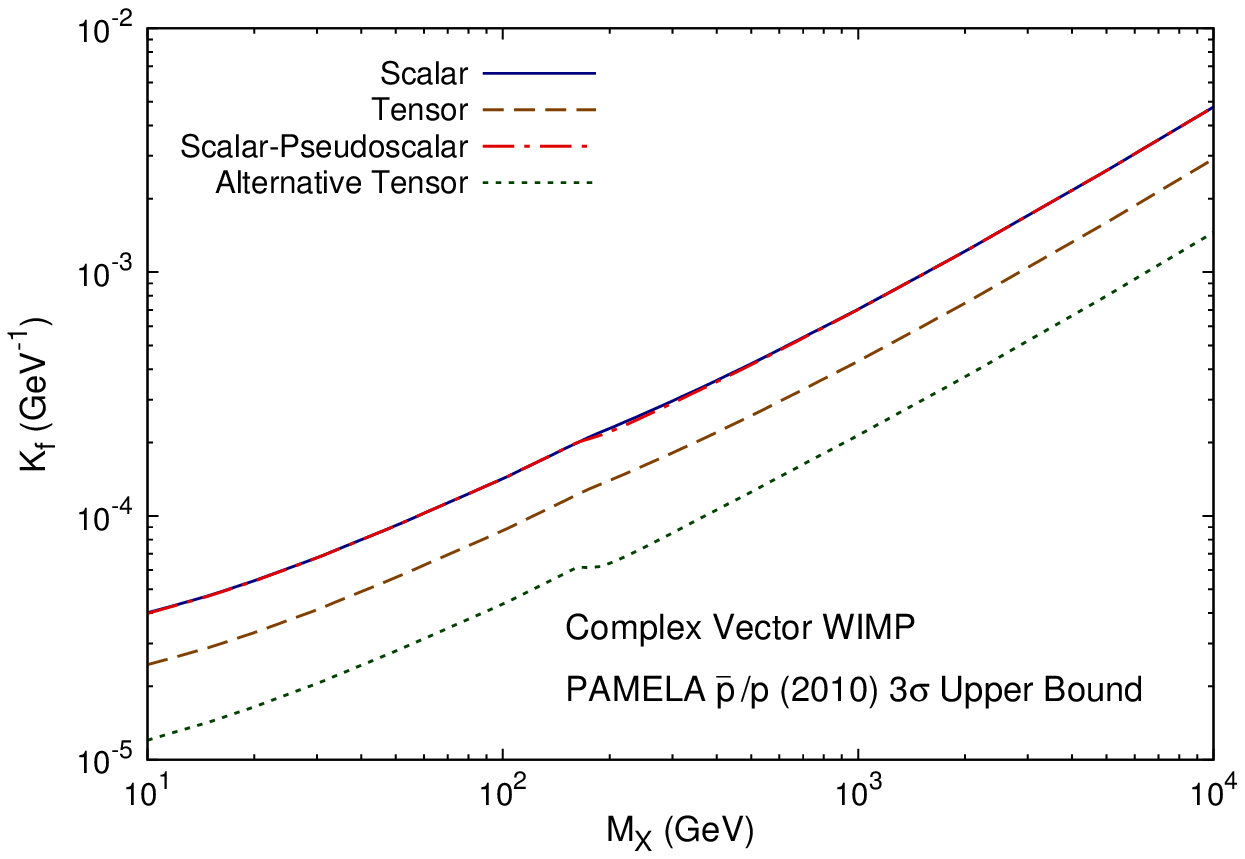}%
\hspace{0.01\textwidth}%
\includegraphics[width=0.44\textwidth]{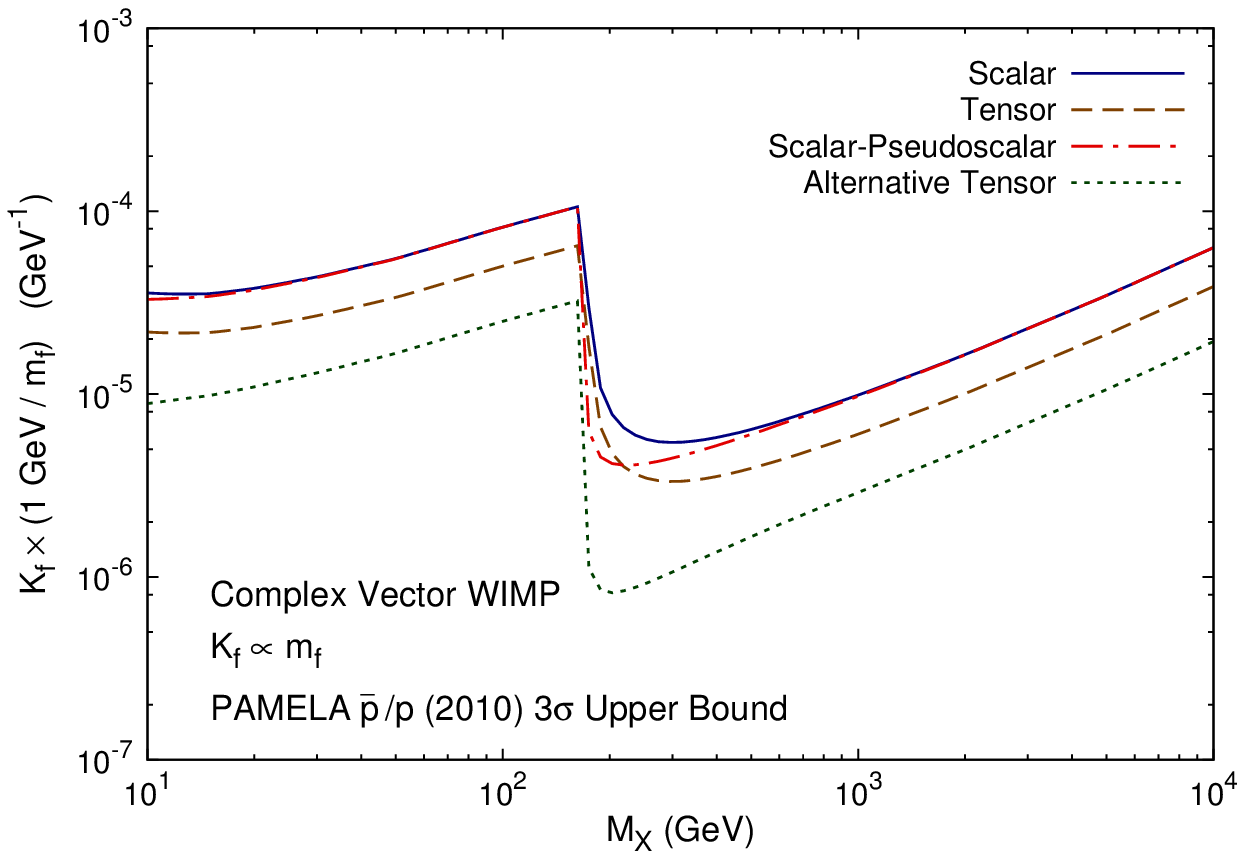}%
\\
\includegraphics[width=0.44\textwidth]{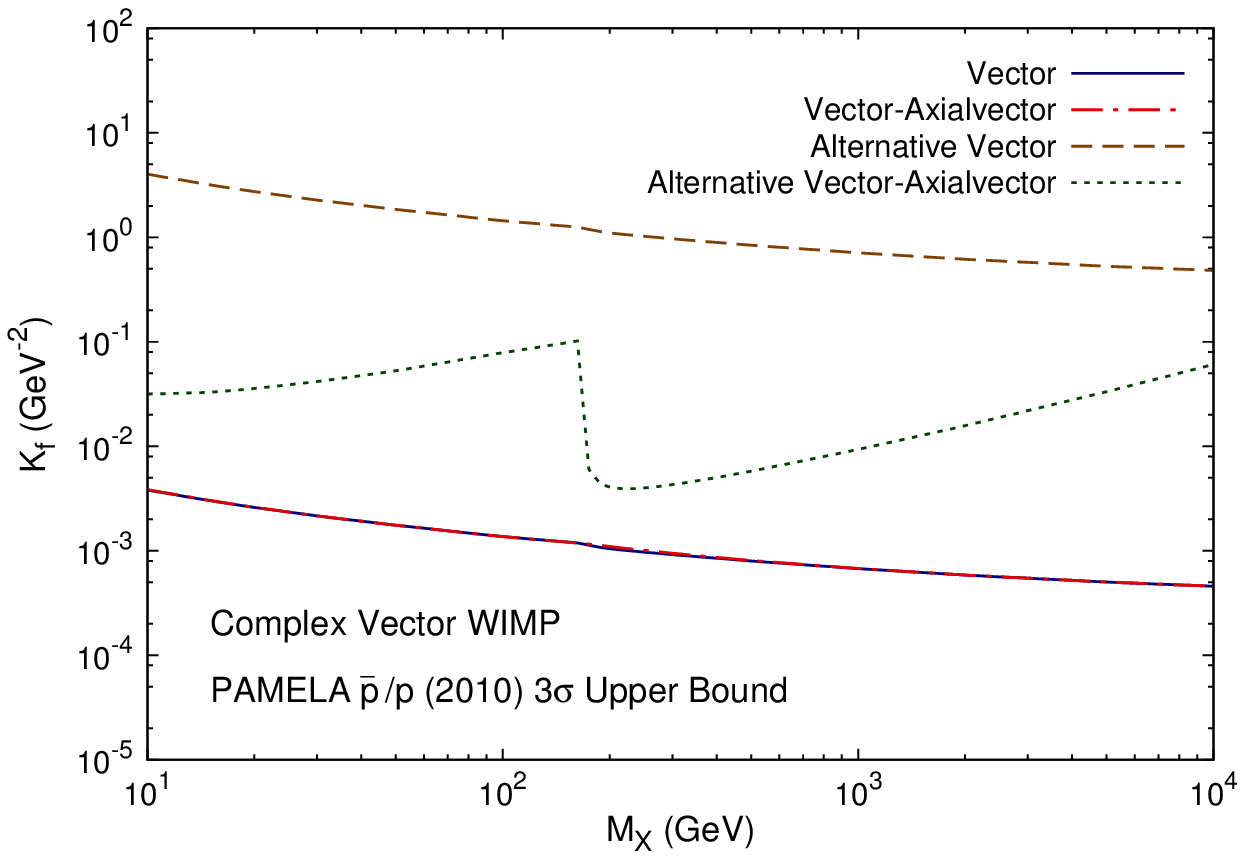}%
\hspace{0.01\textwidth}%
\includegraphics[width=0.44\textwidth]{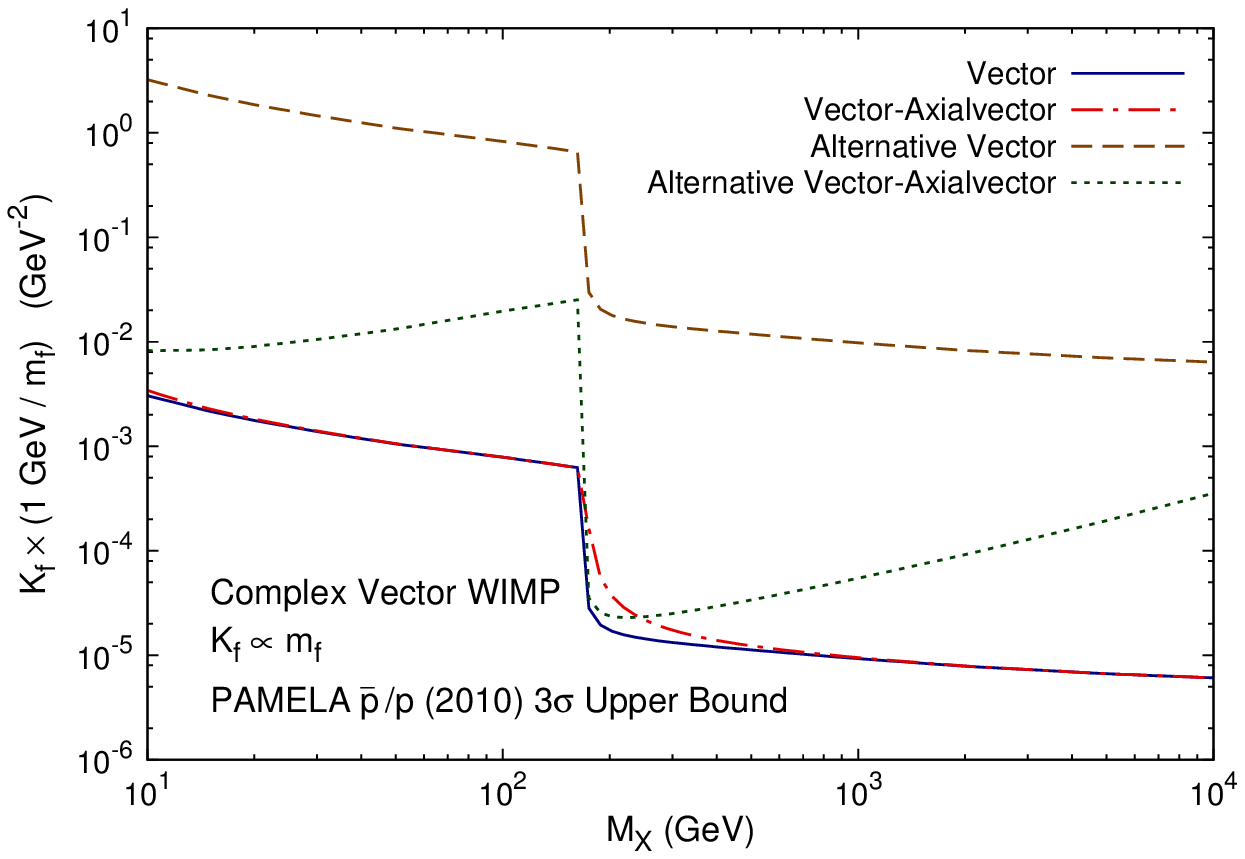}
\caption{The $3\sigma$ upper bounds on the coupling constants $K_f$ from
the PAMELA $\bar p / p$ ratio~\cite{Adriani:2010rc} in each effective
complex vector WIMP model. In the left (right) column, results are given
for universal couplings ($K_f\propto m_f$).} \label{fig:vector:anni:comp_cpl}
\end{figure}

To compare with the searching bounds of direct detection experiments,
we plot the predicted $\sigma_{XN}$ vs. $M_X$ curves with fixed couplings.
In Fig.~\ref{fig:vector:SI}, the spin-dependent WIMP-proton cross sections
(dashed lines) for S and V interactions are shown. In the left (right) column,
results are given for universal couplings ($K_f \propto m_f$).
The upper bounds set by CDMS II~\cite{Ahmed:2009zw} and XENON100
\cite{Aprile:2011hi} are also given in the frames.
In Fig.~\ref{fig:vector:SD}, the spin-dependent (SD) WIMP-neutron cross
sections (dashed lines) for $\widetilde{\mathrm{VA}}$ and T interactions
are shown. In the left (right) column, results are given for
universal couplings ($K_f \propto m_f$). The upper bounds set by
CDMS~\cite{Akerib:2005za} and XENON10~\cite{Angle:2008we} are also
given in the frames.

From Figs.~\ref{fig:vector:SI} and \ref{fig:vector:SD}, we find that
the experimental constraints are much stronger for SI interactions
than those for SD interactions, since the scattering amplitudes between
WIMPs and nucleons in the nuclei add coherently in the case of SI
interactions. Additionally, the curves for dimension-5
operators (S and T) slope downward, while those for dimension-6
operators (V and $\widetilde{\mathrm{VA}}$) remain horizontal in
most range of mass. This is due to an additional factor of $M_X^2$
in Eqs.~\eqref{sigmaN_scat_vector_vect} and \eqref{sigmaN_scat_vector_alt_v_Av}
comparing with Eqs.~\eqref{sigmaN_scat_vector_scal} and
\eqref{sigmaN_scat_vector_tens}.

The calculation of indirect detection constraints of vector WIMPs is
totally similar to that of scalar WIMPs in Subsection \ref{subsec-scal-id}.
In Fig.~\ref{fig:vector:anni:comp_cpl}, we show the $3\sigma$ upper bounds
on the coupling constants $F_f$ from the PAMELA $\bar p / p$
ratio~\cite{Adriani:2010rc} in each effective complex vector WIMP model
for the cases of universal couplings and $K_f \propto m_f$.

Duo to the similar reasons, the bounds in Fig.~\ref{fig:vector:anni:comp_cpl}
have similar properties as the relic density bounds in
Subsection \ref{subsec-vect-rd}:
(a) There are sharp downward bends in the curves at $M_X \sim m_t$
when $\left<\sigma_\mathrm{ann}v\right> \propto m_f^2$ or $\propto m_f^4$.
(b) There are two pairs of interactions that have almost the
identical bounds, S~$\simeq$~SP and V~$\simeq$~VA.
(c) The curves of S and SP interactions lie well above those of
T and $\widetilde{\mathrm{T}}$ interactions, while the curves of
V and VA interactions lie well below those of $\widetilde{\mathrm{V}}$
and $\widetilde{\mathrm{VA}}$ interactions. Again, we can find
the explanations to these features in the annihilation rates
Eqs.~\eqref{sv_vector_scal} -- Eq.~\eqref{sv_vector_alt_tens}.

\subsection{Validity region of effective models and combined constraints
\label{subsec-vect-comb}}

Since we are using effective theory, we can only carry out our analysis
below an energy cutoff scale $\Lambda$, which leads to validity bounds
on the effective couplings $K_f$ if the perturbative theory is assumed
for UV completion, as we discussed in Subsection \ref{subsec-scal-comb}.
Then for the case of universal couplings, we have
\begin{eqnarray}
K_f \ll \frac{2\sqrt{2}\pi}{M_X},&\quad& \text{for S, SP, T
  and $\widetilde{\mathrm{T}}$ interactions}, \label{validity-condi-vect-5}\\
K_f \ll \frac{\sqrt{2}\pi}{M_X^2},&\quad& \text{for V, VA,
  $\widetilde{\mathrm{V}}$ and
  $\widetilde{\mathrm{VA}}$ interactions}. \label{validity-condi-vect-6}
\end{eqnarray}
On the other hand, for the case of $K_f \propto m_f$ we obtain
\begin{eqnarray}
\frac{K_f}{m_f} \ll \frac{\sqrt{2}\pi}{M_X^2},&\quad& \text{for S, SP, T and
  $\widetilde{\mathrm{T}}$ interactions}, \label{validity-condi-vect-mf-5}\\
\frac{K_f}{m_f} \ll \frac{\pi}{\sqrt{2}M_X^3},&\quad& \text{for V, VA,
  $\widetilde{\mathrm{V}}$ and
  $\widetilde{\mathrm{VA}}$ interactions}. \label{validity-condi-vect-mf-6}
\end{eqnarray}

\begin{figure}[!htbp]
\centering
\includegraphics[width=0.44\textwidth]{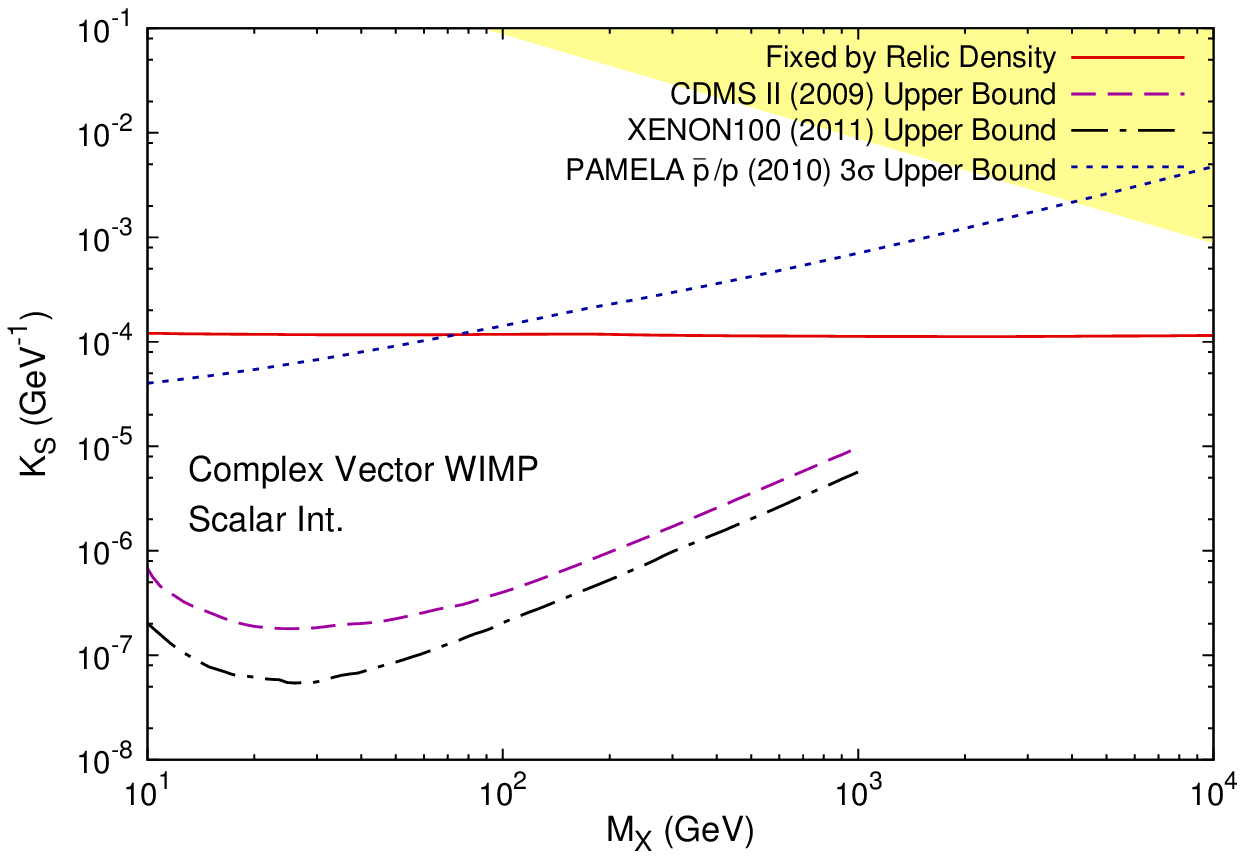}
\hspace{0.01\textwidth}%
\includegraphics[width=0.44\textwidth]{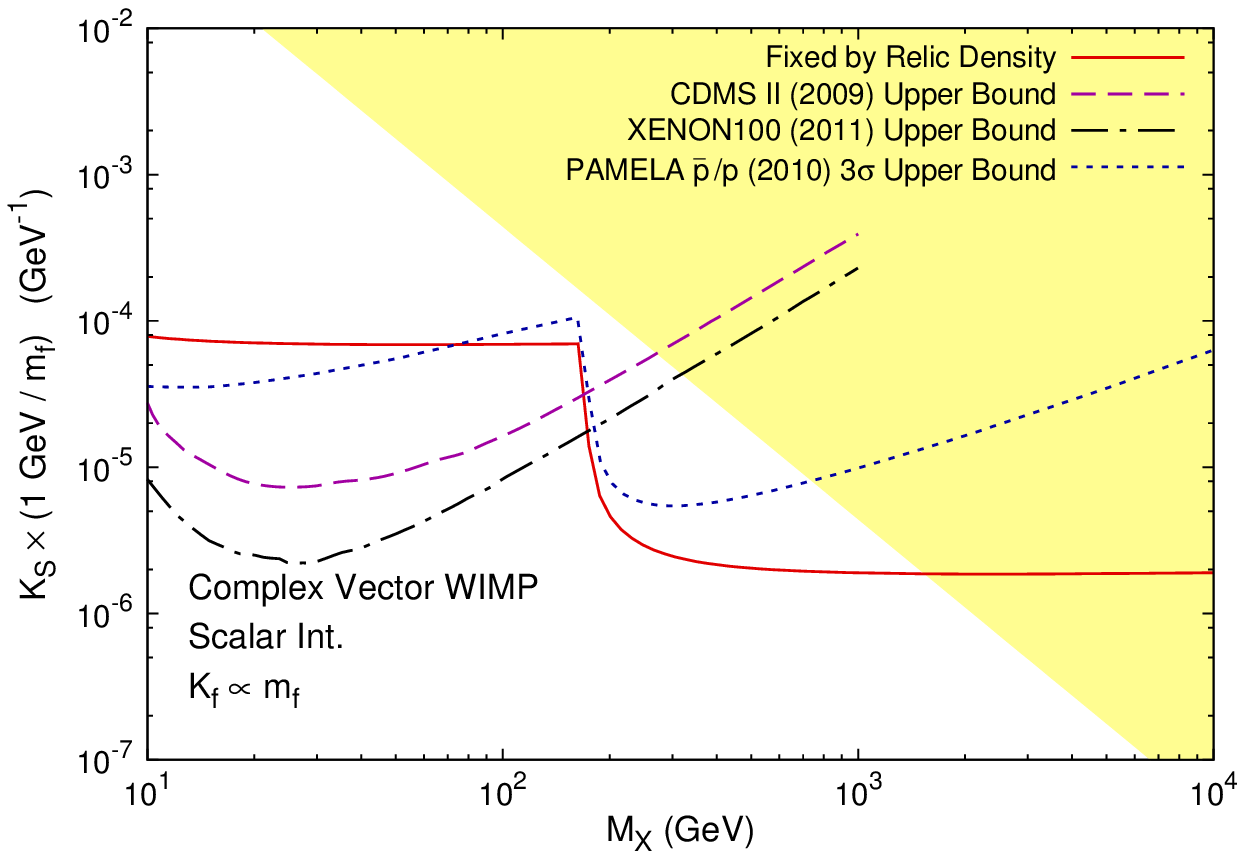}%
\caption{Combined constraints on coupling constants $K_f$ of complex vector
WIMPs with scalar (S) interaction from relic density, direct detection
experiments of CDMS II and XENON100, PAMELA $\bar p / p$ ratio, and
validity of effective theory. The yellow region denotes the invalid parameter
space of effective field theory. The left (right) frame is shown for
the case of universal couplings ($K_f\propto m_f$).
The constraints from relic density, $\bar{p}/p$ ratio
and validity of effective theory for scalar-pseudoscalar (SP) interaction
are very similar to those for scalar interaction, but direct detection
experiments are not sensitive to SP interaction.}
\label{fig-vect-combined-S}
\end{figure}

\begin{figure}[!htbp]
\centering
\includegraphics[width=0.44\textwidth]{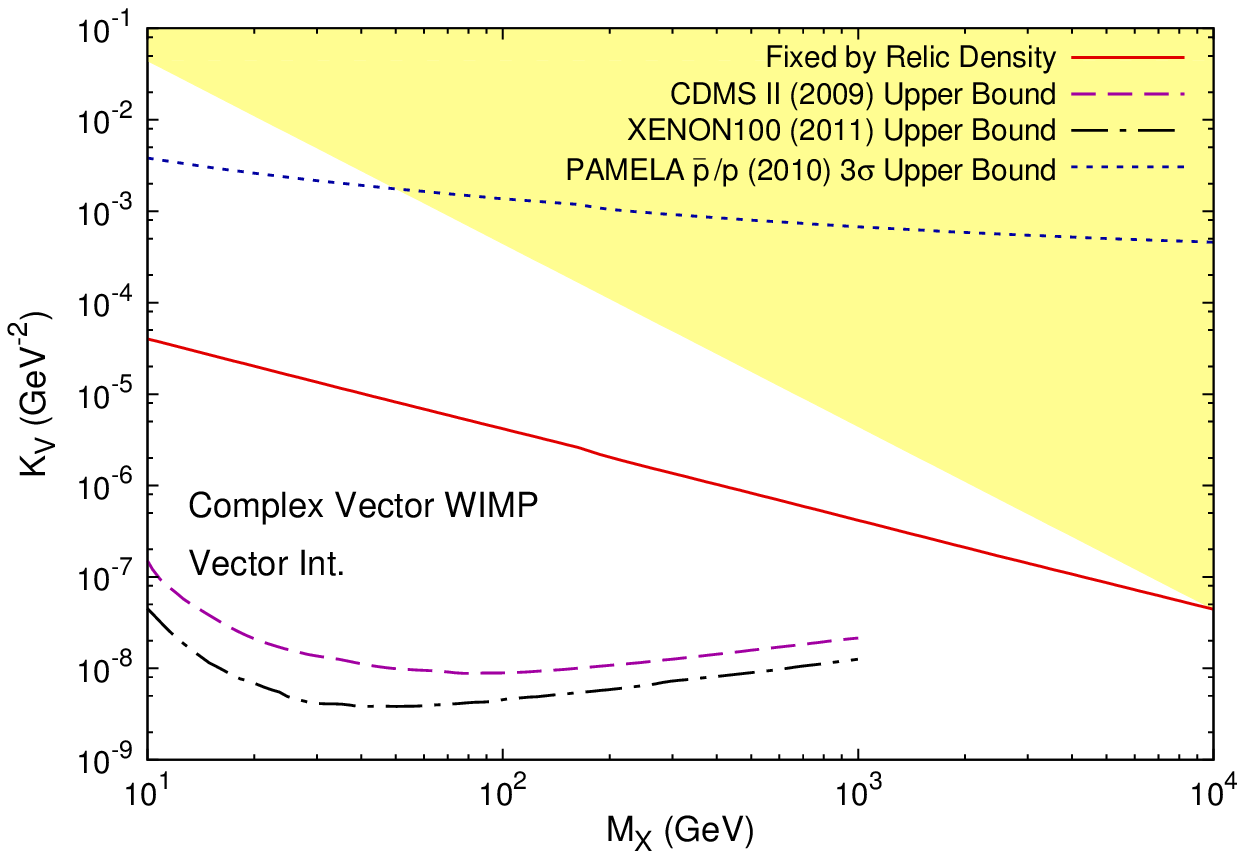}
\hspace{0.01\textwidth}%
\includegraphics[width=0.44\textwidth]{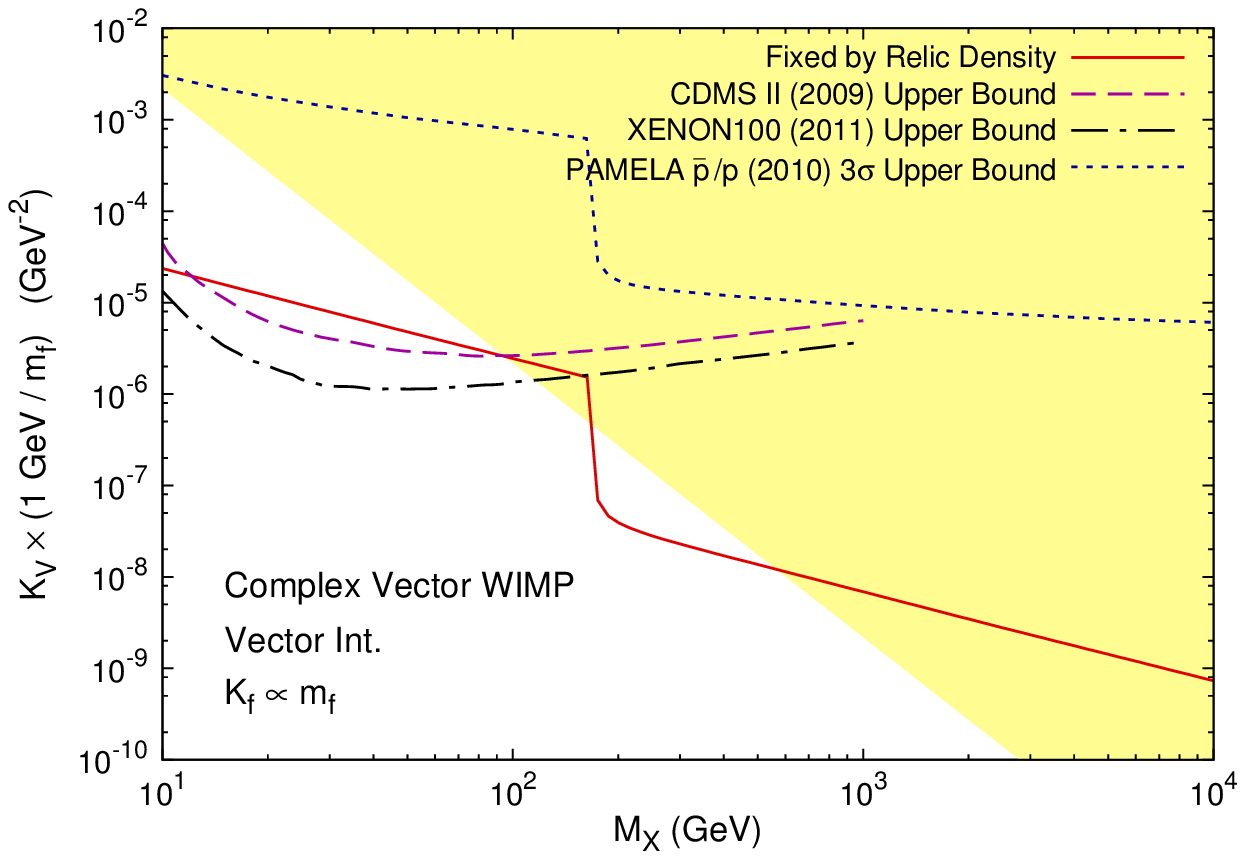}%
\caption{Combined constraints on coupling constants $K_f$ of complex vector
WIMPs with vector (V) interaction from relic density, direct detection
experiments of CDMS II and XENON100, PAMELA $\bar p / p$ ratio, and
validity of effective theory. The constraints from relic density, $\bar{p}/p$
ratio and validity of effective theory for vector-axialvector (VA) interaction
are very similar to those for vector interaction, but direct detection
experiments are not sensitive to VA interaction.}
\label{fig-vect-combined-V}
\end{figure}

\begin{figure}[!htbp]
\centering
\includegraphics[width=0.44\textwidth]{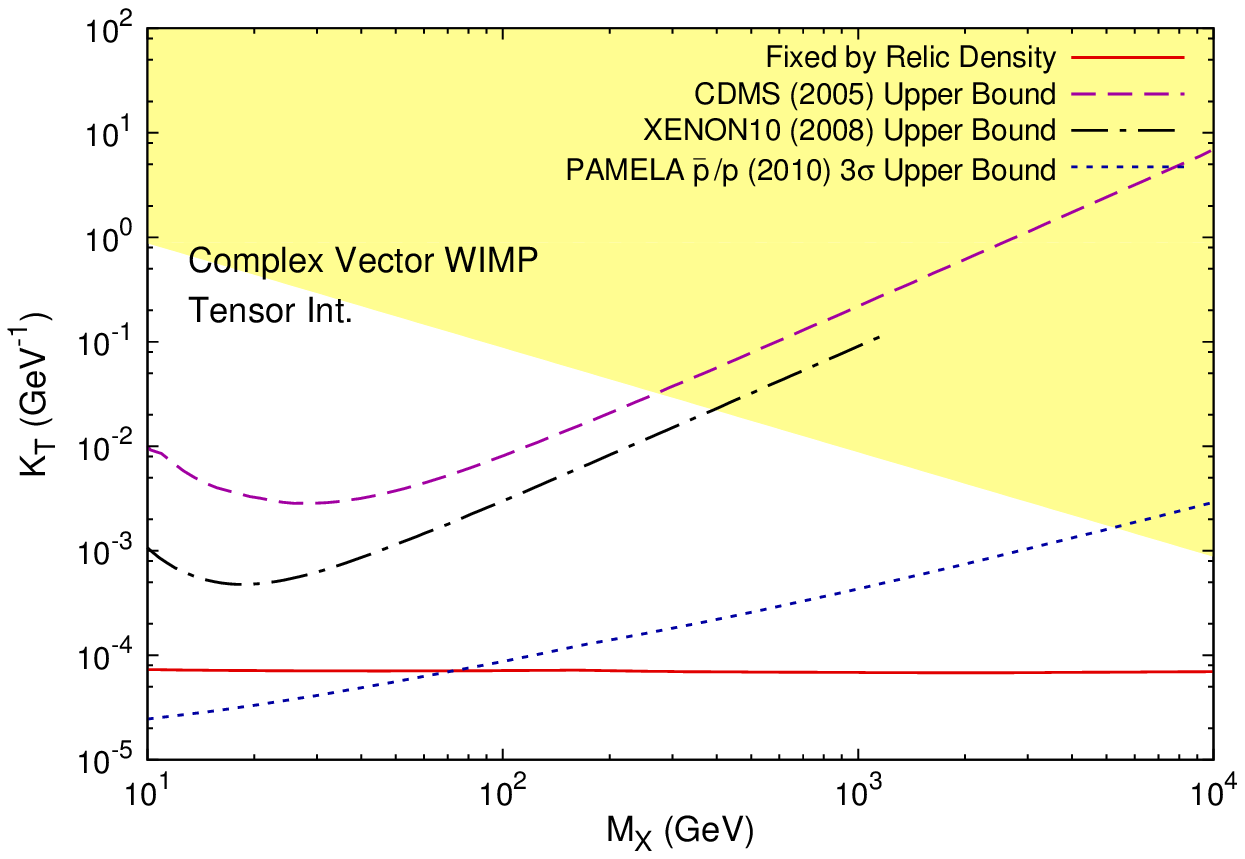}
\hspace{0.01\textwidth}%
\includegraphics[width=0.44\textwidth]{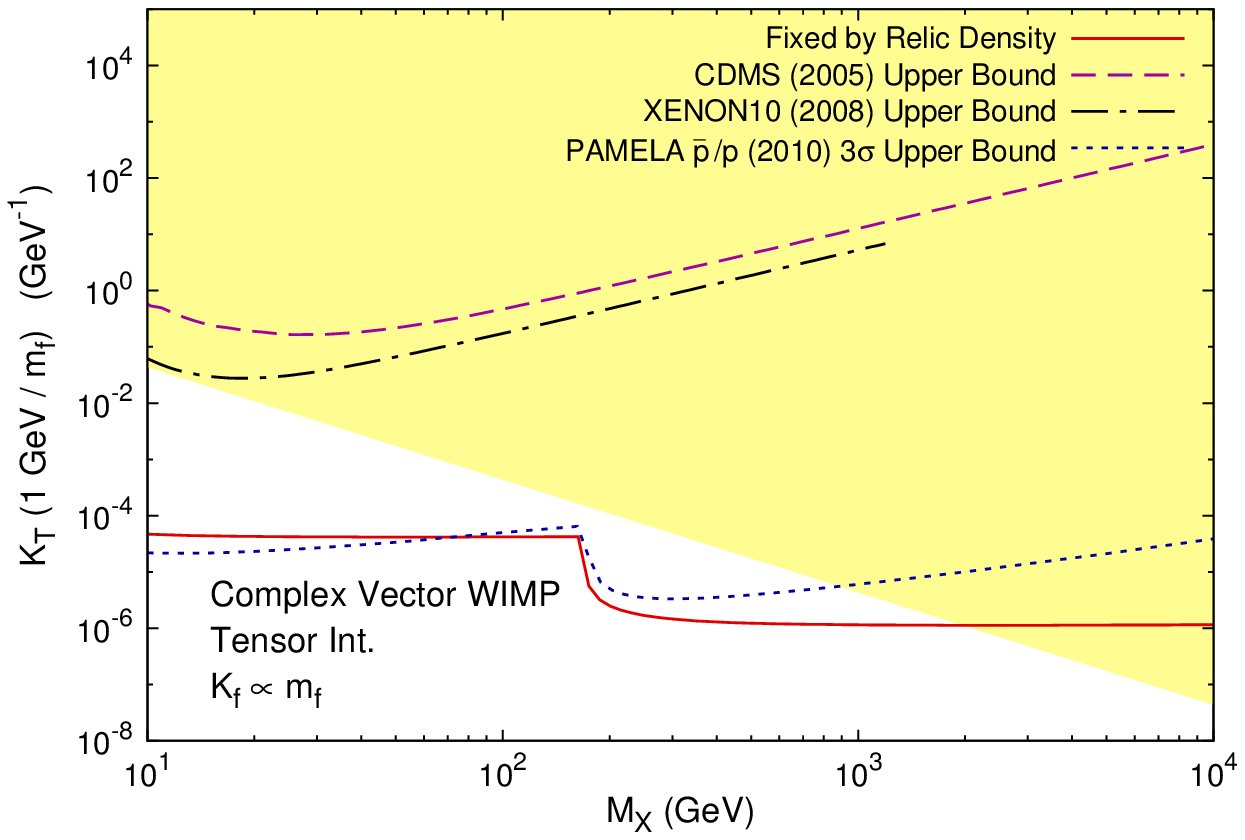}%
\caption{Combined constraints on coupling constants $K_f$ of complex vector
WIMPs with tensor (T) interaction from relic density, direct detection
experiments of CDMS and XENON10, PAMELA $\bar p / p$ ratio, and validity of
effective theory.}
\label{fig-vect-combined-T}
\end{figure}

\begin{figure}[!htbp]
\centering
\includegraphics[width=0.44\textwidth]{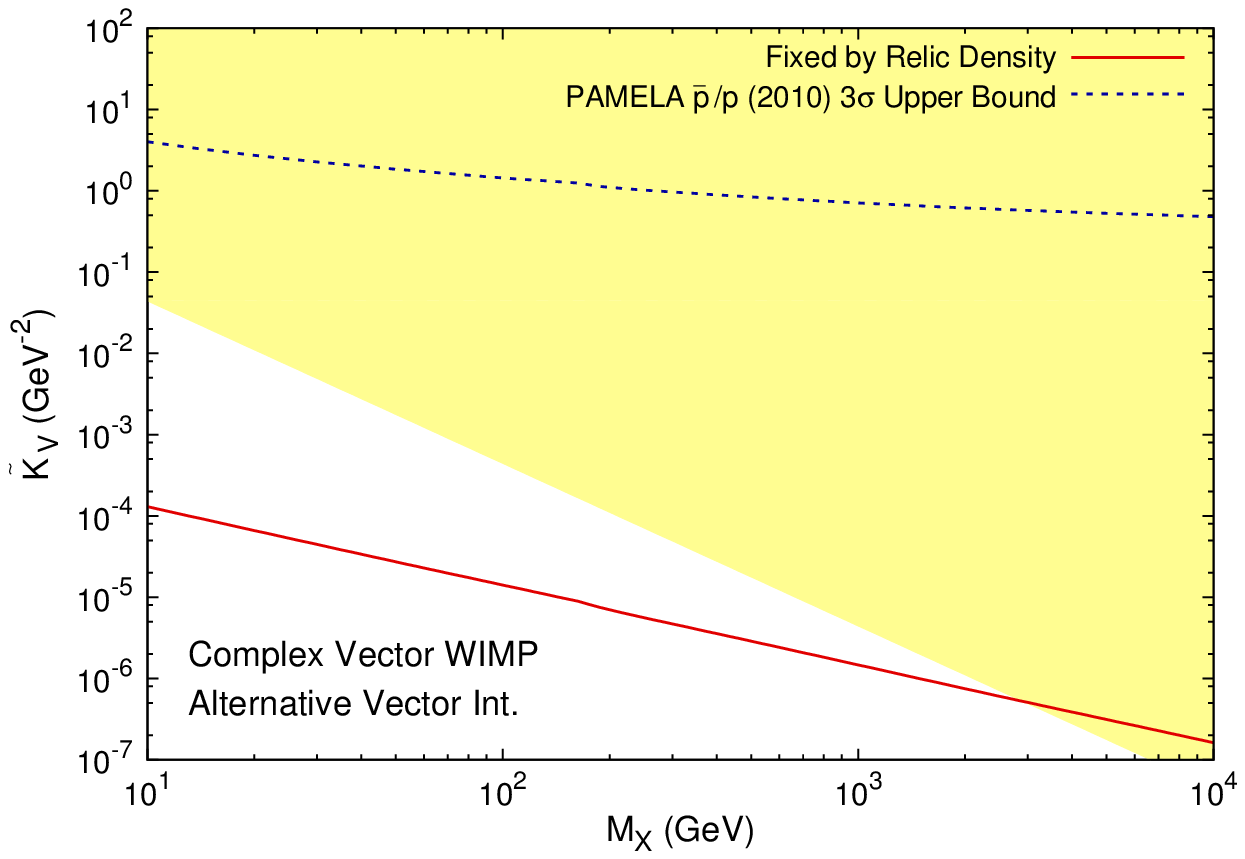}
\hspace{0.01\textwidth}%
\includegraphics[width=0.44\textwidth]{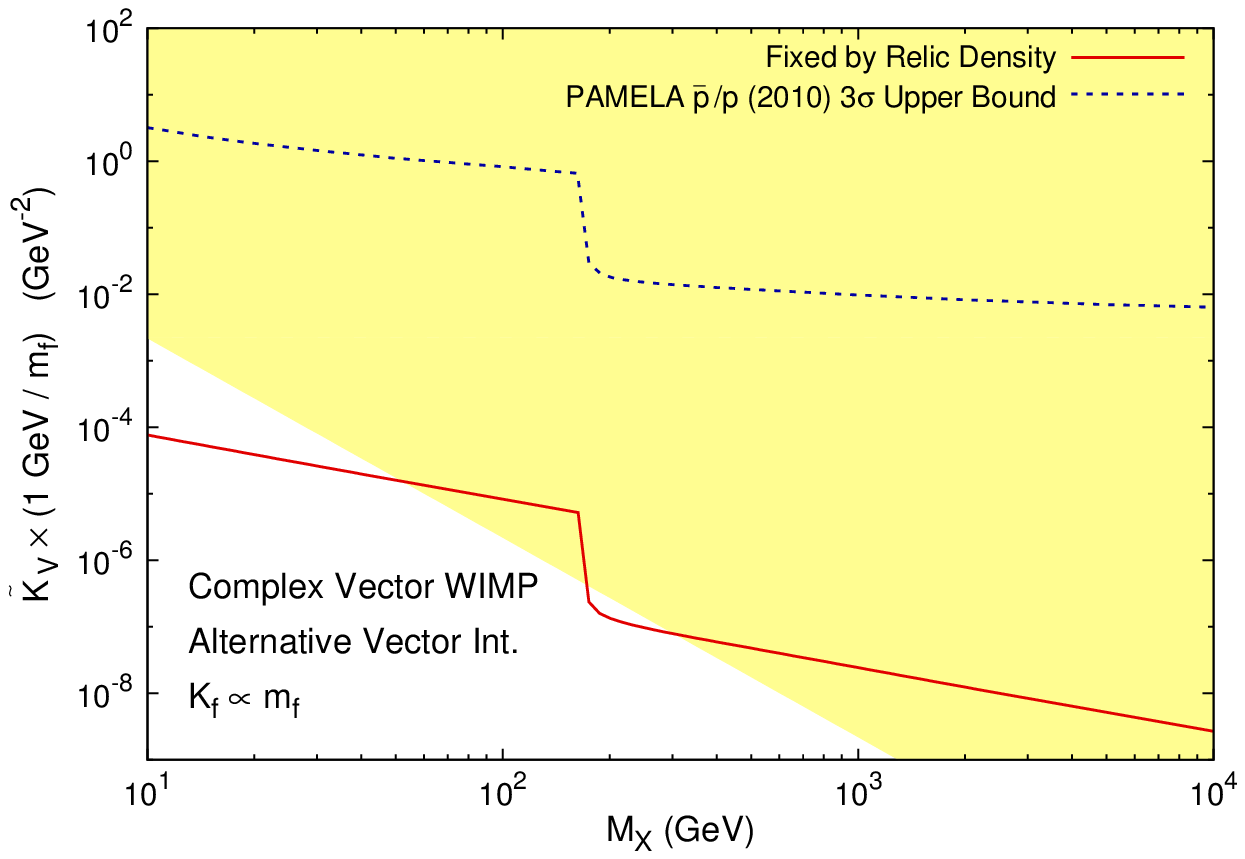}%
\caption{Combined constraints on coupling constants $K_f$ of complex vector
WIMPs with alternative vector ($\widetilde{\mathrm{V}}$) interaction from
relic density, PAMELA $\bar p / p$ ratio, and validity of effective theory.}
\label{fig-vect-combined-alt_vect}
\end{figure}

\begin{figure}[!htbp]
\centering
\includegraphics[width=0.44\textwidth]{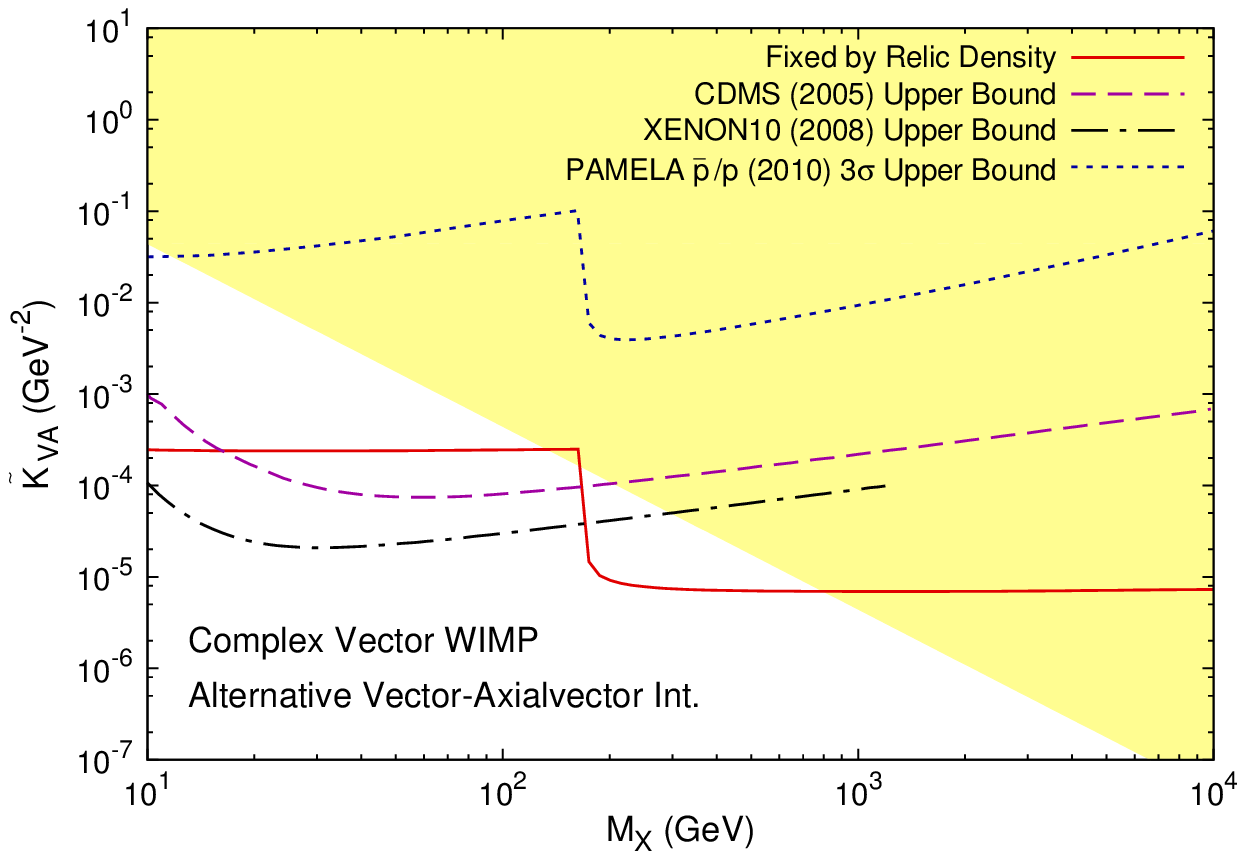}
\hspace{0.01\textwidth}%
\includegraphics[width=0.44\textwidth]{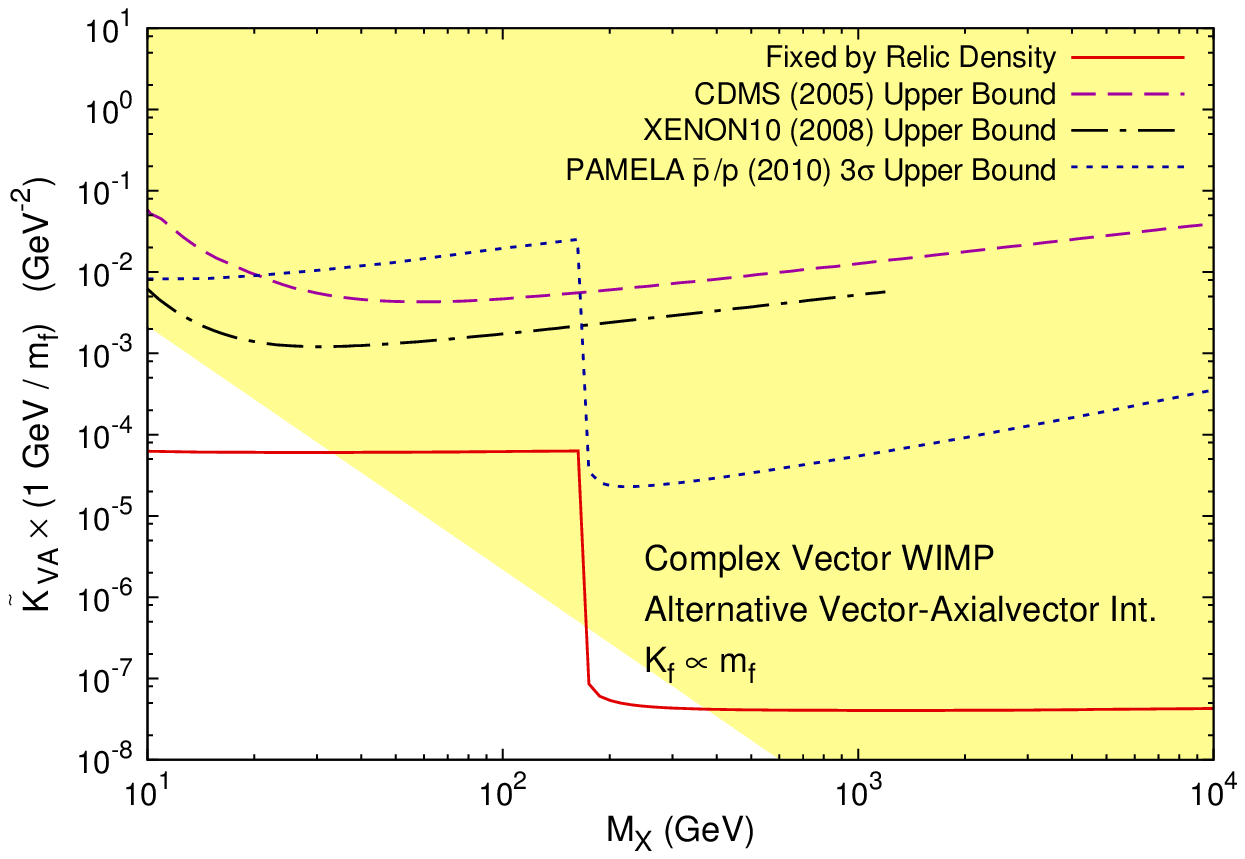}%
\caption{Combined constraints on coupling constants $K_f$ of complex vector
WIMPs with alternative vector-axialvector ($\widetilde{\mathrm{VA}}$)
interaction from relic density, direct detection experiments of CDMS and
XENON10, PAMELA $\bar p / p$ ratio, and validity of effective theory.}
\label{fig-vect-combined-alt_v_Av}
\end{figure}

\begin{figure}[!htbp]
\centering
\includegraphics[width=0.44\textwidth]{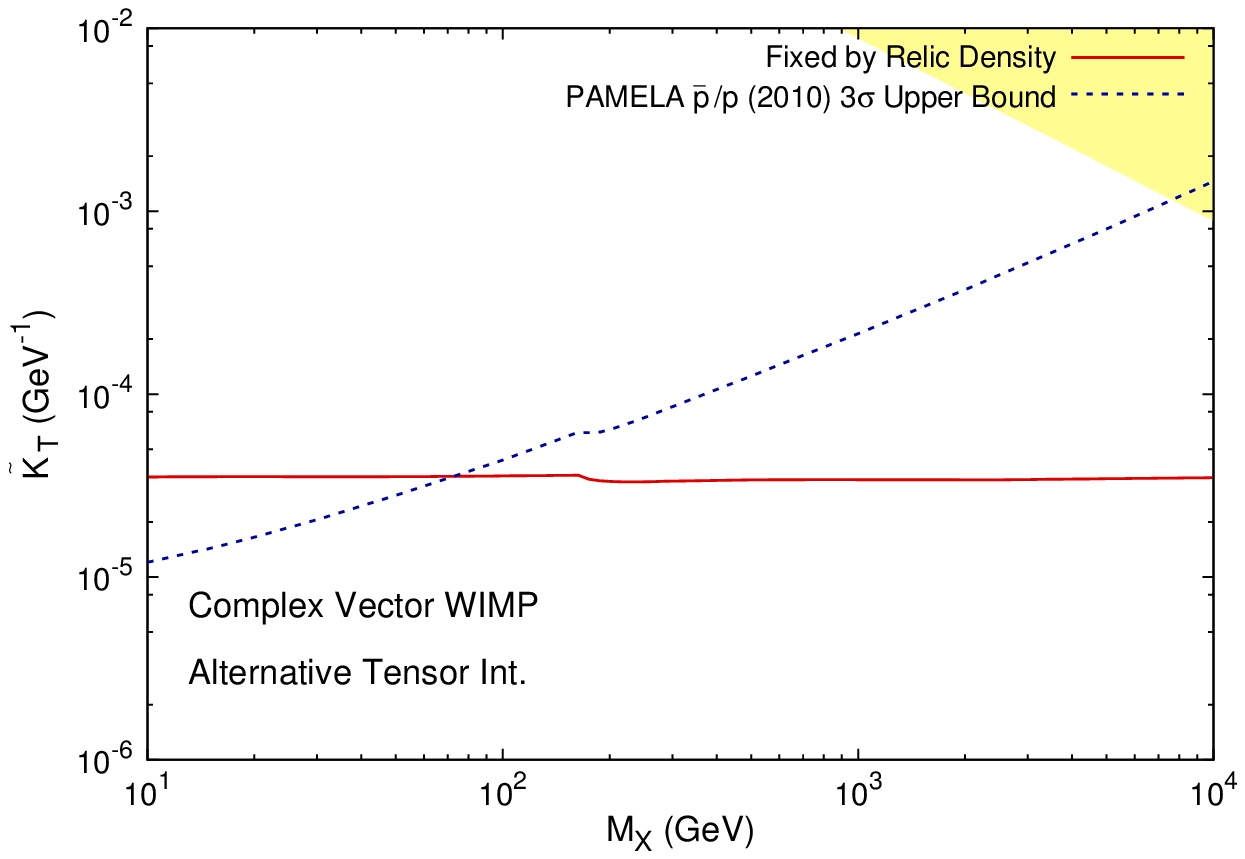}
\hspace{0.01\textwidth}%
\includegraphics[width=0.44\textwidth]{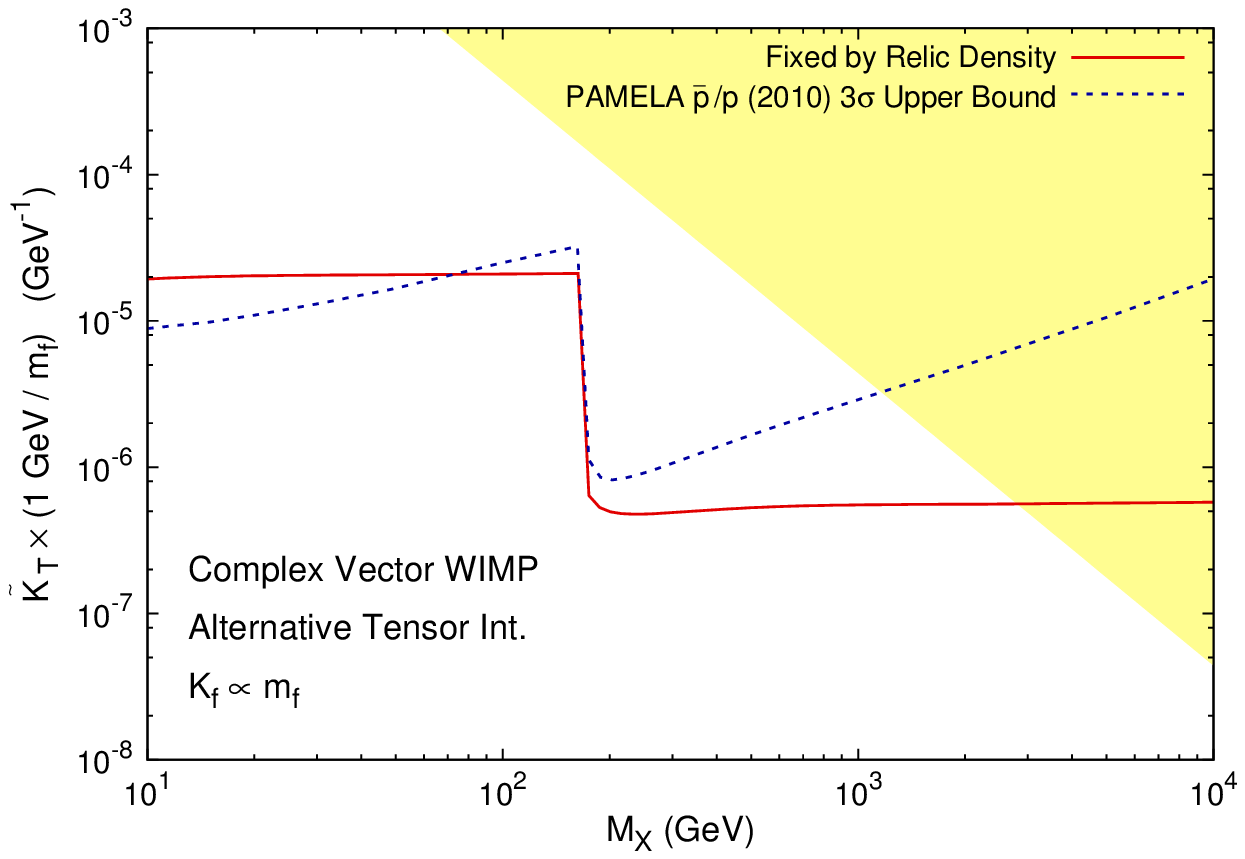}%
\caption{Combined constraints on coupling constants $K_f$ of complex vector
WIMPs with alternative tensor ($\widetilde{\mathrm{T}}$) interaction from
relic density, PAMELA $\bar p / p$ ratio, and validity of effective theory.}
\label{fig-vect-combined-alt_tens}
\end{figure}

Considering the validity conditions \eqref{validity-condi-vect-5} --
\eqref{validity-condi-vect-mf-6} altogether with other phenomenological
constraints, we obtain the combined constraints on the effective vector WIMP
models. In Figs.~\ref{fig-vect-combined-S} --
\ref{fig-vect-combined-alt_tens}, the combined constraints of the effective
models of S, V, T, $\widetilde{\mathrm{V}}$, $\widetilde{\mathrm{VA}}$
and $\widetilde{\mathrm{T}}$ interactions are shown,
with the yellow regions denoting the invalid parameter spaces of
effective field theory.
The constraints on SP and VA interactions are very similar to those on
S and V interactions respectively, except that the SP and VA interactions are
not constrained by direct detection experiments.
S and V interactions with universal couplings are stringently constrained
by direct detection experiments, while the constraints for the case of
$K_f \propto m_f$ are much looser. The direct detection bounds
for SD interactions (T and $\widetilde{\mathrm{VA}}$) is so much looser than
those for SI interactions (S and V) that they only exclude a region
of small mass in the case of $\widetilde{\mathrm{VA}}$ interaction
with universal couplings.
The constraints by PAMELA $\bar{p}/p$ ratio exclude the mass regions of $M_X
\lesssim 70$~GeV for dimension-5 operators (S, SP, T and
$\widetilde{\mathrm{T}}$), while those for dimension-6 operators (V, VA,
$\widetilde{\mathrm{V}}$ and $\widetilde{\mathrm{VA}}$) are much weaker.
The reason is that for dimension-6 operators, the annihilation rates
Eqs.~\eqref{sv_vector_vect}, \eqref{sv_vector_v_Av},
\eqref{sv_vector_alt_vect} and \eqref{sv_vector_alt_v_Av}
are of order $T/M_X$ or $T^2/M_X^2$,
which lead to much looser bounds by indirect detection
because the WIMP temperature today is much lower than that of
the WIMP freeze-out epoch. In the cases of $\widetilde{\mathrm{V}}$
and $\widetilde{\mathrm{VA}}$ interactions with $K_f\propto m_f$,
large parts of the curves set by relic density lie in the invalid regions
of effective theory, which may indicate effects of resonances,
coannihilations, light intermediate states, or final states
other than SM fermion pairs.

\begin{table}[!htb]
\begin{center}
\belowcaptionskip=0.2cm
\caption{A summary for complex vector WIMPs with various effective
interactions. The excluded regions of $M_X$ given by direct and
indirect experiments are indicated.} \label{tab:vector_sum}
\renewcommand{\arraystretch}{1.3}
\footnotesize
\begin{tabular}{ccc}
\hline \hline \multicolumn{3}{c}{Universal coupling} \\
Interaction & Direct detection & PAMELA $\bar p / p$ \\
\hline
S  & Excluded $M_X \simeq 10~\mathrm{GeV} - \mathrm{above}~1~\mathrm{TeV}$
   & Excluded $M_X \simeq 10 - 74~\mathrm{GeV}$ \\
V  & Excluded $M_X \simeq 10~\mathrm{GeV} - \mathrm{above}~1~\mathrm{TeV}$
   & Not sensitive \\
T  & Not sensitive & Excluded $M_X \simeq 10 - 71~\mathrm{GeV}$ \\
SP & Not sensitive & Excluded $M_X \simeq 10 - 73~\mathrm{GeV}$ \\
VA & Not sensitive & Not sensitive \\
$\widetilde{\mathrm{V}}$  & Not sensitive & Not sensitive \\
$\widetilde{\mathrm{VA}}$ & Excluded $M_X \simeq 10 - 170~\mathrm{GeV}$
                          & Not sensitive \\
$\widetilde{\mathrm{T}}$  & Not sensitive
                          & Excluded $M_X \simeq 10 - 72~\mathrm{GeV}$ \\
\hline \hline \multicolumn{3}{c}{$K_f \propto m_f$} \\
Interaction & Direct detection & PAMELA $\bar p / p$ \\
\hline
S  & Excluded $M_X \simeq 10 - 172~\mathrm{GeV}$
   & Excluded $M_X \simeq 10 - 74~\mathrm{GeV}$ \\
V  & Excluded $M_X \simeq 10 - 157~\mathrm{GeV}$ & Not sensitive \\
T  & Not sensitive & Excluded $M_X \simeq 10 - 70~\mathrm{GeV}$ \\
SP & Not sensitive & Excluded $M_X \simeq 10 - 74~\mathrm{GeV}$ \\
VA & Not sensitive & Not sensitive \\
$\widetilde{\mathrm{V}}$  & Not sensitive & Not sensitive \\
$\widetilde{\mathrm{VA}}$ & Not sensitive & Not sensitive \\
$\widetilde{\mathrm{T}}$  & Not sensitive
                          & Excluded $M_X \simeq 10 - 72~\mathrm{GeV}$ \\
\hline \hline
\end{tabular}
\end{center}
\end{table}
As a summary of the study on the vector WIMP,
in Table \ref{tab:vector_sum}, the excluded regions of $M_X$ given by
direct and indirect experiments for complex vector WIMPs with various
effective interactions are shown.

\section{Spin-3/2 dark matter\label{sec-3/2}}

In this section, we will extend our discuss to the case that
DM consists of spin-3/2 fermionic WIMPs.
The spin-3/2 WIMP can be described by a vector-spinor $\chi_\alpha^\mu$
which carries both a vector index $\mu$ and a spinor index $\alpha$.
Then a free WIMP satisfies the rank-1 Rarita-Schwinger equations
\cite{Rarita:1941mf}
\begin{equation}
(i\slashed\partial -M_\chi)\chi^\mu = 0, \quad
\gamma_\mu \chi^\mu = 0,
\label{RS_eqs}
\end{equation}
where the spinor indices have been suppressed. From Eqs.~\eqref{RS_eqs},
it can be seen that each component of $\chi^\mu$ acts as a Dirac spinor,
while all the four components together are constrained by
the subsidiary condition $\gamma_\mu \chi^\mu = 0$.
By introducing an orthogonal spin projection operator,
the spin sum relations in momentum space
can be derived~\cite{Behrends:1957,Fronsdal:1958}, i.e.,
\begin{eqnarray}
\sum\limits_{s=1}^4{u_s^\mu(p)\bar u_s^\nu(p)}
= (\slashed p + M_\chi)\left( P^{\mu\nu} - \frac{1}{3}
P^{\mu\rho}P^{\nu\sigma} \gamma_\rho \gamma_\sigma \right),\label{spin_sum1} \\
\sum\limits_{s=1}^4{v_s^\mu(p)\bar v_s^\nu(p)}
= (\slashed p - M_\chi)\left( P^{\mu\nu} - \frac{1}{3}
P^{\mu\rho}P^{\nu\sigma} \gamma_\rho \gamma_\sigma \right),\label{spin_sum2}
\end{eqnarray}
where $s$ is the spin index, and $u_s^\mu(p)$ ($v_s^\mu(p)$) are
the positive (negative) energy solutions of Eqs.~\eqref{RS_eqs}, and
$P^{\mu\nu} \equiv g^{\mu\nu} - p^\mu p^\nu / p^2$.

Although spin-3/2 field theories suffer from a common problem of
non-renormalizability, it is safe to consider effective field theory
below an appropriate energy cutoff.
Following the assumptions in Sections \ref{sec-scal} and \ref{sec-vect}, we
construct the effective interaction operators of
$\bar\chi^\mu$-$\chi^\nu$-$f$-$f$ type between the WIMPs and the SM fermions
as follows,
\begin{eqnarray}
    \text{Scalar int. (S)}:&&\qquad{} \mathcal{L}_\mathrm{S} =
\sum_f \frac{G_{\mathrm{S},f}}{\sqrt 2} {\bar\chi}^\mu \chi_\mu \bar f f,
\label{lag:3/2:S} \\
    \text{Pseudoscalar int. (P)}:&&\qquad{} \mathcal{L}_\mathrm{P} =
\sum_f \frac{G_{\mathrm{P},f}}{\sqrt 2}
{\bar\chi}^\mu \gamma_5 \chi_\mu \bar f \gamma_5 f,
\label{lag:3/2:P} \\
    \text{Vector int. (V)}:&&\qquad{} \mathcal{L}_\mathrm{V} =
\sum_f \frac{G_{\mathrm{V},f}}{\sqrt 2}
{\bar \chi}^\rho \gamma^\mu \chi_\rho \bar f \gamma_\mu f,
\label{lag:3/2:V} \\
    \text{Axialvector int. (A)}:&&\qquad{} \mathcal{L}_\mathrm{A} =
\sum_f \frac{G_{\mathrm{A},f}}{\sqrt 2}
{\bar\chi}^\rho \gamma^\mu \gamma_5 \chi_\rho \bar f \gamma_\mu \gamma_5 f,
\label{lag:3/2:A} \\
    \text{Tensor int.~1 (T1)}:&&\qquad{} \mathcal{L}_\mathrm{T1} =
\sum_f \frac{G_{\mathrm{T1},f}}{\sqrt 2}
{\bar \chi}^\mu \sigma^{\rho\sigma} \chi_\mu \bar f \sigma_{\rho\sigma} f,
\label{lag:3/2:T1} \\
    \text{Tensor int.~2 (T2)}:&&\qquad{} \mathcal{L}_\mathrm{T2} =
\sum_f \frac{G_{\mathrm{T2},f}}{\sqrt 2}
i({\bar \chi}^\mu \chi^\nu - {\bar \chi}^\nu \chi^\mu) \bar f \sigma_{\mu\nu}f,
\label{lag:3/2:T2} \\
    \text{Tensor int.~3 (T3)}:&&\qquad{} \mathcal{L}_\mathrm{T3} =
\sum_f \frac{G_{\mathrm{T3},f}}{\sqrt 2}
\varepsilon^{\mu\nu\rho\sigma}i({\bar\chi}_\mu \chi_\nu -{\bar\chi }_\nu
\chi_\mu) \bar f \sigma_{\rho\sigma}f,
\label{lag:3/2:T3} \\
    \text{Tensor int.~4 (T4)}:&&\qquad{} \mathcal{L}_\mathrm{T4} =
\sum_f \frac{G_{\mathrm{T4},f}}{\sqrt 2}
\varepsilon^{\mu\nu\rho\sigma}{\bar\chi}^\tau \sigma_{\mu\nu}
\chi_\tau \bar f \sigma_{\rho\sigma} f,
\label{lag:3/2:T4} \\
    \text{Tensor int.~5 (T5)}:&&\qquad{} \mathcal{L}_\mathrm{T5} =
\sum_f \frac{G_{\mathrm{T5},f}}{\sqrt 2}
({\bar\chi}^\mu \gamma_5 \chi^\nu - {\bar\chi}^\nu \gamma_5 \chi^\mu)
\bar f \sigma_{\mu\nu} f,
\label{lag:3/2:T5} \\
    \text{Tensor int.~6 (T6)}:&&\qquad{} \mathcal{L}_\mathrm{T6} =
\sum_f \frac{G_{\mathrm{T6},f}}{\sqrt 2}
\varepsilon^{\mu\nu\rho\sigma}({\bar\chi}_\mu \gamma_5 \chi_\nu
-{\bar\chi}_\nu \gamma_5 \chi_\mu)\bar f \sigma_{\rho\sigma} f,
\label{lag:3/2:T6} \\
    \text{Scalar-Pseudoscalar int. (SP)}:&&\qquad{} \mathcal{L}_\mathrm{SP} =
\sum_f \frac{G_{\mathrm{SP},f}}{\sqrt 2}
{\bar \chi}^\mu \chi_\mu \bar f i\gamma_5 f,
\label{lag:3/2:SP} \\
    \text{Pseudoscalar-Scalar int. (PS)}:&&\qquad{} \mathcal{L}_\mathrm{PS} =
\sum_f \frac{G_{\mathrm{PS},f}}{\sqrt 2}
{\bar \chi}^\mu i\gamma_5 \chi_\mu \bar f f,
\label{lag:3/2:PS} \\
    \text{Vector-Axialvector int. (VA)}:&&\qquad{} \mathcal{L}_\mathrm{VA} =
\sum_f \frac{G_{\mathrm{VA},f}}{\sqrt 2}
{\bar \chi}^\mu \gamma^\nu \chi_\mu \bar f \gamma_\nu \gamma_5 f,
\label{lag:3/2:VA} \\
    \text{Axialvector-Vector int. (AV)}:&&\qquad{} \mathcal{L}_\mathrm{AV} =
\sum_f \frac{G_{\mathrm{AV},f}}{\sqrt 2}
{\bar\chi}^\mu \gamma^\nu \gamma_5 \chi_\mu \bar f \gamma_\nu f,
\label{lag:3/2:AV}
\end{eqnarray}
where the sum of $f$ is over all the SM fermions, and the effective coupling
constants $G_f$ are real numbers with mass dimension of $-2$. Since each
component of $\chi^\mu$ looks like a Dirac spinor,
the S, P, V, A, T1, T4, SP, PS, VA and AV interactions are analogous to
the S, P, V, A, T, $\widetilde{\mathrm{T}}$, SP, PS, VA and AV
interactions defined in our previous work
on Dirac fermionic WIMPs~\cite{Zheng:2010js}, respectively.
Even though the spin-3/2 fermionic WIMP has more degrees of freedom,
it is not important in the calculation of phenomenological constraints.
Thus we can expect the phenomenological constraints of spin-3/2 WIMPs
are also much similar to those of Dirac fermionic WIMPs.
For this reason, we just give a brief
discussion on spin-3/2 WIMPs in the following.

The annihilation cross sections of WIMP-antiWIMP to SM fermion-antifermion pairs
are given by
\begin{eqnarray}
    \sigma_\mathrm{S,\,ann} &=& \frac{1}{576\pi M_\chi^4}\sum\limits_f
\left(\frac{G_{\mathrm{S},f}}{\sqrt 2}\right)^2
c_f \sqrt{\frac{s-4m_f^2}{s-4M_\chi^2}}
\frac{(s - 4m_f^2)(s - 4M_\chi ^2)(s^2 - 6M_\chi ^2 s + 18M_\chi ^4)}{s},
\label{sigma_3/2_S}\\
    \sigma_\mathrm{P,\,ann} &=& \frac{1}{{576\pi M_\chi ^4}}\sum\limits_f
\left(\frac{G_{\mathrm{P},f}}{\sqrt 2}\right)^2
c_f \sqrt{\frac{s - 4m_f^2}
{s - 4M_\chi^2}} s (s^2 - 2sM_\chi^2 + 10M_\chi^4),
\label{sigma_3/2_P}\\
    \sigma_\mathrm{V,\,ann} &=& \frac{1}{432\pi M_\chi^4}\sum\limits_f
\left(\frac{G_{\mathrm{V},f}}{\sqrt 2}\right)^2
c_f \sqrt{\frac{s - 4m_f^2}{s - 4M_\chi ^2}}
\frac{(s + 2m_f^2)(s^3 - 2s^2 M_\chi ^2 - 2sM_\chi^4 + 36M_\chi^6)}{s},
\label{sigma_3/2_V}\\
    \sigma_\mathrm{A,\,ann} &=& \frac{1}{{432\pi M_\chi ^4}}\sum\limits_f
\left(\frac{G_{\mathrm{A},f}}{\sqrt 2}\right)^2
c_f \sqrt{\frac{s - 4m_f^2} {s - 4M_\chi^2}} \nonumber\\
    && \times \left[s^3 - 4(m_f^2 + 2M_\chi^2)s^2 + (44m_f^2 M_\chi^2 + 26
M_\chi^4)s - 128m_f^2 M_\chi^4 - 40M_\chi^6 + 280\frac{m_f^2 M_\chi^6}s\right],
\label{sigma_3/2_A}\\
    \sigma_\mathrm{T1,\,ann} &=& \frac{1}{216\pi M_\chi^4}\sum\limits_f
\left(\frac{G_{\mathrm{T1},f}}{\sqrt 2}\right)^2
c_f \sqrt{\frac{{s - 4m_f^2}}{s - 4M_\chi^2}} \nonumber\\
    && \times \left[s^3 + 2(m_f^2 - M_\chi^2)s^2 + 2M_\chi^2
(M_\chi ^2 + 10m_f^2)s + 20M_\chi^6 - 92m_f^2M_\chi ^4
+ 400\frac{m_f^2 M_\chi^6}s \right],
\label{sigma_3/2_T1}\\
    \sigma_\mathrm{T2,\,ann} &=& \frac{1}{432\pi M_\chi^4} \sum\limits_f
\left(\frac{G_{\mathrm{T2},f}}{\sqrt 2}\right)^2
c_f \sqrt{\frac{{s - 4m_f^2}}
{s - 4M_\chi^2}} \left(s - 4M_\chi^2\right) \nonumber\\
    && \times \left[s^2 + 4(2m_f^2 + M_\chi^2)s
- 2M_\chi^2(2m_f^2 + 5M_\chi^2) - 200\frac{m_f^2 M_\chi^4}s \right],
\label{sigma_3/2_T2}\\
    \sigma_\mathrm{T3,\,ann} &=& \frac{1}{{108\pi M_\chi^4}} \sum\limits_f
\left(\frac{G_{\mathrm{T3},f}}{\sqrt 2}\right)^2
c_f \sqrt{\frac{s - 4m_f^2}{s - 4M_\chi^2}}(s - 4M_\chi^2) \nonumber\\
    && \times \left[s^2 + 4(M_\chi^2 - m_f^2)s
+ 10(2m_f^2 - M_\chi^2)M_\chi ^2 + 160\frac{m_f^2 M_\chi^4}s\right],
\label{sigma_3/2_T3}\\
    \sigma_\mathrm{T4,\,ann} &=& \frac{1}{54\pi M_\chi^4}\sum\limits_f
\left(\frac{G_{\mathrm{T4},f}}{\sqrt 2}\right)^2
c_f \sqrt{\frac{s - 4m_f^2} {s - 4M_\chi^2}} \nonumber \\
    && \times \left[s^3 + 2(m_f^2 - M_\chi^2)s^2 + 2M_\chi^2(M_\chi^2 - 14
m_f^2)s + 20M_\chi^6 + 100m_f^2M_\chi^4 - 320\frac{m_f^2 M_\chi^6}{s} \right],
\label{sigma_3/2_T4}\\
    \sigma_\mathrm{T5,\,ann} &=& \frac{1}{432\pi M_\chi^4} \sum\limits_f
\left(\frac{G_{\mathrm{T5},f}}{\sqrt 2} \right)^2
c_f \sqrt{\frac{s - 4m_f^2}{s - 4M_\chi ^2}} \nonumber \\
    && \times \left[s^3 + 4(2m_f^2 + M_\chi^2)s^2
- 2M_\chi^2(14m_f^2 + M_\chi^2)s - 136m_f^2 M_\chi^4 \right],
\label{sigma_3/2_T5}\\
    \sigma_\mathrm{T6,\,ann} &=& \frac{1}{108\pi M_\chi^4}\sum\limits_f
\left(\frac{G_{\mathrm{T6},f}}{\sqrt 2}\right)^2
c_f \sqrt{\frac{s - 4m_f^2}{s - 4M_\chi^2}} \\
    && \times \left[s^3 + 4(M_\chi^2 - m_f^2) s^2 + 2M_\chi^2
(22m_f^2 - M_\chi^2)s + 128m_f^2 M_\chi^4 \right],
\label{sigma_3/2_T6}\\
    \sigma_\mathrm{SP,\,ann} &=& \frac{1}{576\pi M_\chi^4}\sum\limits_f
\left(\frac{G_{\mathrm{SP},f}}{\sqrt 2}\right)^2
c_f \sqrt{\frac{s - 4m_f^2}
{s - 4M_\chi ^2}} (s - 4M_\chi^2)(s^2 - 6M_\chi ^2s + 18M_\chi^4),
\label{sigma_3/2_SP}\\
    \sigma_\mathrm{PS,\,ann} &=& \frac{1}{576\pi M_\chi^4}\sum\limits_f
\left(\frac{G_{\mathrm{PS},f}}{\sqrt 2}\right)^2
c_f \sqrt{\frac{s - 4m_f^2}
{s - 4M_\chi ^2}} (s - 4m_f^2)(s^2 - 2M_\chi^2 s + 10M_\chi^4),
\label{sigma_3/2_PS}\\
    \sigma_\mathrm{VA,\,ann} &=& \frac{1}{432\pi M_\chi^4}\sum\limits_f
\left(\frac{G_{\mathrm{VA},f}}{\sqrt 2}\right)^2
c_f \sqrt{\frac{s - 4m_f^2}{s - 4M_\chi ^2}}
\frac{(s - 4m_f^2)(s^3 - 2M_\chi^2 s^2 - 2M_\chi^4 s + 36M_\chi^6)}{s},
\label{sigma_3/2_VA}\\
    \sigma_\mathrm{AV,\,ann} &=& \frac{1}{432\pi M_\chi^4} \sum\limits_f
\left(\frac{G_{\mathrm{AV},f}}{\sqrt 2}\right)^2
c_f \sqrt{\frac{s - 4m_f^2}{s - 4M_\chi^2}}
\frac{(s - 4M_\chi^2)(s + 2m_f^2)(s^2 - 4M_\chi^2 s + 10M_\chi^4)}{s}.
\label{sigma_3/2_AV}
\end{eqnarray}
Taking the thermal average, we obtain
\begin{eqnarray}
    \left<\sigma_\mathrm{S,\,ann}v\right> &\simeq& \frac{5}{24\pi}\sum\limits_f
\left(\frac{G_{\mathrm{S},f}}{\sqrt 2}\right)^2
c_f \left(1-\frac{m_f^2}{M_\chi ^2}\right)^{3/2} M_\chi^2 \frac{T}{M_\chi},
\label{sv_3/2_S}\\
    \left<\sigma_\mathrm{P,\,ann}v\right> &\simeq& \frac{1}{4\pi}\sum\limits_f
\left(\frac{G_{\mathrm{P},f}}{\sqrt 2}\right)^2
c_f \sqrt{1 - \frac{m_f^2}{M_\chi^2}} M_\chi^2 \left[1 +
\frac{8 - 5m_f^2/M_\chi^2}{4 (1 - m_f^2/M_\chi^2)}\frac{T}{M_\chi}\right],
\label{sv_3/2_P}\\
    \left<\sigma_\mathrm{V,\,ann}v\right> &\simeq& \frac{5}{18\pi}\sum\limits_f
\left(\frac{G_{\mathrm{V},f}}{\sqrt 2}\right)^2
c_f \sqrt{1 - \frac{m_f^2}{M_\chi^2}}M_\chi^2\left[\left(1 +
\frac{1}{2}\frac{m_f^2}{M_\chi ^2}\right) + \frac{3(4 - 2m_f^2/M_\chi^2
+ m_f^4/M_\chi ^4)}{8(1 - m_f^2/M_\chi ^2)} \frac{T}{M_\chi}\right],
\label{sv_3/2_V}\\
    \left<\sigma_\mathrm{A,\,ann}v \right> &\simeq& \frac{1}{4\pi}\sum\limits_f
\left(\frac{G_{\mathrm{A},f}}{\sqrt 2}\right)^2
c_f \sqrt {1 - \frac{{m_f^2}}{M_\chi ^2}} M_\chi ^2\left[\frac{m_f^2}{M_\chi ^2}
+ \frac{40 - 116m_f^2/M_\chi ^2 + 103m_f^4/M_\chi ^4}{36(1 - m_f^2/M_\chi ^2)}
\frac{T}{M_\chi} \right],
\label{sv_3/2_A}\\
    \left<\sigma_\mathrm{T1,\,ann}v\right> &\simeq& \frac{5}{9\pi}\sum\limits_f
\left(\frac{G_{\mathrm{T1},f}}{\sqrt 2}\right)^2
c_f \sqrt{1 - \frac{{m_f^2}}{M_\chi^2}}M_\chi^2 \left[\left(1 + 2\frac{m_f^2}
{M_\chi^2}\right) + \frac{38 - 61m_f^2/M_\chi^2 + 68m_f^4/M_\chi ^4}
{20(1 - m_f^2/M_\chi^2)} \frac{T}{M_\chi}\right],
\label{sv_3/2_T1}\\
    \left<\sigma_\mathrm{T2,\,ann}v\right> &\simeq&\frac{11}{18\pi}\sum\limits_f
\left(\frac{G_{\mathrm{T2},f}}{\sqrt 2}\right)^2
c_f \left(1 - \frac{m_f^2}{M_\chi ^2}\right)^{3/2} M_\chi^2 \frac{T}{M_\chi},
\label{sv_3/2_T2}\\
    \left<\sigma_\mathrm{T3,\,ann}v\right> &\simeq& \frac{22}{9\pi}\sum\limits_f
\left(\frac{G_{\mathrm{T3},f}}{\sqrt 2}\right)^2
c_f \sqrt{1 - \frac{m_f^2}{M_\chi^2}}
M_\chi^2 \left(1 + 2\frac{m_f^2}{M_\chi^2}\right) \frac{T}{M_\chi},
\label{sv_3/2_T3}\\
    \left<\sigma_\mathrm{T4,\,ann}v\right> &\simeq& \frac{20}{9\pi}\sum\limits_f
\left(\frac{G_{\mathrm{T4},f}}{\sqrt 2}\right)^2
c_f \left({1 - \frac{m_f^2}{M_\chi^2}}\right)^{3/2} M_\chi^2 \left[1 +
\frac{38 + 61m_f^2/M_\chi^2}{20(1 - m_f^2/M_\chi^2)}\frac{T}{M_\chi}\right],
\label{sv_3/2_T4}\\
    \left<\sigma_\mathrm{T5,\,ann}v\right> &\simeq& \frac{5}{9\pi}\sum\limits_f
\left(\frac{G_{\mathrm{T5},f}}{\sqrt 2}\right)^2
c_f \left(1 - \frac{m_f^2}{M_\chi^2}\right)^{3/2} M_\chi^2 \left[1 +
\frac{3(16 + 27m_f^2/M_\chi^2)}{20(1 - m_f^2/M_\chi^2)}\frac{T}{M_\chi}\right],
\label{sv_3/2_T5}\\
    \left<\sigma_\mathrm{T6,\,ann}v\right> &\simeq& \frac{20}{9\pi}\sum\limits_f
\left(\frac{G_{\mathrm{T6},f}}{\sqrt 2}\right)^2
c_f \sqrt{1 - \frac{m_f^2}{M_\chi^2}} M_\chi^2 \left[\left(1 + 2\frac{m_f^2}
{M_\chi^2}\right) + \frac{3(16 - 27m_f^2/M_\chi^2 + 26m_f^4/M_\chi^4)}
{20(1 - m_f^2/M_\chi ^2)}\frac{T}{M_\chi} \right],
\label{sv_3/2_T6}\\
    \left<\sigma_\mathrm{SP,\,ann}v\right> &\simeq& \frac{5}{24\pi}\sum\limits_f
\left(\frac{G_{\mathrm{SP},f}}{\sqrt 2}\right)^2
c_f \sqrt{1 - \frac{m_f^2}{M_\chi^2}} M_\chi^2\frac{T}{M_\chi},
\label{sv_3/2_SP}\\
    \left<\sigma_\mathrm{PS,\,ann}v\right> &\simeq& \frac{1}{4\pi}\sum\limits_f
\left(\frac{G_{\mathrm{PS},f}}{\sqrt 2}\right)^2
c_f \left(1 - \frac{m_f^2}{M_\chi ^2}\right)^{3/2} M_\chi^2 \left[1 +
\frac{8 + m_f^2/M_\chi^2}{4(1 - m_f^2/M_\chi^2)}\frac{T}{M_\chi} \right],
\label{sv_3/2_PS}\\
    \left<\sigma_\mathrm{VA,\,ann}v\right> &\simeq& \frac{5}{18\pi}\sum\limits_f
\left(\frac{G_{\mathrm{VA},f}}{\sqrt 2}\right)^2
c_f \left(1 - \frac{m_f^2}{M_\chi^2}\right)^{3/2} M_\chi^2 \left[1 +
\frac{3(2 + m_f^2/M_\chi^2)}{4(1 - m_f^2/M_\chi^2)}\frac{T}{{M_\chi}} \right],
\label{sv_3/2_VA}\\
    \left<\sigma_\mathrm{AV,\,ann}v\right> &\simeq& \frac{5}{18\pi}\sum\limits_f
\left(\frac{G_{\mathrm{AV},f}}{\sqrt 2}\right)^2
c_f \sqrt{1 - \frac{m_f^2}{M_\chi^2}} M_\chi^2 \left(1 +
\frac{1}{2}\frac{m_f^2}{M_\chi^2}\right)\frac{T}{M_\chi}.
\label{sv_3/2_AV}
\end{eqnarray}
When calculating the relic density of spin-3/2 WIMPs, it is worth noting that
the degree of freedom for a spin-3/2 fermionic WIMP is $g=4$.

In the low velocity limit, the scattering cross sections of P, T3, T4,
T5, T6, SP, PS, VA and AV interactions between WIMPs and nuclei vanish.
In the remaining interactions, the S and V interactions are spin-independent,
while the A, T1 and T2 interactions are spin-dependent.
The WIMP-nucleon cross sections are given by
\begin{eqnarray}
\text{Scalar int.}:\qquad && \sigma_{\mathrm{S},\,\chi N} =
\frac{m_N^2 M_\chi^2}{\pi(M_\chi +m_N)^2}
\left(\frac{G_{\mathrm{S},N}}{\sqrt 2}\right)^2,
\label{sigmaN_scat_3/2_S}\\
\text{Vector int.}:\qquad && \sigma_{\mathrm{V},\,\chi N} =
\frac{m_N^2 M_\chi^2}{\pi(M_\chi +m_N)^2}
\left(\frac{G_{\mathrm{V},N}}{\sqrt 2}\right)^2,
\label{sigmaN_scat_3/2_V}\\
\text{Axialvector int.}:\qquad && \sigma_{\mathrm{A},\,\chi N} =
\frac{5 m_N^2 M_\chi^2}{3\pi(M_\chi +m_N)^2}
\left(\frac{G_{\mathrm{A},N}}{\sqrt 2}\right)^2,
\label{sigmaN_scat_3/2_A}\\
\text{Tensor int.~1}:\qquad && \sigma_{\mathrm{T1},\,\chi N} =
\frac{20 m_N^2 M_\chi^2}{3\pi(M_\chi +m_N)^2}
\left(\frac{G_{\mathrm{T1},N}}{\sqrt 2}\right)^2,
\label{sigmaN_scat_3/2_T1}\\
\text{Tensor int.~2}:\qquad && \sigma_{\mathrm{T2},\,\chi N} =
\frac{20 m_N^2 M_\chi^2}{3\pi(M_\chi +m_N)^2}
\left(\frac{G_{\mathrm{T2},N}}{\sqrt 2}\right)^2,
\label{sigmaN_scat_3/2_T2}
\end{eqnarray}
where $G_N$ ($N=p,n$) are the induced couplings of effective WIMP-nucleon
interactions, related to the couplings of WIMP-quark $G_q$ by form factors.
For S, V, A and T1 (or T2) interactions, the relations between $G_N$ and $G_q$
are similar to Eqs.~\eqref{form-S}, \eqref{form-V}, \eqref{form-A} and
\eqref{form-T}, respectively. Since all the spin-3/2 WIMP effective models
\eqref{lag:3/2:S} -- \eqref{lag:3/2:AV}
have effective couplings $G_f$ with mass dimension
of $-2$, their validity regions are similar to those indicated in
Eqs.~\eqref{validity-condi-scal-6} and \eqref{validity-condi-scal-mf-6}.

\begin{figure}[!htbp]
\centering
\includegraphics[width=0.44\textwidth]{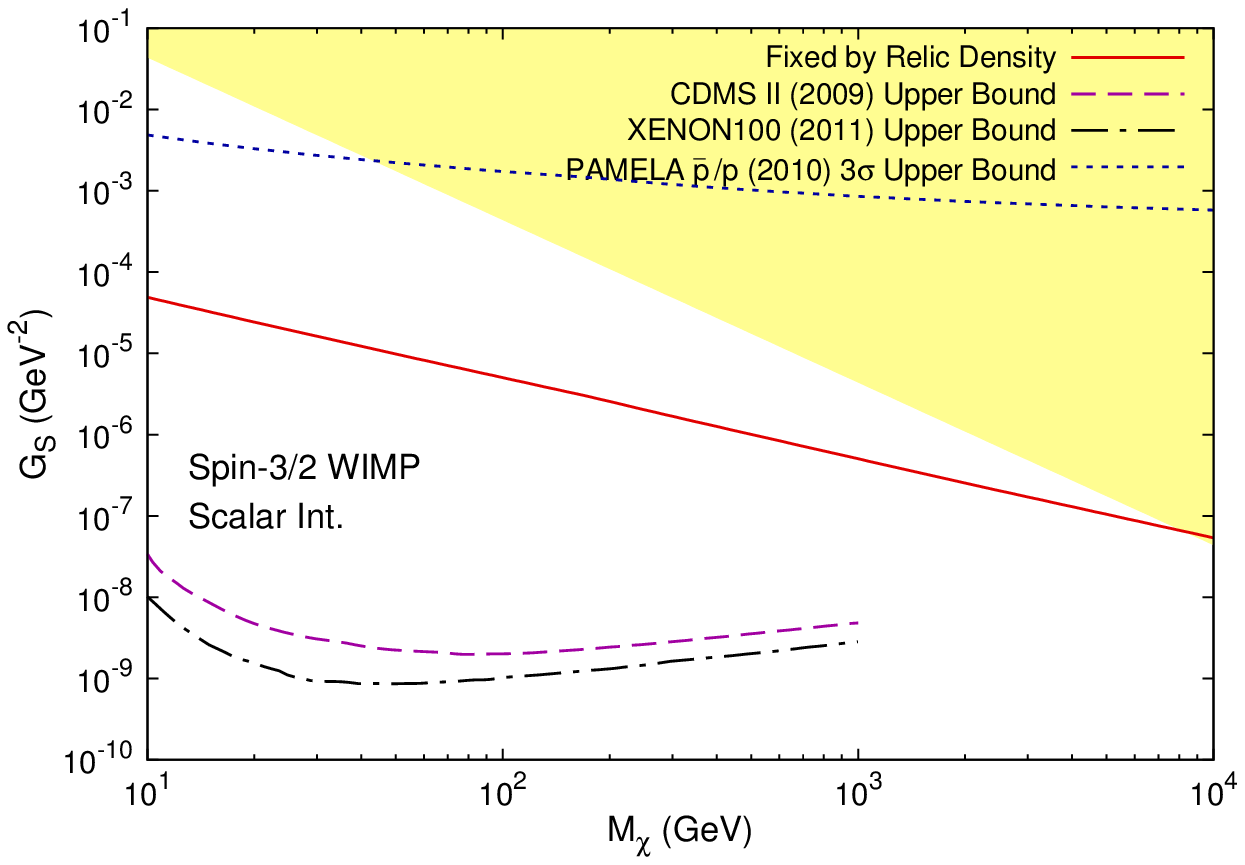}
\hspace{0.01\textwidth}%
\includegraphics[width=0.44\textwidth]{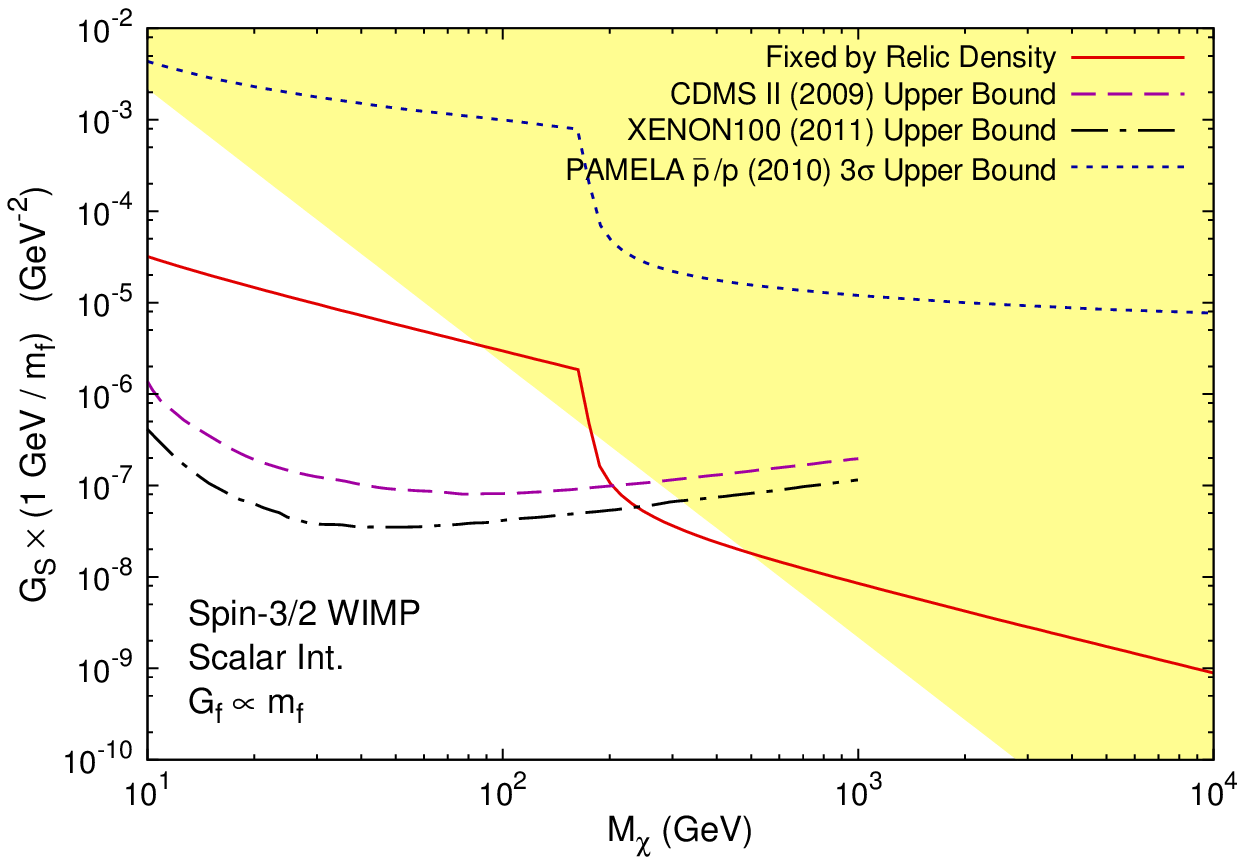}%
\caption{Combined constraints on coupling constants $G_f$ of spin-3/2
WIMPs with scalar (S) interaction from relic density~\cite{Komatsu:2010fb},
direct detection experiments of CDMS II~\cite{Ahmed:2009zw}
and XENON100~\cite{Aprile:2011hi},
PAMELA $\bar p / p$ ratio~\cite{Adriani:2010rc}, and
validity of effective theory. The yellow region denotes the invalid parameter
space of effective field theory. The left (right) frame is shown for
the case of universal couplings ($G_f\propto m_f$).
The constraints from relic density, $\bar{p}/p$ ratio
and validity of effective theory for scalar-pseudoscalar (SP) interaction
are very similar to those for scalar interaction, but direct detection
experiments are not sensitive to SP interaction.}
\label{fig-3/2-combined-S}
\end{figure}
\begin{figure}[!htbp]
\centering
\includegraphics[width=0.44\textwidth]{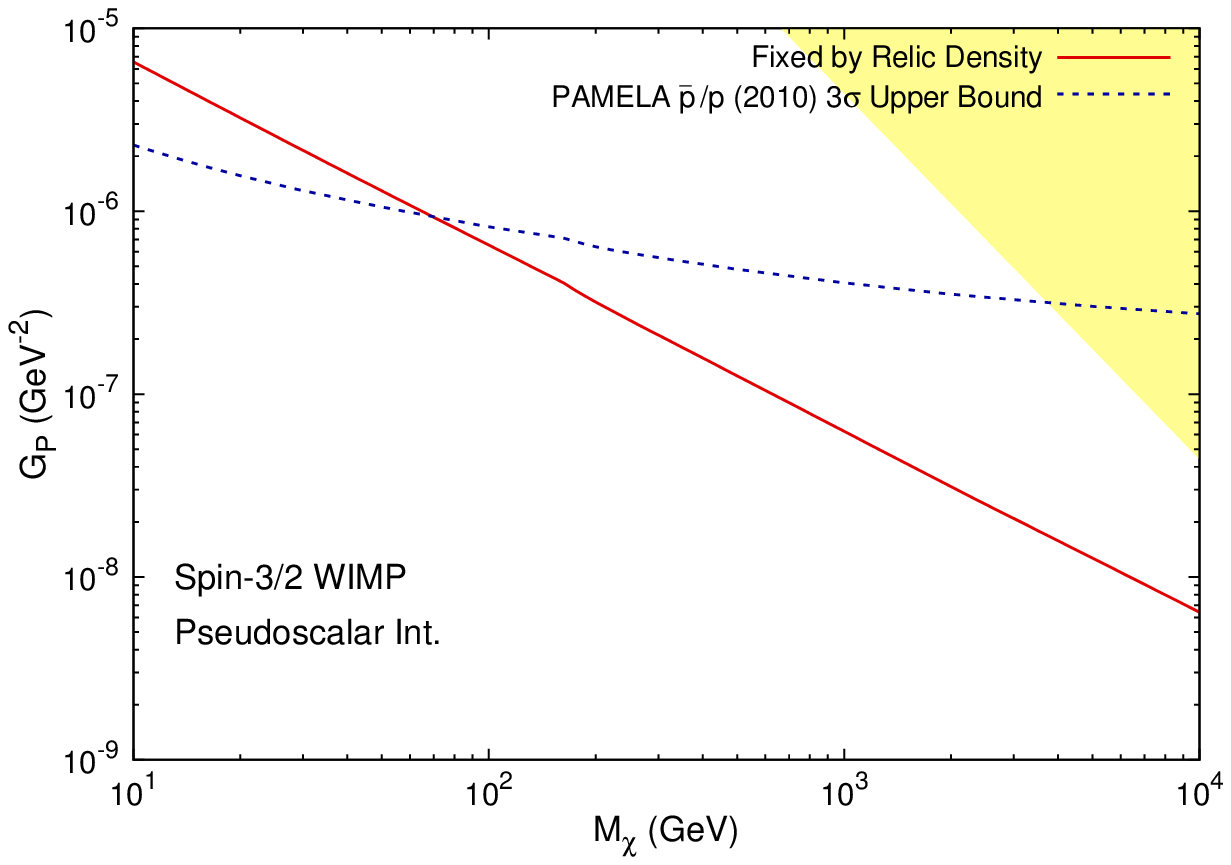}
\hspace{0.01\textwidth}%
\includegraphics[width=0.44\textwidth]{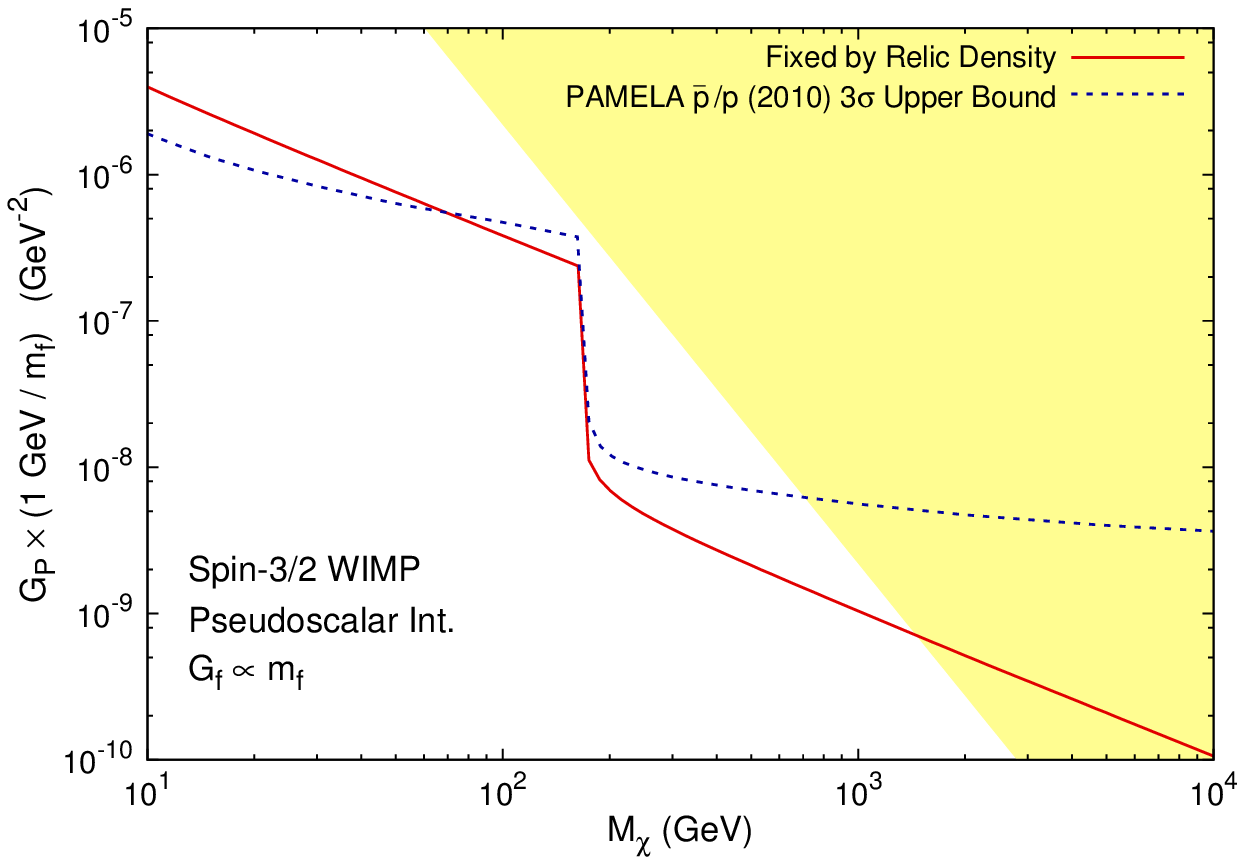}%
\caption{Combined constraints on coupling constants $G_f$ of spin-3/2
WIMPs with pseudoscalar (P) interaction from relic density,
PAMELA $\bar p / p$ ratio, and
validity of effective theory. The constraints from relic density, $\bar{p}/p$
ratio and validity of effective theory for pseudoscalar-scalar (PS)
interaction are very similar to those for pseudoscalar interaction.}
\label{fig-3/2-combined-P}
\end{figure}
\begin{figure}[!htbp]
\centering
\includegraphics[width=0.44\textwidth]{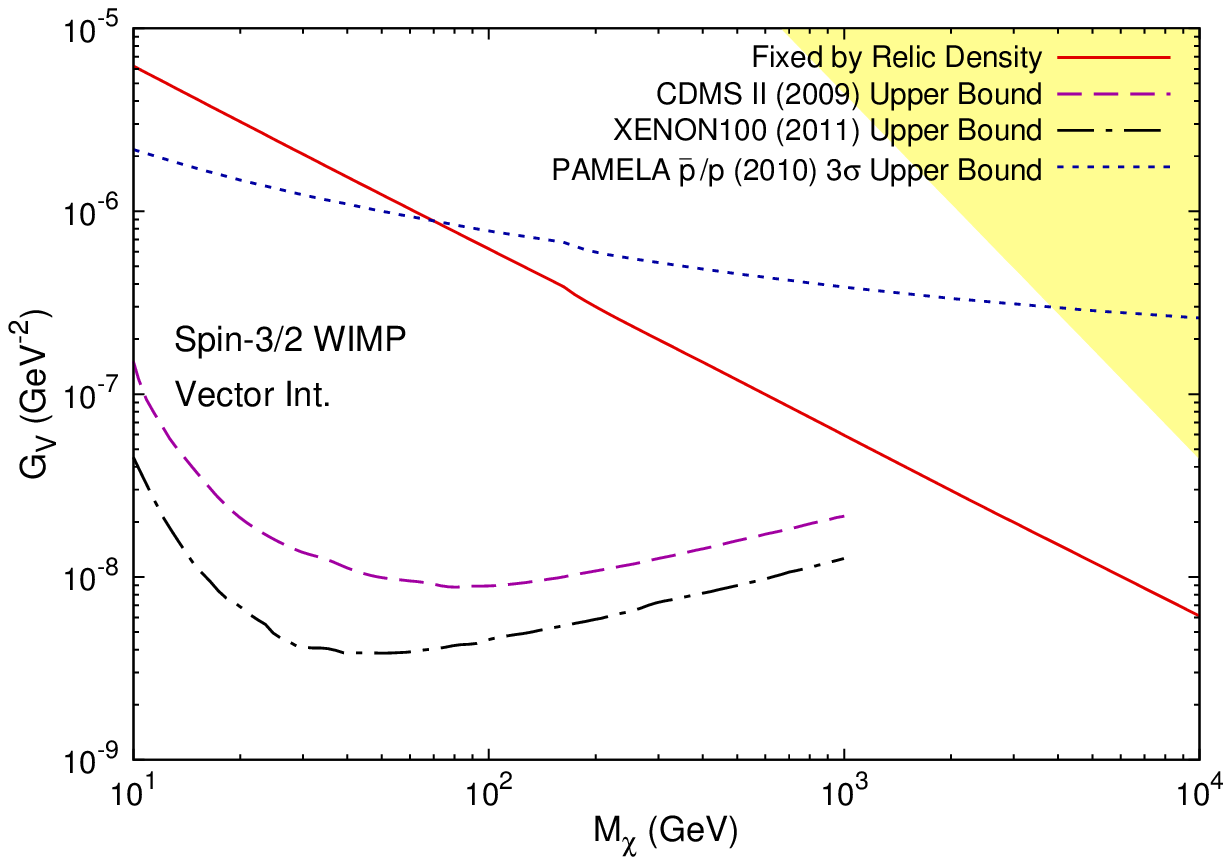}
\hspace{0.01\textwidth}%
\includegraphics[width=0.44\textwidth]{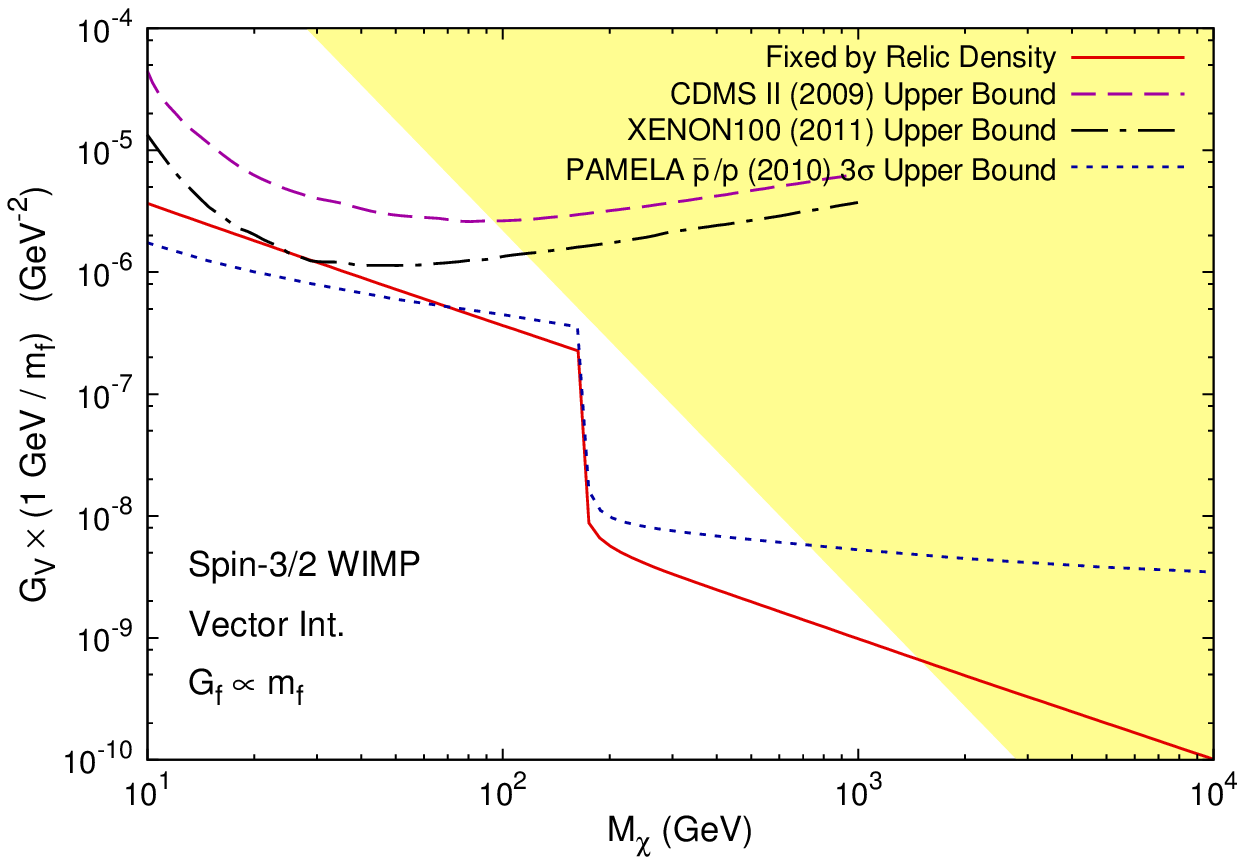}%
\caption{Combined constraints on coupling constants $G_f$ of spin-3/2
WIMPs with vector (V) interaction from relic density, direct detection
experiments of CDMS II and XENON100, PAMELA $\bar p / p$ ratio, and
validity of effective theory. The constraints from relic density, $\bar{p}/p$
ratio and validity of effective theory for vector-axialvector (VA) interaction
are very similar to those for vector interaction, but direct detection
experiments are not sensitive to VA interaction.}
\label{fig-3/2-combined-V}
\end{figure}
\begin{figure}[!htbp]
\centering
\includegraphics[width=0.44\textwidth]{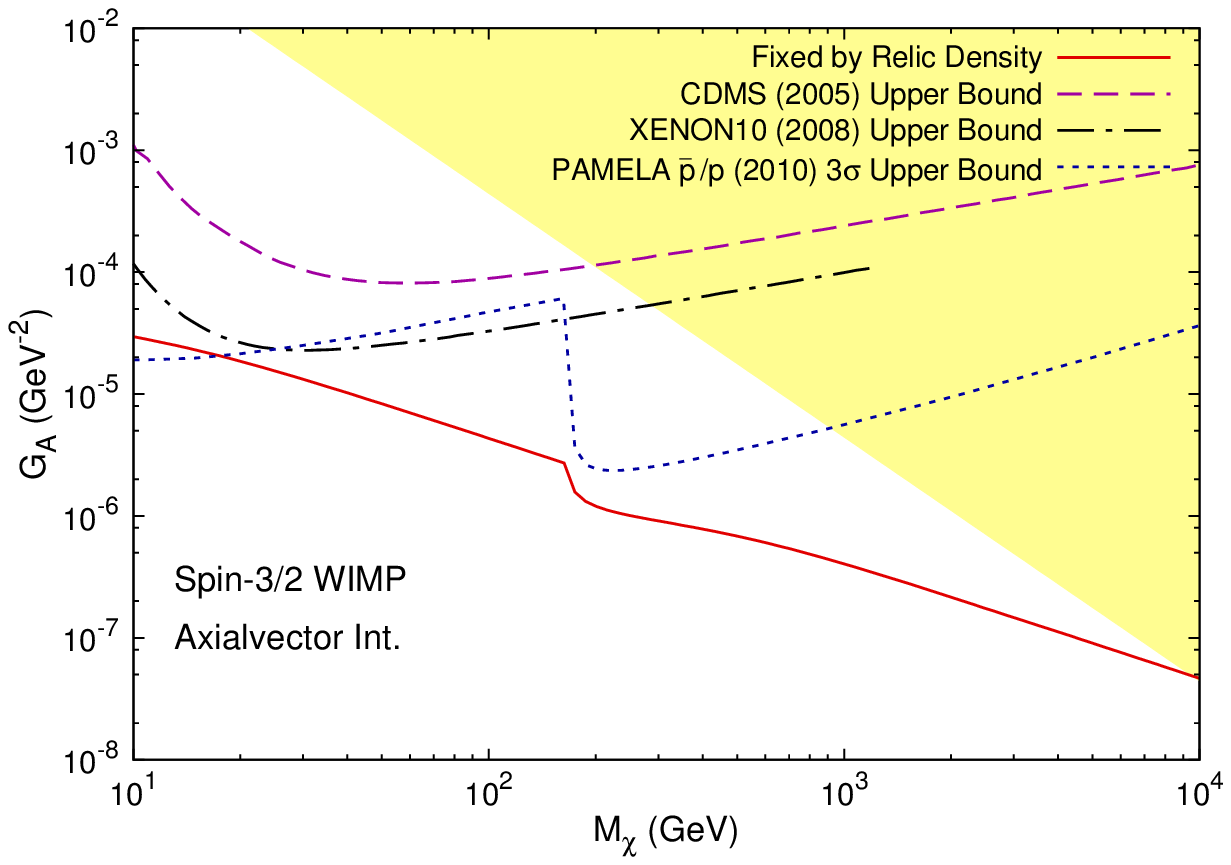}
\hspace{0.01\textwidth}%
\includegraphics[width=0.44\textwidth]{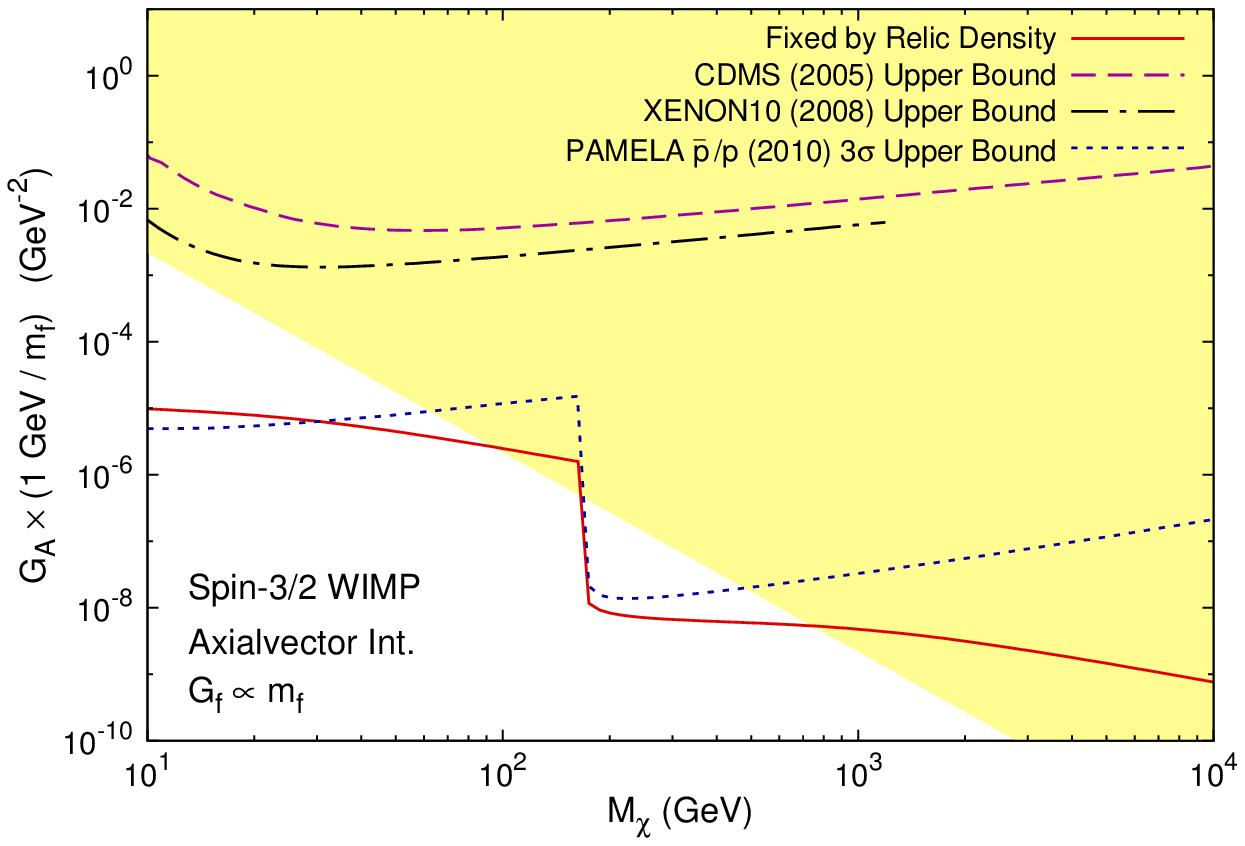}%
\caption{Combined constraints on coupling constants $G_f$ of spin-3/2
WIMPs with axialvector (A) interaction from relic density, direct detection
experiments of CDMS~\cite{Akerib:2005za} and XENON10~\cite{Angle:2008we},
PAMELA $\bar p / p$ ratio, and validity of effective theory.}
\label{fig-3/2-combined-A}
\end{figure}
\begin{figure}[!htbp]
\centering
\includegraphics[width=0.44\textwidth]{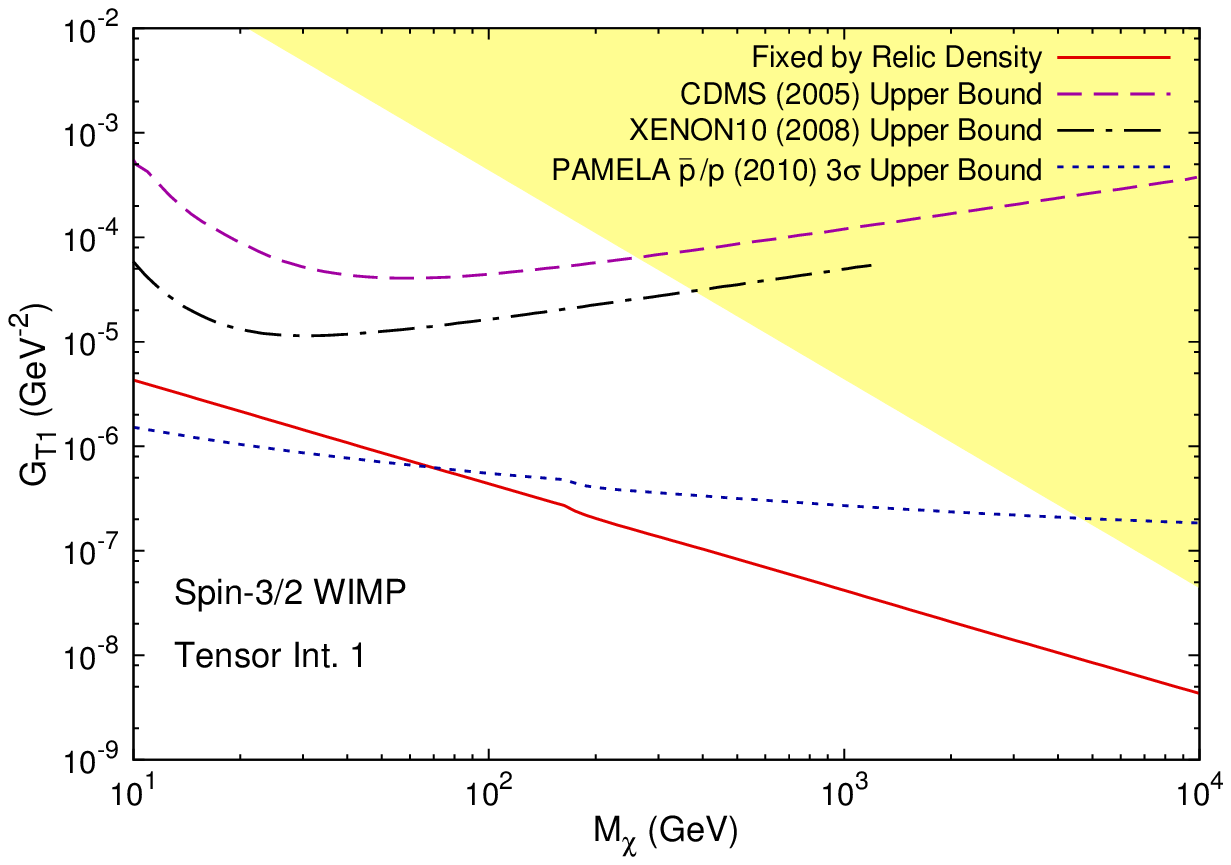}
\hspace{0.01\textwidth}%
\includegraphics[width=0.44\textwidth]{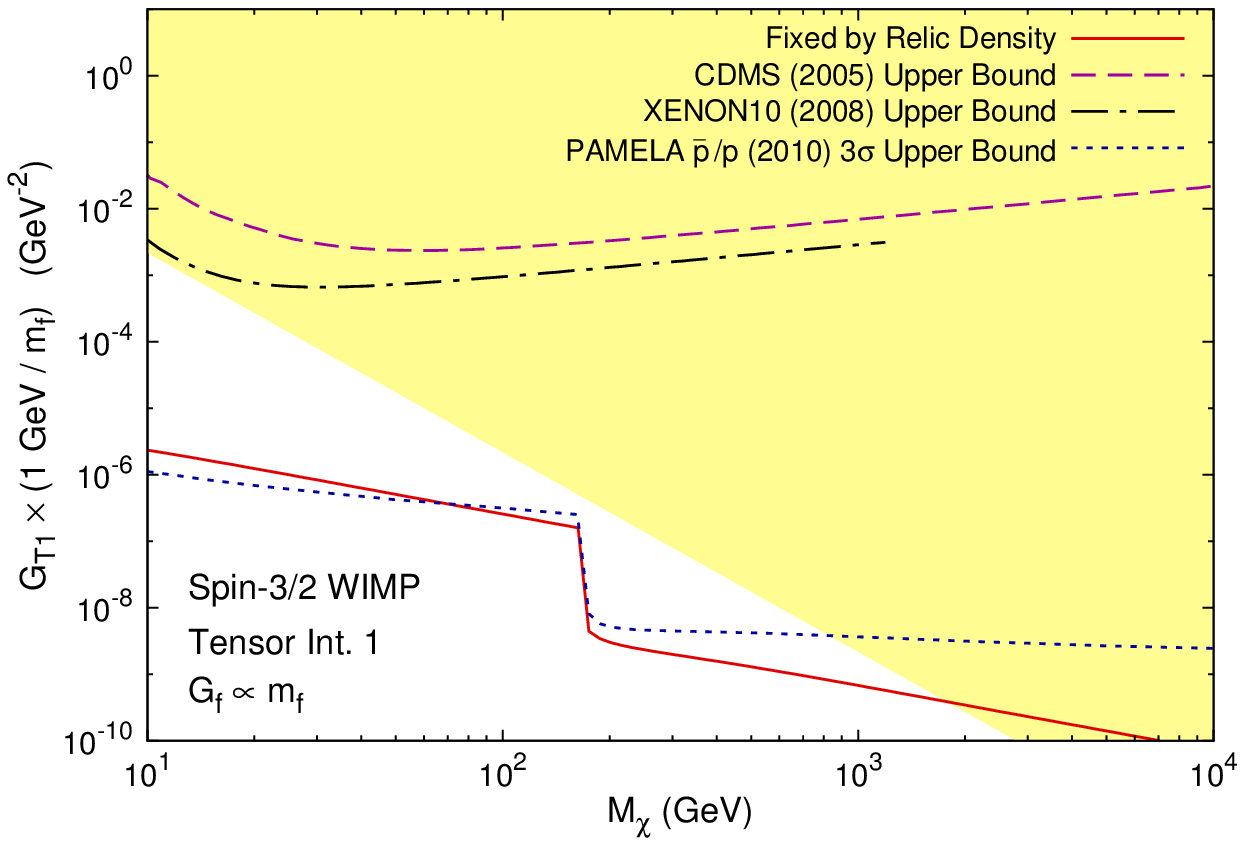}%
\caption{Combined constraints on coupling constants $G_f$ of spin-3/2
WIMPs with tensor interaction 1 (T1) from relic density, direct detection
experiments of CDMS and XENON10,
PAMELA $\bar p / p$ ratio, and validity of effective theory.
The constraints from relic density, $\bar{p}/p$ ratio
and validity of effective theory for tensor interaction 5 (T5)
are very similar to those for T1 interaction, but direct detection
experiments are not sensitive to T5 interaction.}
\label{fig-3/2-combined-T1}
\end{figure}
\begin{figure}[!htbp]
\centering
\includegraphics[width=0.44\textwidth]{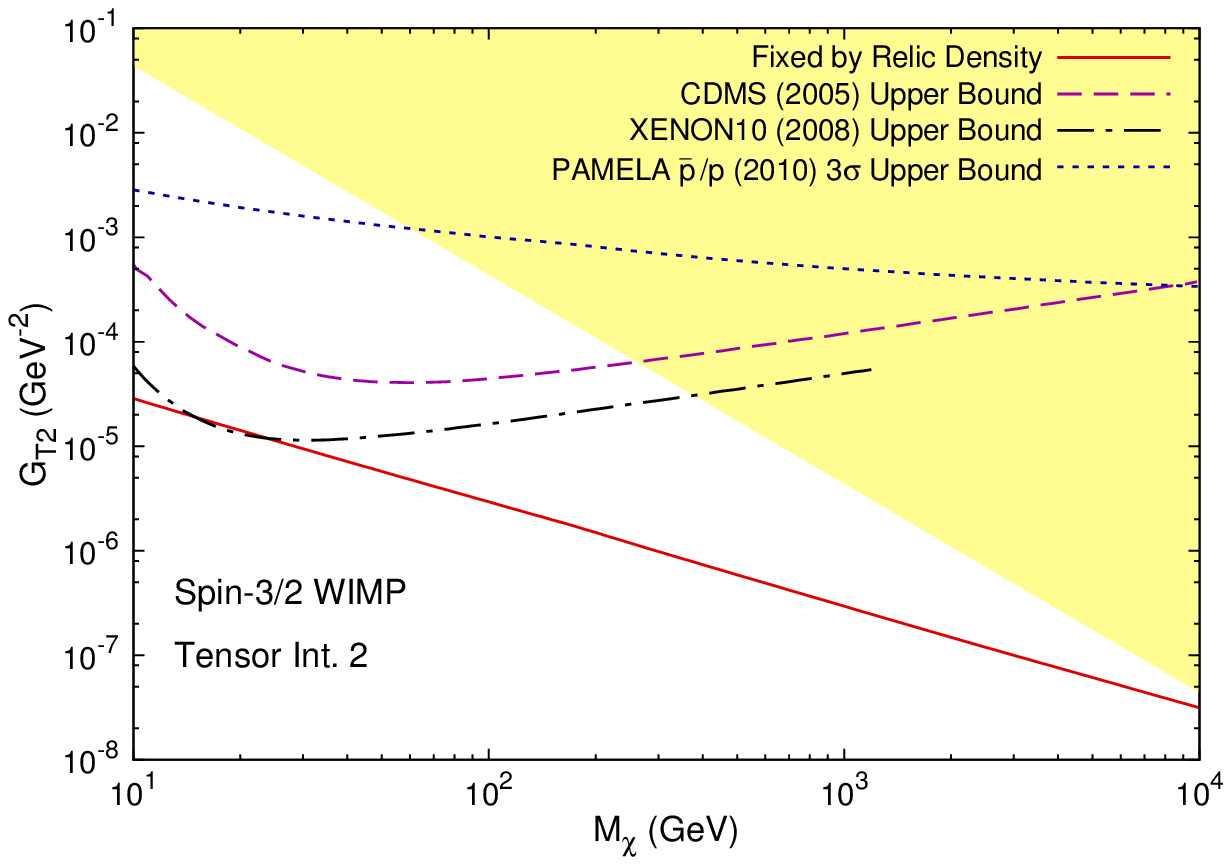}
\hspace{0.01\textwidth}%
\includegraphics[width=0.44\textwidth]{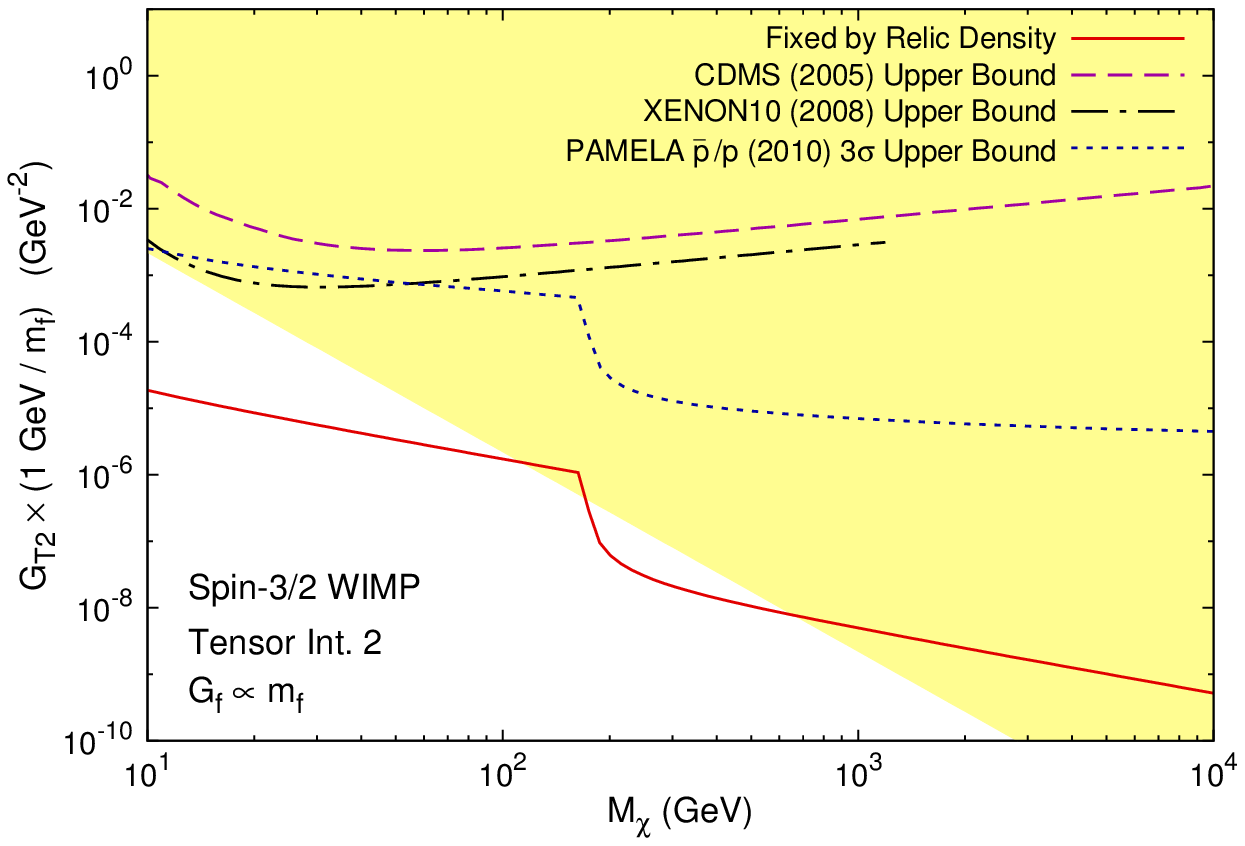}%
\caption{Combined constraints on coupling constants $G_f$ of spin-3/2
WIMPs with tensor interaction 2 (T2) from relic density, direct detection
experiments of CDMS and XENON10,
PAMELA $\bar p / p$ ratio, and validity of effective theory.}
\label{fig-3/2-combined-T2}
\end{figure}
\begin{figure}[!htbp]
\centering
\includegraphics[width=0.44\textwidth]{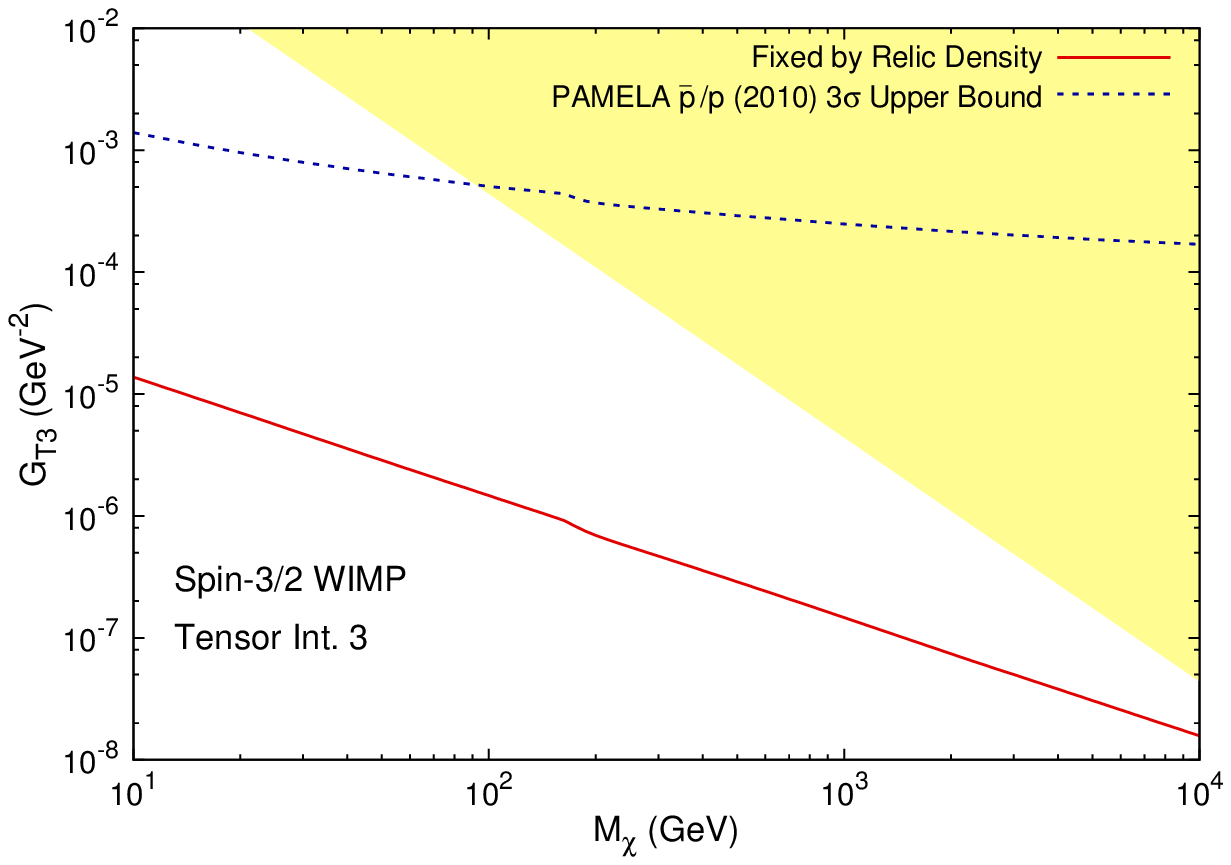}
\hspace{0.01\textwidth}%
\includegraphics[width=0.44\textwidth]{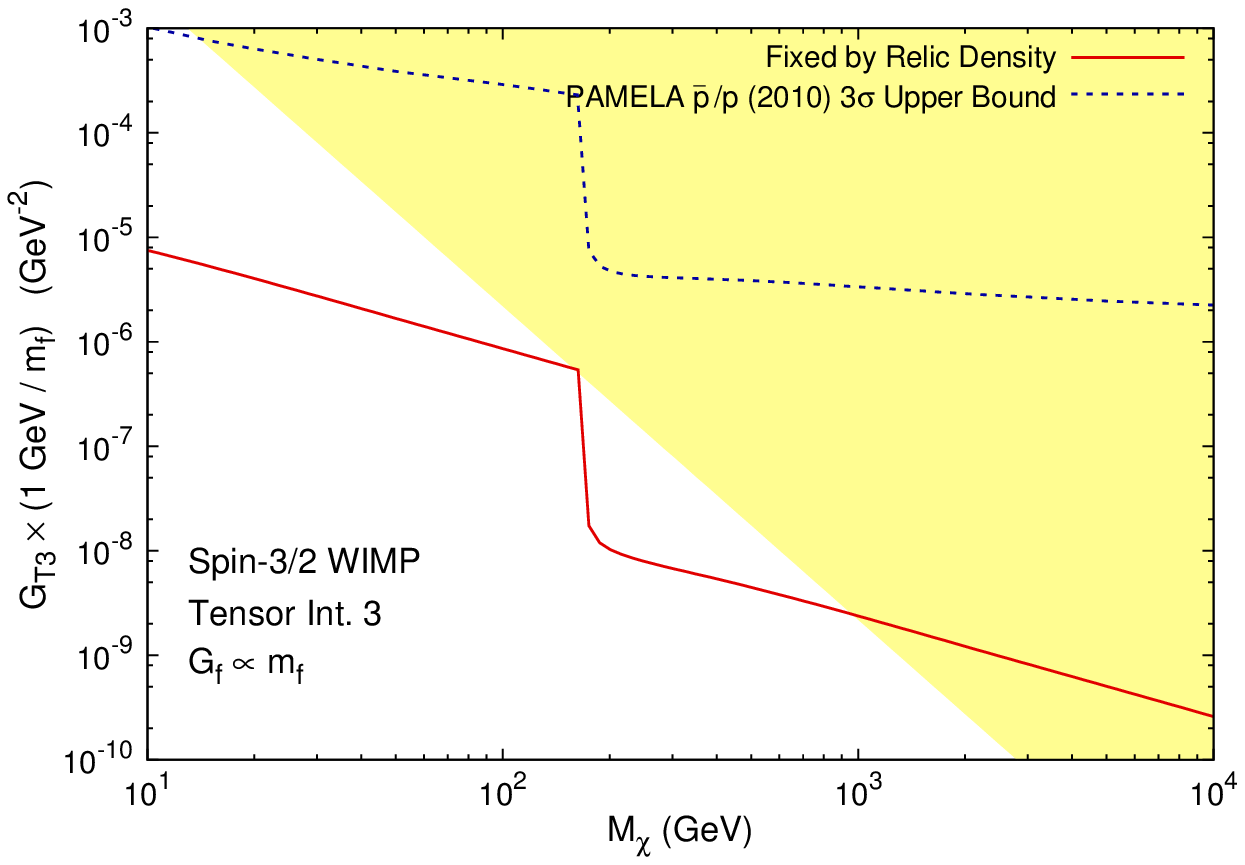}%
\caption{Combined constraints on coupling constants $G_f$ of spin-3/2
WIMPs with tensor interaction 3 (T3) from relic density,
PAMELA $\bar p / p$ ratio, and validity of effective theory.}
\label{fig-3/2-combined-T3}
\end{figure}
\begin{figure}[!htbp]
\centering
\includegraphics[width=0.44\textwidth]{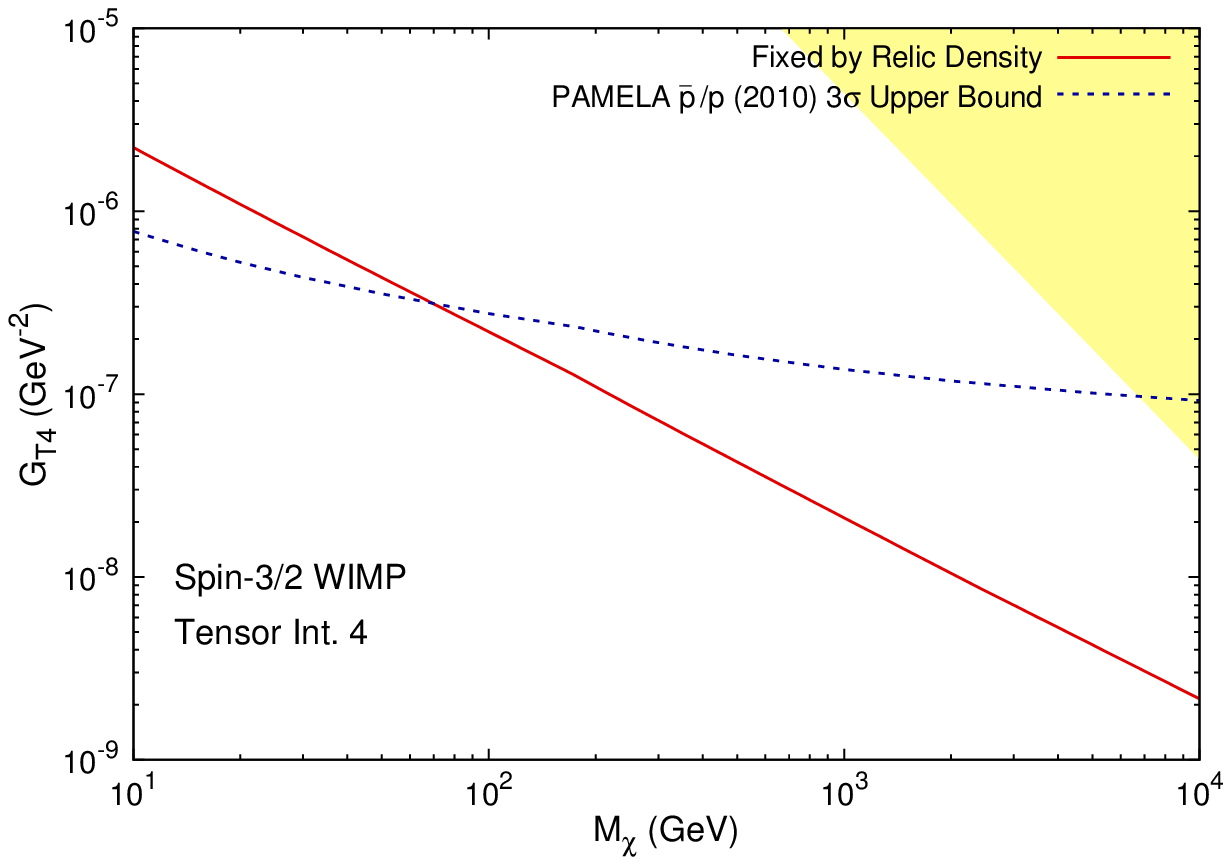}
\hspace{0.01\textwidth}%
\includegraphics[width=0.44\textwidth]{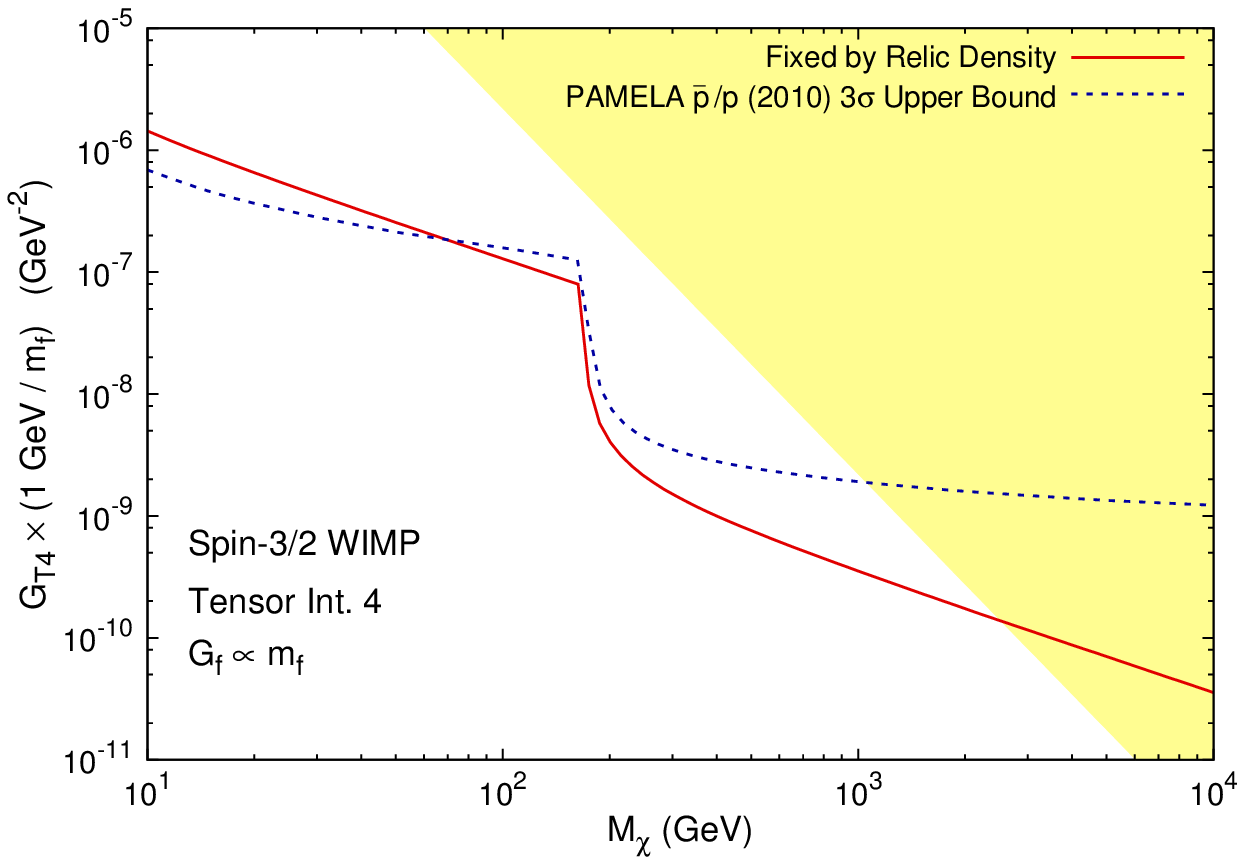}%
\caption{Combined constraints on coupling constants $G_f$ of spin-3/2
WIMPs with tensor interaction 4 (T4) from relic density,
PAMELA $\bar p / p$ ratio, and validity of effective theory.
The constraints from relic density, $\bar{p}/p$ ratio
and validity of effective theory for tensor interaction 6 (T6)
are very similar to those for T4 interaction.}
\label{fig-3/2-combined-T4}
\end{figure}
\begin{figure}[!htbp]
\centering
\includegraphics[width=0.44\textwidth]{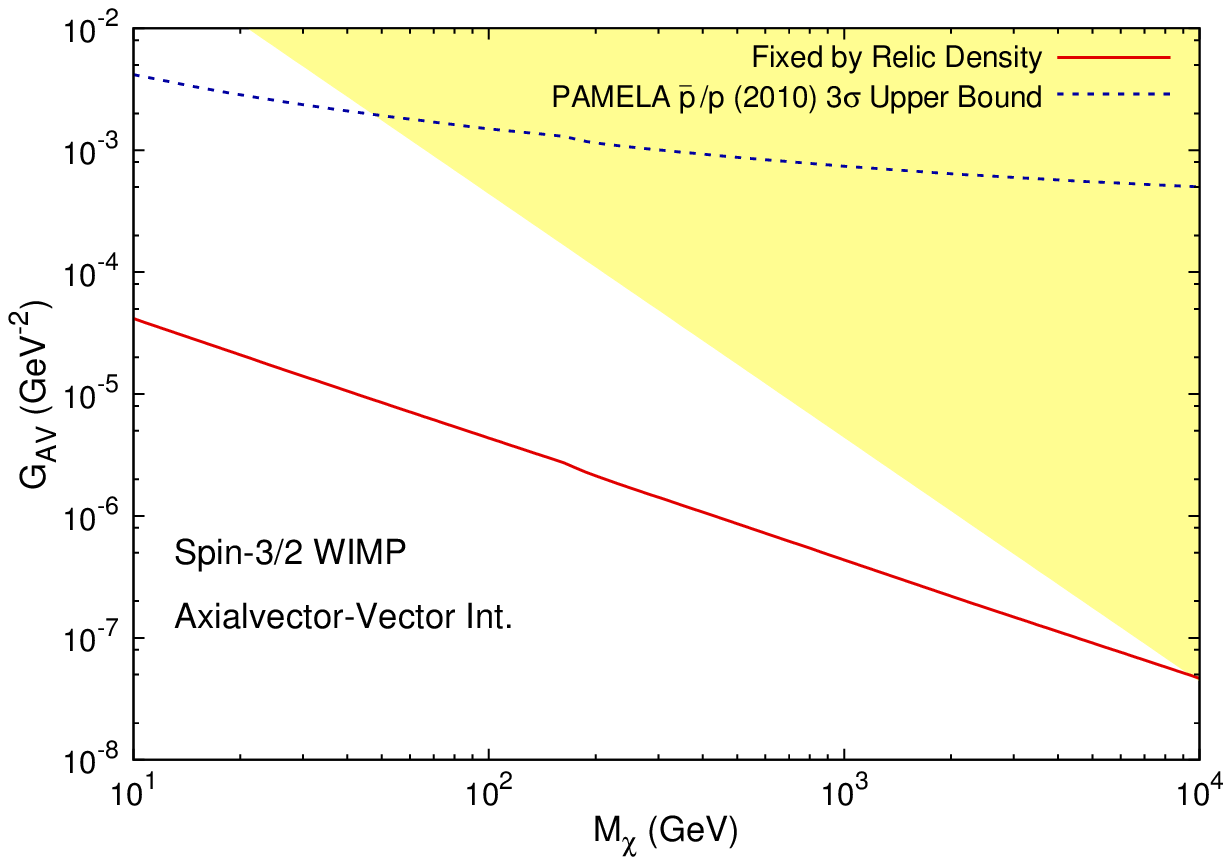}
\hspace{0.01\textwidth}%
\includegraphics[width=0.44\textwidth]{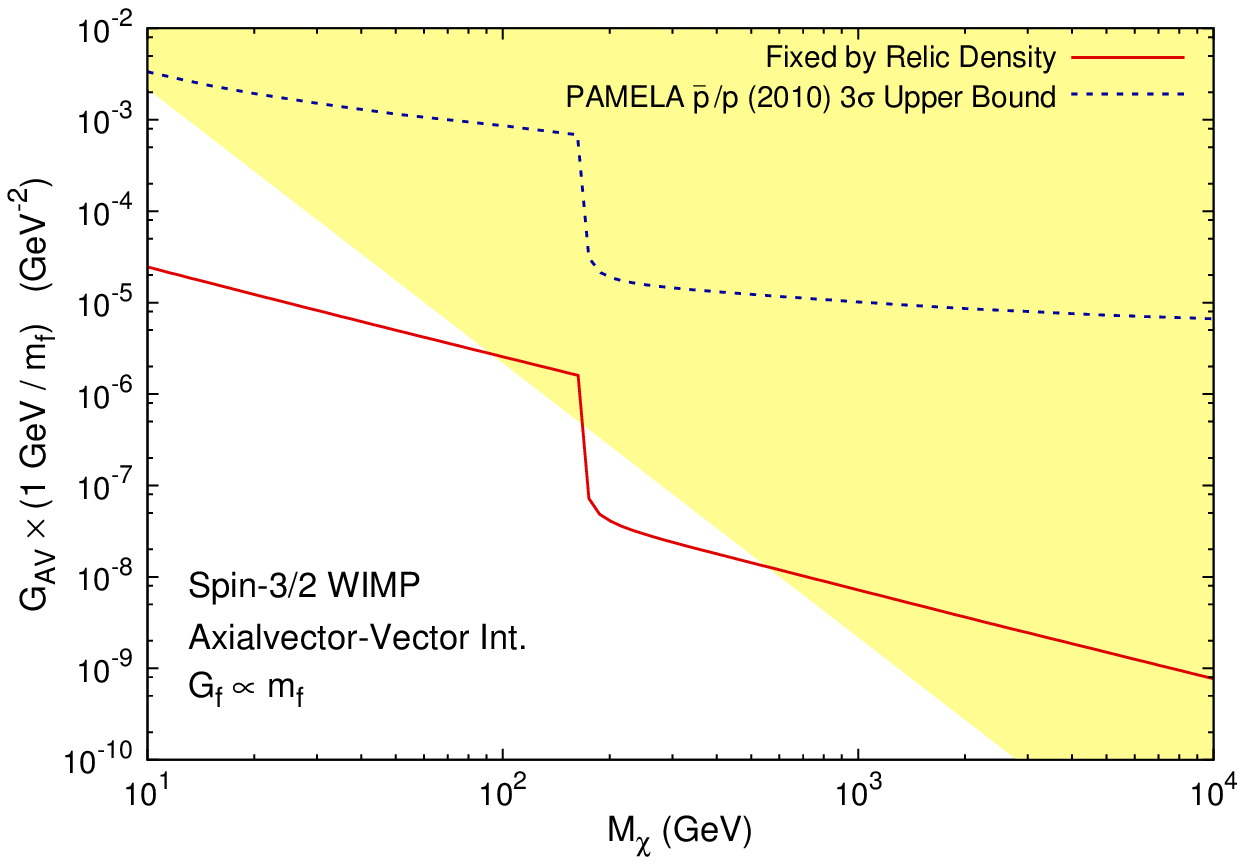}%
\caption{Combined constraints on coupling constants $G_f$ of spin-3/2
WIMPs with axialvector-vector (AV) interaction from relic density,
PAMELA $\bar p / p$ ratio, and validity of effective theory.}
\label{fig-3/2-combined-AV}
\end{figure}

\newpage

The combined constraints on coupling constants $G_f$ of the spin-3/2
WIMP effective models of S, P, V, A, T1, T2, T3, T4 and AV interactions
from relic density~\cite{Komatsu:2010fb},
direct detection experiments of CDMS~\cite{Ahmed:2009zw,Akerib:2005za}
and XENON~\cite{Aprile:2011hi,Angle:2008we},
and PAMELA $\bar p / p$ ratio~\cite{Adriani:2010rc}
are shown in Figs.~\ref{fig-3/2-combined-S} -- \ref{fig-3/2-combined-AV},
with the yellow regions denoting the invalid parameter spaces of
effective field theory.
The constraints on SP, PS, T5, T6 and VA interactions are very
similar to those on S, P, T1, T4 and V interactions respectively,
except that the SP, T5 and VA interactions
are not constrained by direct detection experiments.
As stated above, the constraints on S, P, V, A, T1, T4, SP, PS, VA and AV
interactions are much similar to those on the analogous interactions
of Dirac fermionic WIMPs in our previous work~\cite{Zheng:2010js},
respectively.
In this work, however, we adopt the latest constraint of XENON100 SI direct
detection \cite{Aprile:2011hi}, which is stronger than that adopted
in our previous work~\cite{Zheng:2010js}. Thus the constraints on
S and V interactions of spin-3/2 WIMPs are more stringent than
those for Dirac fermionic WIMPs we calculated before.
On the other hand, the T2, T3, T5 and T6 interactions have no analogue of
Dirac fermionic WIMPs. From Figs.~\ref{fig-3/2-combined-T2} and
\ref{fig-3/2-combined-T3}, we can see that the direct detection constraints
on T2 and T3 interactions are so loose that only a little region
of $M_\chi$ is excluded by SD direct detection in the case of
T2 interaction with universal couplings.
In addition, the constraints by PAMELA $\bar p / p$ ratio exclude
the mass regions of $M_\chi \lesssim 70$~GeV for T5 and T6 interactions,
while it is not sensitive to T2 and T3 interactions
because their annihilation rates \eqref{sv_3/2_T2} and \eqref{sv_3/2_T3}
are both of order $T/M_\chi$.
As a summary of the study on the spin-3/2 WIMP,
in Table \ref{tab:3/2_sum}, the excluded regions of $M_\chi$ given by direct
and indirect detection experiments for spin-3/2 fermionic WIMPs with
various effective interactions are shown.

\begin{table}[!htb]
\begin{center}
\belowcaptionskip=0.2cm
\caption{A summary for spin-3/2 WIMPs with various effective interactions.
The excluded regions of $M_\chi$ given by direct and indirect experiments
are indicated.} \label{tab:3/2_sum}
\renewcommand{\arraystretch}{1.3}
\footnotesize
\begin{tabular}{ccc}
\hline \hline \multicolumn{3}{c}{Universal couplings} \\
Interaction & Direct detection & PAMELA $\bar p / p$ \\
\hline
S  & Excluded $M_\chi \simeq 10~\mathrm{GeV} - \mathrm{above}~1~\mathrm{TeV}$
   & Not sensitive \\
P  & Not sensitive & Excluded $M_\chi \simeq 10 - 70~\mathrm{GeV}$ \\
V  & Excluded $M_\chi \simeq 10~\mathrm{GeV} - \mathrm{above}~1~\mathrm{TeV}$
   & Excluded $M_\chi \simeq 10 - 70~\mathrm{GeV}$ \\
A  & Not sensitive & Excluded $M_\chi \simeq 10 - 17~\mathrm{GeV}$ \\
T1 & Not sensitive & Excluded $M_\chi \simeq 10 - 70~\mathrm{GeV}$ \\
T2 & Excluded $M_\chi \simeq 15 - 24~\mathrm{GeV}$ & Not sensitive \\
T3 & Not sensitive & Not sensitive \\
T4 & Not sensitive & Excluded $M_\chi \simeq 10 - 70~\mathrm{GeV}$ \\
T5 & Not sensitive & Excluded $M_\chi \simeq 10 - 70~\mathrm{GeV}$ \\
T6 & Not sensitive & Excluded $M_\chi \simeq 10 - 70~\mathrm{GeV}$ \\
SP & Not sensitive & Not sensitive \\
PS & Not sensitive & Excluded $M_\chi \simeq 10 - 70~\mathrm{GeV}$ \\
VA & Not sensitive & Excluded $M_\chi \simeq 10 - 70~\mathrm{GeV}$ \\
AV & Not sensitive & Not sensitive \\
\hline \hline \multicolumn{3}{c}{$G_f \propto m_f$} \\
Interaction & Direct detection & PAMELA $\bar p / p$ \\
\hline
S  & Excluded $M_\chi \simeq 10 - 240~\mathrm{GeV}$ & Not sensitive \\
P  & Not sensitive & Excluded $M_\chi \simeq 10 - 70~\mathrm{GeV}$ \\
V  & Excluded $M_\chi \simeq 25 - 29~\mathrm{GeV}$
   & Excluded $M_\chi \simeq 10 - 70~\mathrm{GeV}$ \\
A  & Not sensitive & Excluded $M_\chi \simeq 10 - 30~\mathrm{GeV}$ \\
T1 & Not sensitive & Excluded $M_\chi \simeq 10 - 70~\mathrm{GeV}$ \\
T2 & Not sensitive & Not sensitive \\
T3 & Not sensitive & Not sensitive \\
T4 & Not sensitive & Excluded $M_\chi \simeq 10 - 70~\mathrm{GeV}$ \\
T5 & Not sensitive & Excluded $M_\chi \simeq 10 - 70~\mathrm{GeV}$ \\
T6 & Not sensitive & Excluded $M_\chi \simeq 10 - 70~\mathrm{GeV}$ \\
SP & Not sensitive & Not sensitive \\
PS & Not sensitive & Excluded $M_\chi \simeq 10 - 70~\mathrm{GeV}$ \\
VA & Not sensitive & Excluded $M_\chi \simeq 10 - 70~\mathrm{GeV}$ \\
AV & Not sensitive & Not sensitive \\
\hline \hline
\end{tabular}
\end{center}
\end{table}

\section{Conclusions\label{sec-con}}

In this work, we give a general analysis of the 4-particle interaction
between SM particles and DM which consists of scalar,
vector or spin-3/2 WIMPs.
The most general forms of the 4-particle operators
up to dimension 6 have been considered.
We find that for scalar, vector and spin-3/2 DM, the constraints from DM relic
density, DM direct and indirect detection are complementary to each other.
Thus the comparison among different kinds of experimental results gives us
a complete picture about the current DM searches.
In general, the constraints from SI direct detection are the most stringent,
while those from SD direct detection are quite weak. On the other hand,
for light DM (whose mass $\lesssim 70$~GeV) the cosmic-ray $\bar p/p$
data can be more sensitive than direct detection or relic density
in some cases.

Assuming one operator dominates the effective interaction between
DM and SM fermions, we find that in some cases the constraints
are so strong that the Universe will be overclosed by DM thermal production.
If the standard cosmology is still retained, the DM models in such cases
should be excluded, which are indicated in Tables \ref{tab:scalar_sum},
\ref{tab:vector_sum} and \ref{tab:3/2_sum}.
In the case of scalar DM, recent direct detection
experiments exclude some $M_\phi$ regions for S and V interactions,
while the PAMELA $\bar p/p$ ratio excludes some small $M_\phi$ regions
($\lesssim 70$~GeV) for S and SP interactions.
In the case of vector DM, recent direct detection experiments exclude
some $M_X$ regions for S and V interactions,
and for $\widetilde{\mathrm{VA}}$ interaction only with universal couplings.
The PAMELA $\bar p/p$ ratio, however, excludes some small $M_X$ regions
($\lesssim 70$~GeV) for S, T, SP and $\widetilde{\mathrm{T}}$ interactions,
and most of these interactions cannot be excluded by direct detection.
In the case of spin-3/2 DM, the constraints on most interactions
(S, P, V, A, T1, T4, SP, PS, VA and AV)
are much similar to those on their analogous interactions of
Dirac fermionic DM in our previous work \cite{Zheng:2010js}.
Besides, the constraints on T2 and T3 interactions
are so weak that only direct detection can exclude a little region
of $M_\chi$ in the case of T2 interaction with universal couplings.
For T5 and T6 interactions, some small $M_\chi$ regions
($\lesssim 70$~GeV) are excluded by the PAMELA $\bar p/p$ ratio.
Among the scalar, vector and spin-3/2 DM effective models,
there are still some effective interactions to which
the recent DM direct and indirect search experiments are not sensitive at all.

\begin{acknowledgments}
This work is supported in part by the National Natural Science Foundation
of China (NSFC) under Grant Nos. 10773011, 11005163, 11074310 and 11075169, the
Specialized Research Fund for the Doctoral Program of Higher
Education (SRFDP) under Grant No. 200805581030, the Fundamental
Research Funds for the Central Universities, the 973 project under Grant
No. 2010CB833000, the Project of Knowledge Innovation Program (PKIP)
of Chinese Academy of Sciences under Grant No. KJCX2.YW.W10, and Sun Yat-Sen
University Science Foundation.
\end{acknowledgments}

\end{document}